\documentclass[11pt, oneside]{article}
\pdfoutput=1

\usepackage{jheppub}

\usepackage{amsmath}
\usepackage{amssymb}
\usepackage{amsfonts}
\usepackage{array}
\usepackage{booktabs}
\usepackage{placeins}
\usepackage{float}

\newenvironment{lyxcode}
	{\par\begin{list}{}{
		\setlength{\rightmargin}{\leftmargin}
		\setlength{\listparindent}{0pt}
		\raggedright
		\setlength{\itemsep}{0pt}
		\setlength{\parsep}{0pt}
		\normalfont\ttfamily}%
	 \item[]}
	{\end{list}}

\usepackage[usenames,dvipsnames]{xcolor}

\usepackage{subcaption}
\usepackage{adjustbox}

\newcommand\lr[3]{\left#1#2\right#3}
\newcommand\p[1]{\lr({#1})}
\newcommand\tl[1]{\widetilde{#1}}
\renewcommand\bar[1]{\overline{#1}}
\renewcommand\hat[1]{\widehat{#1}}
\newcommand\<{\langle}
\renewcommand\>{\rangle}
\newcommand\x{\times}
\newcommand\oo{\infty}
\newcommand\ptl{\partial}
\newcommand\half{\frac 1 2}
\newcommand\thalf{\tfrac 1 2}
\renewcommand\a{\alpha}
\renewcommand\b{\beta}
\newcommand\g{\gamma}
\newcommand\G{\Gamma}
\newcommand\de{\delta}
\newcommand\De{\Delta}
\newcommand\e{\epsilon}
\renewcommand\l{\lambda}
\renewcommand\L{\Lambda}
\renewcommand\r{\rho}
\newcommand\f{\phi}
\newcommand\s{\sigma}
\newcommand\w{\omega}
\renewcommand\th{\theta}
\newcommand\ka{\kappa}
\newcommand\R{\mathbb{R}}
\newcommand\Z{\mathbb{Z}}
\renewcommand\Re{\mathop{\mathrm{Re}}}
\newcommand\cA{\mathcal{A}}
\newcommand\cB{\mathcal{B}}
\newcommand\cC{\mathcal{C}}
\newcommand\cF{\mathcal{F}}
\newcommand\cK{\mathcal{K}}
\newcommand\cN{\mathcal{N}}
\newcommand\cO{\mathcal{O}}
\newcommand\cP{\mathcal{P}}
\newcommand\cT{\mathcal{T}}
\def\be#1\ee{\begin{align}#1\end{align}}
\newcommand\nn{\nonumber}

\newcommand{\vb}{\texttt{virasoro\_blocks}}
\newcommand{\vbs}{\texttt{virasoro\_bootstrap}}
\newcommand{\sdpb}{\texttt{SDPB}}
\newcommand{\gap}{\mathrm{gap}}
\newcommand{\hb}{\bar h}
\newcommand{\zb}{\bar z}
\newcommand{\qb}{\bar q}
\newcommand{\yb}{\bar y}
\newcommand{\vac}{\mathbf{1}}

\makeatletter
\def\@fpheader{\ }
\makeatother

\title{\vb: Software for the 2d bootstrap}

\author{David Poland}

\affiliation{Department of Physics, Yale University, New Haven, CT 06520, USA \\
}

\emailAdd{david.poland@yale.edu}

\abstract{We introduce \vb, software that computes Virasoro conformal blocks at any central charge in a form suitable for the numerical bootstrap. At each finite order in the elliptic nome, Zamolodchikov's recursion gives the block as a positive prefactor times a rational function of the exchanged weight. However, its many Kac poles make it difficult to use in semidefinite programming approaches to the bootstrap. In this software we use optimal interpolation to reduce the number of poles while keeping high accuracy. The software works in arbitrary precision and handles singular central charges, including the minimal models and $c=1$. It also computes global blocks, $\cN=1$ superconformal blocks, torus characters, and torus one-point blocks. We illustrate it with two applications. Assuming only scalar Virasoro primaries, we compute bounds on the scalar gap that approach the window of Liouville solutions between $(c-1)/16$ and $(c-1)/12$. We also bound the average exchanged dimension in four-point functions of identical scalars at several central charges. These bounds have a rich structure, and many known 2d CFTs lie on or near their boundary.} \date{}
\preprint{}

\begin{document}

\maketitle
\pagenumbering{roman}\setcounter{page}{2}

\newpage
\pagenumbering{arabic}
\section{Introduction}

The numerical conformal bootstrap~\cite{Rattazzi:2008pe} uses crossing symmetry and unitarity to constrain the spectrum and OPE coefficients of conformal field theories (CFTs); see~\cite{Poland:2018epd} for a review. These constraints can be formulated as semidefinite programs~\cite{Poland:2011ey,Simmons-Duffin:2015qma,Landry:2019qug}, which require accurate representations of conformal block derivatives over the allowed spectrum of exchanged dimensions. In $d>2$, general-purpose software such as \texttt{scalar\_blocks}~\cite{scalarblocks} and \texttt{blocks\_3d}~\cite{Erramilli:2020rlr} supplies this input and has made such computations widely accessible.

In two dimensions, the enhanced local conformal symmetry organizes four-point functions into Virasoro blocks, each accounting for an entire Virasoro representation~\cite{Belavin:1984vu}. For torus partition functions, the modular bootstrap~\cite{Hellerman:2009bu,Friedan:2013cba,Collier:2016cls,Afkhami-Jeddi:2019zci,Hartman:2019pcd,Benjamin:2026lbj} uses characters known in closed form. On the sphere, Zamolodchikov's recursion relation~\cite{Zamolodchikov:1984eqp,Zamolodchikov:1987avt} computes the blocks efficiently as a power series in the elliptic nome $q$, with coefficients rational in the exchanged weight $h$. Numerical studies using these blocks include non-unitary theories~\cite{Esterlis:2016psv,Huang:2025qkk} and Potts and loop models~\cite{Picco:2016ilr,He:2020rfk,Nivesvivat:2023kfp}. Their use for rigorous bounds through convex optimization remains comparatively limited, with applications to superconformal spectra~\cite{Lin:2015wcg,Lin:2016gcl} and spectral functions~\cite{Collier:2017shs}. Related bounds at generic $c>1$ follow from positivity of the pillow expansion~\cite{Besken:2021eps}.

The obstacle for Virasoro blocks is the number and location of the Kac poles. Semidefinite programming requires block derivatives as polynomials in the exchanged dimension times a positive prefactor. At order $N$ in the nome, the elliptic part $H$ is rational in $h$, with candidate poles at the degenerate Kac weights $h_{r,s}$ for $rs\leq N$ (see~\cite{Ribault:2014hia,Ribault:2024rvk} for reviews,~\cite{Perlmutter:2015iya} for closed forms, and~\cite{Cho:2017oxl} for recursions on the sphere and torus). The number of poles grows like $N\log N$, so the resulting polynomials have degree in the hundreds at $N\gtrsim60$. In $d>2$ the poles are real and sparse, and ``pole shifting''~\cite{Kos:2013tga} retains only a handful. For $1<c<25$ most Kac poles are complex, and there are many of them with no natural ordering. Keeping every pole leads to ill-conditioned bases. In~\cite{Collier:2017shs} this limited the semidefinite programs to low derivative orders, since derivative order $25$ already requires the $q$-expansion to order $100$ and polynomials of degree near $700$.

Optimal interpolation, introduced in~\cite{Chang:2025mwt}, replaces the exact rational function by a low-degree interpolant sampled at near-optimal nodes. For Virasoro blocks, a few dozen well-chosen poles reproduce the leading Taylor coefficient to $10^{-40}$ or better (section~\ref{sec:interp}), at polynomial degrees comparable to those used in three dimensions. We implement this approach in \vb.

Concrete targets for a Virasoro bootstrap have recently emerged. The infrared fixed points of $N$ coupled unitary minimal models at large $m$~\cite{Antunes:2022vtb,Antunes:2024mfb}, building on~\cite{Ludwig:1987gs,Dotsenko:1998gyp,Fendley:2008fp} and generalized to other finite groups in~\cite{Antunes:2025erb}, are expected to be compact, unitary, irrational CFTs whose only chiral algebra is Virasoro, with a positive twist gap and an $S_N$ symmetry, at central charges just below $c=N$. A lattice realization of a candidate theory of this kind, $N\geq3$ coupled Fibonacci anyon chains, was proposed in~\cite{Antunes:2025huk}.

The main output of \vb\ is a table of derivatives around the crossing-symmetric point $z=\zb=\half$. Given a central charge $c$, external weights $(h_i,\hb_i)$, spins $\ell$, and a derivative cutoff $\L$, it approximates the Taylor coefficients in the ``damped rational'' form
\be
\frac{1}{m!\,n!}\ptl_z^m\ptl_{\zb}^n G_{\De,\ell}(z,\zb)\Big|_{z=\zb=\half} \approx \chi_\ell(\De)\, P_{\ell}^{m,n}(\De),\qquad \chi_\ell(\De)=\frac{e^{-\pi \De}}{\prod_{p}(\De-\De_{p})},
\ee
where $G_{\De,\ell}$ is the block of a scalar four-point function in the default normalization of section~\ref{sec:preliminaries}, $P^{m,n}_\ell$ are polynomials, and the $\De_p$ are retained Kac poles $2h_{r,s}\pm\ell$. The interpolation nodes follow the construction of~\cite{Chang:2025mwt}. Our implementation selects poles by estimating the error caused by dropping them, extends the interpolation to complex Kac poles, and handles pole collisions at singular central charges. The resulting tables provide input to the semidefinite program solver \sdpb~\cite{Simmons-Duffin:2015qma,Landry:2019qug}.

\vb\ uses arbitrary-precision arithmetic in \texttt{Python}, with \texttt{mpmath}~\cite{mpmath} accelerated by \texttt{FLINT}/\texttt{Arb}~\cite{Hart:2010flint,Johansson:2017arb,pythonflint}. It handles the singular central charges at which the recursion coefficients have poles (the minimal models, $c=1$, $c=25$, and the $c>25$ values with rational $b^2$) by exact Laurent expansion with merged Kac poles, and besides sphere Virasoro blocks it computes global $\mathrm{SL}(2)\x\mathrm{SL}(2)$ blocks, $\cN=1$ superconformal blocks in the Neveu--Schwarz and Ramond sectors~\cite{Hadasz:2006qb,Hadasz:2007nt,Belavin:2007gz,Hadasz:2008dt}, and torus characters and one-point blocks~\cite{Hadasz:2009db,Poghossian:2009mk,Cho:2017oxl}. A companion package, \vbs, assembles crossing equations and semidefinite programs for a number of standard bootstrap problems. The exact moments of the known 2d CFTs to which we compare our bounds were computed with separate research code and are distributed as data (appendix~\ref{app:models}).

We illustrate the software with two applications. First, we bound the scalar gap when all Virasoro primaries are scalars, the setting in which Liouville theory was argued to be the unique solution~\cite{Collier:2017shs}. Taking the external scalar to be the lightest operator, we find an allowed window with lower bound $\De=(c-1)/16$ and upper bound just above the Liouville threshold $(c-1)/12$, which it appears to approach as $\L$ increases.

Second, following~\cite{Chiang:2026nmd,Poland:2024hvb}, we bound the average exchanged dimension $\langle\De\rangle$ in $\langle\f\f\f\f\rangle$ at several central charges. The bounds exhibit a rich structure of kinks, spikes, and dips, many saturated by known rational and $c=1$ theories. As $c\to1$, narrow features develop at the degenerate external dimensions. Near these dimensions the dependence on $\L$ suggests that the limits $c\to1$ and $\L\to\oo$ do not commute, so theories just above $c=1$ can be much more strongly constrained there. We also study $1<c<8/7$, where no compact unitary CFT is known but candidate modular-invariant spectra were recently found~\cite{Benjamin:2026lbj}.

Section~\ref{sec:preliminaries} fixes conventions and reviews the recursion. Section~\ref{sec:package} describes the interpolation and its accuracy, the supported block families, and the software interface and implementation. Section~\ref{sec:results} presents the two applications, and section~\ref{sec:outlook} some future directions. The appendices give the code location, the numerical parameters, and the computation of the exact moments of the known theories.

\section{Mathematical preliminaries}
\label{sec:preliminaries}
\label{sec:recursion}

We consider four-point functions of Virasoro primary operators $\cO_i$ with weights $(h_i,\hb_i)$, so that the scaling dimension of $\cO_i$ is $\De_i=h_i+\hb_i$. Using global conformal transformations we place the operators at
\be
\<\cO_1(0)\cO_2(z,\zb)\cO_3(1)\cO_4(\oo)\> \equiv \lim_{w\to\oo}w^{2h_4}\bar w^{2\hb_4}\<\cO_1(0)\cO_2(z,\zb)\cO_3(1)\cO_4(w,\bar w)\>,
\ee
so that $z=\frac{z_{12}z_{34}}{z_{13}z_{24}}$ in terms of the positions $z_{ij}=z_i-z_j$. In the $s$-channel ($\cO_1\x\cO_2\to\cO\to\cO_3\x\cO_4$) the four-point function decomposes into a sum over Virasoro primaries $\cO$ of weights $(h,\hb)$,
\be
\label{eq:blockdecomp}
\<\cO_1(0)\cO_2(z,\zb)\cO_3(1)\cO_4(\oo)\>=\sum_{\cO}C_{12\cO}C_{43\cO}\,\cF(h_i;h;c;z)\,\cF(\hb_i;\hb;c;\zb),
\ee
where $\cF(h_i;h;c;z)$ is the holomorphic Virasoro block and the $C_{ij\cO}$ are OPE coefficients, with $C_{ji\cO}=(-1)^\ell C_{ij\cO}$ for scalar $\cO_i$, $\cO_j$ and a primary $\cO$ of spin $\ell$, normalized so that the two-point functions of the $\cO_i$ and $\cO$ are unit normalized. Our conventions for the chiral block, its recursive representation, and the resulting normalization of OPE coefficients follow~\cite{Cho:2017oxl}.

Summing over the descendants of $\cO$ gives the series $\cF(h_i;h;c;z)=z^{h-h_1-h_2}\sum_{n\ge0}\cF_n z^n$, with $\cF_n$ the level-$n$ contribution. Zamolodchikov~\cite{Zamolodchikov:1987avt} showed that the block has a much better-behaved expansion in terms of the elliptic nome
\be
\label{eq:nome}
q(z)=\exp\p{-\pi\frac{K(1-z)}{K(z)}},\qquad K(z)=\int_0^{1}\frac{dt}{\sqrt{(1-t^2)(1-zt^2)}}, \qquad z(q)=\p{\frac{\th_2(q)}{\th_3(q)}}^4,
\ee
with $\th_2(q)=\sum_{n\in\Z}q^{(n+1/2)^2}$ and $\th_3(q)=\sum_{n\in\Z}q^{n^2}$. Near $z=0$, the nome behaves as $q\approx z/16$, while the crossing-symmetric point $z=\half$ maps to $q=e^{-\pi}\approx 0.0432$.

The elliptic representation isolates the exponential dependence on the exchanged weight in the factor $(16q)^h$:
\be
\label{eq:elliptic}
\cF(h_i;h;c;z)=(16q)^{h-\frac{c-1}{24}}\,z^{\frac{c-1}{24}-h_1-h_2}\,(1-z)^{\frac{c-1}{24}-h_2-h_3}\,\th_3(q)^{\frac{c-1}{2}-4\sum_{i}h_i}\,H(h_i;h;c;q),
\ee
where $H(h_i;h;c;q)=1+O(q)$ is a power series in $q$. Since $|q|<1$ throughout the cut $z$-plane, an expansion in $q$ that converges in the unit disk covers all of cross-ratio space. For the four-point function of a unitary theory this follows from the positivity of its $q$-expansion~\cite{Maldacena:2015iua}. For individual blocks it is a long-standing expectation~\cite{Zamolodchikov:1987avt} that has been proven in special cases~\cite{Ghosal:2020sis}, and in practice we evaluate the series near $q(\half)=e^{-\pi}$, where it converges rapidly.

To state the recursion for $H$, we use the Liouville parametrization of the central charge and weights,
\be
c=1+6Q^2,\qquad Q=b+\frac1b,\qquad h=\frac{Q^2-\l^2}{4},\qquad h_i=\frac{Q^2-\l_i^2}{4}.
\ee
The library solves $c=1+6(b+1/b)^2$ for $b^2$, choosing the root with $|b^2|\le 1$ and $\Re\, b\geq0$. Thus $b$ is real for $c\ge 25$, lies on the unit circle for $1<c<25$, and is purely imaginary for $c<1$. All recursion coefficients depend only on $b^2$. The degenerate Kac weights~\cite{Kac:1978ge,Feigin:1988se} are
\be
h_{r,s}=\frac{Q^2}{4}-\frac{(rb+s/b)^2}{4},\qquad r,s\in\Z_{\ge 1}.
\ee

As a function of $h$, $H$ is meromorphic, with simple poles at these degenerate weights $h=h_{r,s}$ and residues proportional to blocks of weight $h_{r,s}+rs$, corresponding to the null descendant. With $H\to1$ as $h\to\oo$, this gives the recursion relation~\cite{Zamolodchikov:1987avt,Cho:2017oxl}
\be
\label{eq:hrecursion}
H(h_i;h;c;q)=1+\sum_{r,s\ge1}\frac{(16q)^{rs}\,R_{r,s}(h_i;c)}{h-h_{r,s}}\,H(h_i;h_{r,s}+rs;c;q),
\ee
with residues
\be
R_{r,s}(h_i;c)=A_{r,s}(c)\,P_{r,s}[h_1,h_2]\,P_{r,s}[h_4,h_3],
\ee
where
\be
\label{eq:Ars}
A_{r,s}(c)&=\frac{1}{2}\prod_{m=1-r}^{r}\prod_{n=1-s}^{s}{}^{\prime}\frac{1}{mb+n/b},\\
\label{eq:Prs}
P_{r,s}[h_1,h_2]&=\prod_{\substack{p=1-r\\ p\in 2\Z+1-r}}^{r-1}\ \prod_{\substack{k=1-s\\ k\in 2\Z+1-s}}^{s-1}\frac{\l_1+\l_2+pb+k/b}{2}\cdot\frac{\l_1-\l_2+pb+k/b}{2}.
\ee
The order of the arguments matters: $P_{r,s}[h_3,h_4]=(-1)^{rs}P_{r,s}[h_4,h_3]$, and with this order $R_{1,1}=\frac12(h_2-h_1)(h_3-h_4)$ reproduces the level-one coefficient $\frac{(h+h_2-h_1)(h+h_3-h_4)}{2h}$ of the block. The prime on the product in~\eqref{eq:Ars} indicates that the pairs $(m,n)=(0,0)$ and $(m,n)=(r,s)$ are omitted, and the products in~\eqref{eq:Prs} run over $p$ and $k$ in steps of 2. The fusion polynomial $P_{r,s}[h_1,h_2]$ vanishes precisely when the fusion rule of the degenerate representation $h_{r,s}$ with $\cO_1$ into $\cO_2$ is satisfied, i.e.\ when the null descendant decouples from the three-point function. In particular, for pairwise identical external operators, $P_{r,s}[h,h]=0$ whenever $r$ and $s$ are both odd, so that the pole at $h_{1,1}=0$ corresponding to the identity is absent.

Writing $H=\sum_{k\ge0}H_k(h)\,q^k$, equation~\eqref{eq:hrecursion} determines the coefficients level by level,
\be
\label{eq:levelrec}
H_k(h)=\de_{k,0}+\sum_{\substack{r,s\ge 1\\ rs\le k}}\frac{16^{rs}R_{r,s}}{h-h_{r,s}}\,H_{k-rs}(h_{r,s}+rs),
\ee
so that to compute $H$ to order $q^N$ one needs the coefficients $H_{k}(h_{r,s}+rs)$ for all $(r,s)$ with $rs\le N$ and $k\le N-rs$. This is a triangular table of size $O(N\cdot D(N))$, where $D(N)\sim N\log N$ is the number of pairs $(r,s)$ with $rs\le N$. Because $q(\half)=e^{-\pi}$, the nome factor alone suppresses each additional order by roughly $1.36$ decimal digits near the crossing-symmetric point, although the actual gain depends on the coefficients and derivatives, and orders $N\sim 60$--$120$ suffice for the derivative tables used in this paper. \vb\ refers to the truncation order $N$ as \texttt{order}.

At fixed order $N$, equation~\eqref{eq:levelrec} gives $H^{(N)}(h,q)=\sum_{k\le N}H_k(h)q^k$ as an exact rational function of $h$. At generic central charge this takes the form
\be
\label{eq:rationalH}
H^{(N)}(h,q)=\frac{\sum_{k=0}^N \cN_k(h)\,q^k}{D_N(h)},\qquad D_N(h)=\prod_{\substack{rs\le N\\ R_{r,s}\ne 0}}(h-h_{r,s}),
\ee
with polynomial numerators $\cN_k(h)$. The library removes from $D_N$ any pole whose residue vanishes identically for the given external weights. At the recursion-singular central charges (rational $b^2$) individual residues diverge and Kac weights collide. There the library expands the recursion in a local variable around the singular value (section~\ref{sec:implementation}), takes the constant term, which is the finite limit, and writes it as a sum over merged pole locations, each at the order that survives the cancellations, so that $D_N$ is replaced by the product of the merged poles. The library can compute the numerators $\cN_k(h)$ explicitly at arbitrary precision, or run the recursion directly at a fixed numerical value of $h$. The latter is faster and is used to generate the interpolated tables of section~\ref{sec:interp}.

For bootstrap computations, the library rescales~\eqref{eq:elliptic} to a chiral block $F$ with user-controlled normalization,
\be
\label{eq:libnorm}
F(h_i;h;c;z)=a_0^{-h}\,z^{h_1+h_2+\de_z}\,\cF(h_i;h;c;z),
\ee
so that $F=a_0^{-h}z^{h+\de_z}(1+O(z))$. The default choice is $a_0=16$ (\texttt{--chiral-norm}), for which the exponential prefactor at the crossing-symmetric point becomes $(16q(\half)/a_0)^{h}=e^{-\pi h}$. This normalization plays the same role as the factor $(4/a_0)^\De$ in~\cite{Erramilli:2020rlr}.

The choice $a_0=16$ follows~\cite{Chang:2025mwt}, where it was argued that the blocks should be approximated, and the semidefinite program set up, in a normalization in which the coefficients of the blocks in a solution to crossing, here $a_0^{\De}$ times the squared OPE coefficients, carry no leading exponential in $\De$. For global blocks this singles out $4^{-\De}g_{\De,\ell}$, the normalization in which rescaled mean-field squared OPE coefficients are power-law. Reference~\cite{Chang:2025mwt} also motivates this rescaling by the $4^\De$ exponential growth of analytic dispersive functionals acting on global blocks. The library nevertheless uses $a_0=16$ for global blocks by default, so that global and Virasoro tables share one normalization of the OPE coefficients, and \texttt{--chiral-norm 4} gives the normalization singled out above.

For Virasoro blocks the continuum OPE density of Virasoro mean field theory~\cite{Collier:2018exn} and the squared DOZZ structure constants~\cite{Dorn:1994xn,Zamolodchikov:1995aa} of Liouville theory both decay as $16^{-\De}$ at large $\De$, which is what $a_0=16$ removes. With $h=Q^2/4+P^2$, the rescaled Liouville density is then power-law in $P$, while the chiral mean-field density retains a subleading exponential $e^{\pi QP}$. The rescaling leaves the crossing equation unchanged and changes only the weight in which the interpolation error is measured and the conditioning of the semidefinite program. The rescaled blocks retain the damping $e^{-\pi\De}$ at the crossing-symmetric point, which becomes the factor $E^\De$ of $\chi_\ell$ in section~\ref{sec:interp}.

The shift $\de_z$ (\texttt{--z-shift}) defaults to $0$, so that the leading power is $z^{h}$ rather than the kinematical $z^{h-h_1-h_2}$. For four identical external scalars $\f$ with $h_i=\hb_i=\De_\f/2$, the choice $\de_z=\de_{\zb}=-\De_\f$ (\texttt{--u-shift}) makes the crossing equation equate the block expansions at $(z,\zb)$ and at $(1-z,1-\zb)$ with the same coefficients and no additional prefactor. In this normalization, the contribution of a primary to~\eqref{eq:blockdecomp} is
\[
C_{12\cO}C_{43\cO}\,a_0^{\De}\,z^{-h_1-h_2-\de_z}\zb^{-\hb_1-\hb_2-\de_{\zb}}\,F(h,z)F(\hb,\zb),
\]
so the coefficient of an ordered block is $a_0^{\De}C_{12\cO}C_{43\cO}$, which is $a_0^{\De}$ times a squared OPE coefficient for identical external operators. The paired basis below requires an additional multiplicity convention (appendix~\ref{app:models}). Since this is a positive rescaling, it does not affect feasibility problems, but it must be taken into account when comparing OPE coefficients or moments extracted from bootstrap computations to exact values.

Multiplying the two chiral factors gives the full block for a primary of weights $(h,\hb)$, $B(h,\hb;z,\zb)=F(h_i;h;c;z)F(\hb_i;\hb;c;\zb)$ (the \texttt{ordered\_B} basis). For a four-point function of scalars $\cO_i$ ($h_i=\hb_i$), it is convenient to define a parity-paired combination at dimension $\De$ and spin $\ell$,
\be
\label{eq:scalarG}
G_{\De,\ell}(z,\zb)=\frac{B(h,\hb;z,\zb)+B(\hb,h;z,\zb)}{1+\de_{\ell,0}},\qquad h=\frac{\De+\ell}{2},\quad \hb=\frac{\De-\ell}{2},
\ee
which is the \texttt{scalar\_G} basis used in the examples of section~\ref{sec:results}. At $\ell=0$ the two terms coincide and $G_{\De,0}=B(h,h;z,\zb)$. This basis pairs the two spin orientations, while the \texttt{ordered\_B} basis retains them separately, including for scalar externals. A parity-even four-point function of scalars is symmetric under $z\leftrightarrow\zb$, so the two orientations enter with equal weight. The blocks carry no overall factor $(-1)^\ell$, so in a mixed correlator the coefficient $C_{12\cO}C_{43\cO}$ is $C_{12\cO}^2$ in $\<\f_1\f_2\f_2\f_1\>$ and $(-1)^\ell C_{12\cO}^2$ in $\<\f_1\f_2\f_1\f_2\>$. For spinning external operators the library provides the parity-projected basis $B_\pm=B\pm\eta\,\cP[B]$, where $\cP$ exchanges $h\leftrightarrow\hb$ and $h_i\leftrightarrow\hb_i$ and $\eta$ is a product of external and internal parity phases in $\{\pm1,\pm i\}$. The precise conventions, and the other block families, are given in section~\ref{sec:families}. The exchanged spin $\ell=h-\hb$ is an integer in the \texttt{scalar\_G} basis, with $G_{\De,-\ell}=G_{\De,\ell}$, and may be half-integer in the other two bases.

In a unitary theory, the sum in~\eqref{eq:blockdecomp} is restricted to $h,\hb\geq0$ for $c\geq 1$, so that $\De\geq|\ell|$, with the identity at $h=\hb=0$. For $c<1$ unitarity restricts to the minimal models~\cite{Friedan:1983xq,Goddard:1986ee}, where the spectrum is discrete. \vb\ nevertheless computes blocks for arbitrary $c$ and weights, and our bootstrap problems assume the continuous unitarity bound $\De\geq|\ell|$.

\section{The \vb\ package}
\label{sec:package}

\subsection{Optimal interpolation}
\label{sec:interp}

The recursion gives an exact rational representation at each order in $q$. The task is to reduce its polynomial degree while preserving its accuracy on the allowed spectrum. At fixed spin, consider the Taylor coefficients around $z=\zb=\half$,
\be
\label{eq:taylortable}
T^{m,n}_{\ell}(\De)\equiv\frac{1}{m!\,n!}\,\ptl_z^m\ptl_{\zb}^n\,G_{\De,\ell}(z,\zb)\Big|_{z=\zb=\half}.
\ee
Inserting the rational representation~\eqref{eq:rationalH} into~\eqref{eq:elliptic} and expanding the $\De$-independent geometric factors around $z=\half$, one finds that every entry is exactly of the form
\be
\label{eq:exactpoleform}
T^{m,n}_\ell(\De)=\frac{E^{\De}\,\tl P^{m,n}_\ell(\De)}{D_N\p{\frac{\De+\ell}{2}}D_N\p{\frac{\De-\ell}{2}}},\qquad E=\frac{16\,q(\half)}{a_0}=\frac{16\,e^{-\pi}}{a_0},
\ee
where $\tl P^{m,n}_\ell$ is a polynomial of degree at most $2\deg D_N+m+n$. The factor $E^\De$ comes from the prefactor $(16q/a_0)^{h+\hb}$, and the remaining $\De$-dependence of the Taylor coefficients of $(16q(z)/a_0)^h$ is polynomial in $h$. The poles sit at $\De=2h_{r,s}\pm\ell$. For $1<c<25$ those with $r\neq s$ form complex-conjugate pairs and those with $r=s$ are real and lie at or below the unitarity bound, $x\le0$, with $h_{1,1}=0$ giving the poles at $\De=\pm\ell$, while for $c>25$ all poles are real and at or below the unitarity bound. The library can output this exact representation (\texttt{--npoles all}), but it has $2D(N)$ poles, several hundred at $N=60$. Semidefinite programs with polynomials of such degree are slow, as found in~\cite{Collier:2017shs}, and poorly conditioned.

The idea is to retain the poles whose effects are hardest to approximate and absorb the rest into a polynomial. In $d>2$, ``pole shifting''~\cite{Kos:2013tga} does this by matching asymptotics at $\De\to\oo$ and at the unitarity bound. Optimal interpolation~\cite{Chang:2025mwt} instead fixes the polynomial by sampling at near-optimal nodes. That work performs the interpolation in \sdpb\ 3.1 from the exact pole list, whereas \vb\ performs it when building the tables. We seek
\be
\label{eq:interpform}
T^{m,n}_\ell(\De)\approx \chi_\ell(x)\,P^{m,n}_\ell(x),\qquad \chi_\ell(x)=\frac{E^{x}}{\prod_{i\in\cK}(x-x_i)},\qquad x=\De-\De_0(\ell),
\ee
where $\De_0(\ell)$ is the unitarity bound $|\ell|$ (raised by any assumed gap, \texttt{--gap}) and $\cK$ labels the $n_{\rm poles}$ retained Kac poles, counted with multiplicity. Each nonzero entry is fitted on its own degree-matched grid. Its degree allowance is $d_{m,n}=m+n+|\cK|$ for a generic fixed-spin entry, matching the large-$\De$ power $T\sim E^\De\De^{m+n}$, and leading cancellations can lower it. The polynomial is fixed by Lagrange interpolation at $d_{m,n}+1$ sample points $x_k$. Two choices determine the accuracy: where to place these points and which poles to retain.

\begin{figure}[t]
\centering
\includegraphics[width=0.8\textwidth]{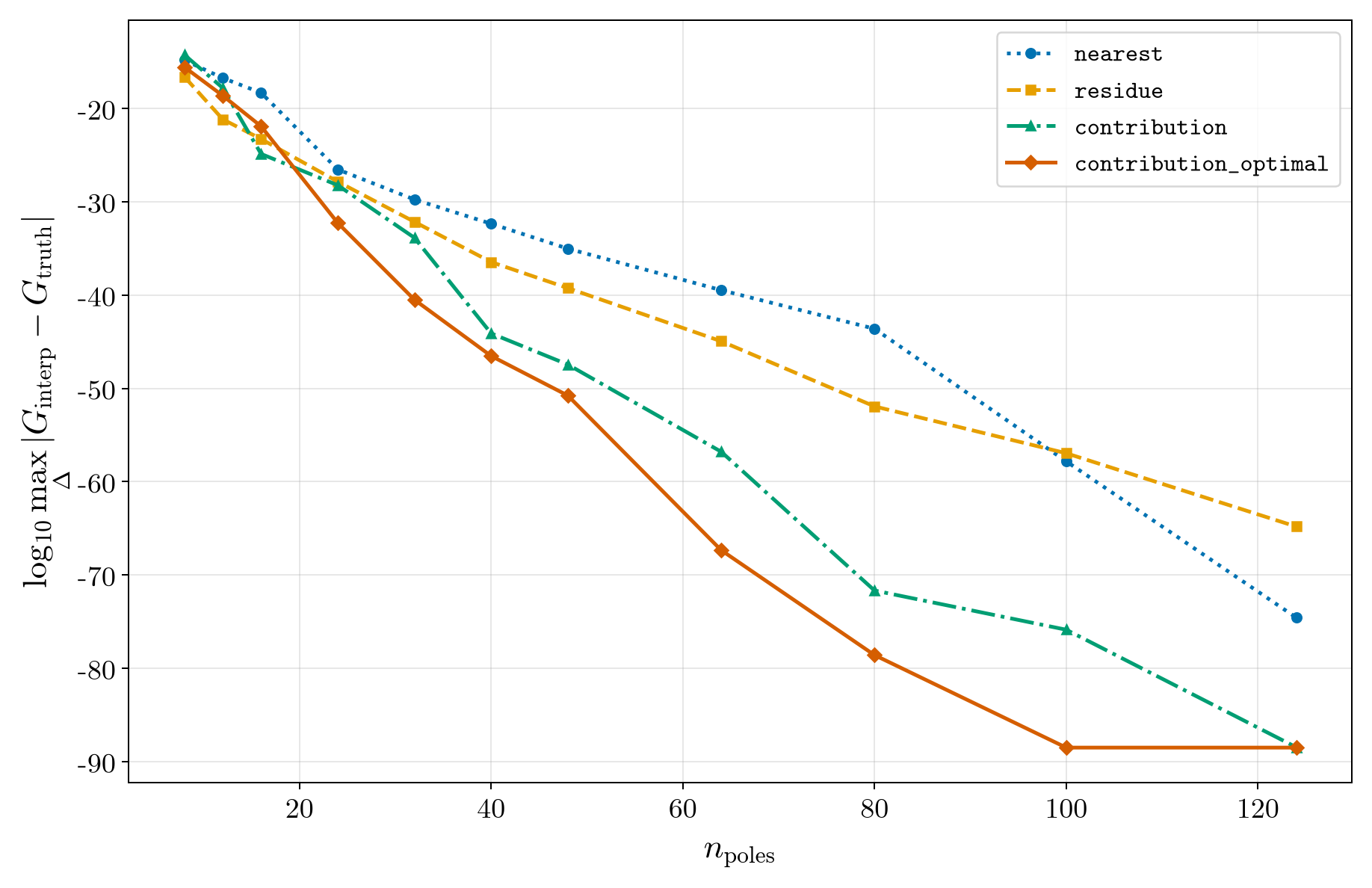}
\caption{Maximum absolute error over $\De$ of the interpolated $(m,n)=(0,0)$ entry of the spin-4 block at $c=2$, $\De_\f=\half$, recursion order $N=60$, as a function of the number of retained poles, for the four pole-selection rules discussed in the text. The reference is the exact rational table at order $N=100$, computed at 300 digits, so that the floor at $10^{-89}$ is the truncation error of the order-60 recursion. The default \texttt{contribution\_optimal} rule is the most accurate for $n_{\rm poles}\geq24$ until the floor is reached.}
\label{fig:selectors}
\end{figure}

\begin{figure}[t]
\centering
\includegraphics[width=0.49\textwidth]{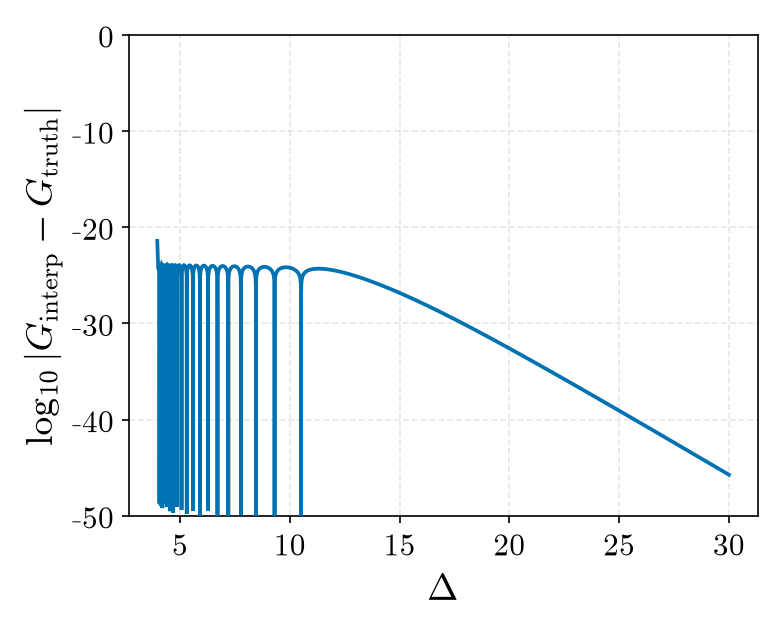}
\includegraphics[width=0.49\textwidth]{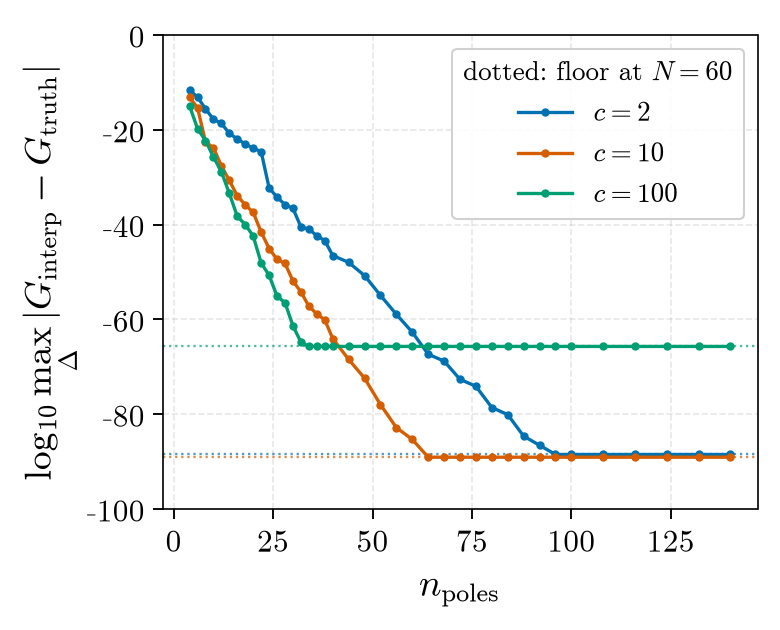}
\caption{Left: absolute error $|G_{\rm interp}-G_{\rm truth}|$ of the interpolated $(0,0)$ entry of the spin-4 block at $c=2$ with $\De_\f=\half$, $N=60$, and $20$ retained poles, as a function of $\De$, with the reference $G_{\rm truth}$ computed at order $100$ as in figure~\ref{fig:selectors}. The zeros are the interpolation nodes, and the error is uniform across the region where the nodes lie and decays exponentially beyond it. Right: the maximum of this error over $\De$ as a function of $n_{\rm poles}$ at $c=2$, $10$, $100$. The dotted lines are the truncation floors of the order-60 recursion at each $c$.}
\label{fig:interperror}
\end{figure}

Following~\cite{Chang:2025mwt}, for a degree-$n$ interpolant the nodes are chosen from the equilibrium density $\r(x)$ of $n+1$ unit charges on $x\geq0$ with logarithmic repulsion in the external potential $-2\log\chi_\ell(x)$~\cite{SaffTotik:1997}, the one-dimensional charge-balance problem familiar from large-$n$ matrix models, whose solution is known in closed form~\cite{Chang:2025mwt}. For a prefactor $\chi(x)=E^x/\prod_p(x+k_p)$ with real poles at $-k_p\le0$ the density is supported on $[0,b]$ and reads
\be
\label{eq:optdensity}
\r(x)&=\frac{1}{(n+1)\pi}\sqrt{\frac{b-x}{x}}\p{-\log E+\sum_p\sqrt{\frac{k_p}{b+k_p}}\,\frac{1}{x+k_p}},\nn\\
-\frac b2\log E&+\sum_p\p{1-\sqrt{\frac{k_p}{b+k_p}}}=n+1,
\ee
where the second equation fixes the endpoint $b$. The library extends these formulas to complex-conjugate pairs by summing the real parts of their contributions. When every retained pole has $\Re x_i\le0$ each term of the density is positive, but complex pairs with $\Re x_i>0$, which occur for $1<c<25$, can in principle make it negative. Writing $\w(x)=(n+1)\int_0^x\r$, the nodes solve the Bohr--Sommerfeld-like condition $\w(x_k)=k+\half$. The construction removes the leading exponential growth of the weighted Lebesgue constant. The residual growth is logarithmic for the real-pole examples of~\cite{Chang:2025mwt}, and we assess the complex-pole extension numerically in figures~\ref{fig:interperror} and~\ref{fig:selectors}.

The block values at the nodes are computed by evaluating the recursion directly at the corresponding weights $h=(x_k+\De_0\pm\ell)/2$ and Taylor expanding the resulting numerical series (\texttt{--sample-via numeric}), which avoids the catastrophic cancellations that occur when the exact rational form~\eqref{eq:exactpoleform} is evaluated at high orders.

\paragraph{Pole selection.} The accuracy of~\eqref{eq:interpform} at fixed $n_{\rm poles}$ depends strongly on which poles are retained (\texttt{--pole-selection}). Simple rules are not satisfactory. On the complex strip of Kac poles at $1<c<25$, keeping the poles closest to the physical region $x\geq0$ (\texttt{nearest}) retains poles that barely contribute, while keeping those with the largest residue (\texttt{residue}) or the largest principal part at $x=0$ (\texttt{contribution}) chases distant poles whose effect the polynomial would absorb anyway. Distant poles also widen the support $[0,b]$ of the node density~\eqref{eq:optdensity}, and since the weight at $x\sim b$ is $E^b$, a wide support produces a badly conditioned semidefinite program.

The default rule, \texttt{contribution\_optimal}, estimates the error caused by dropping each pole. If a provisional set $\cK$ has been retained and a simple pole at $x_p\notin\cK$ with principal-part score $s_p$ is dropped, the reduced function $T/\chi_\cK$ acquires a pole at $x_p$ with residue enhanced by $\prod_{q\in\cK}(x_p-x_q)$. The remainder of interpolating a simple pole at the nodes $x_k$ is $\chi_\cK(x)W(x)/[(x-x_p)W(x_p)]$ times the residue, with $W(x)=\prod_k(x-x_k)$. The cost of dropping $x_p$ is therefore taken to be
\be
\label{eq:dropcost}
\cC_p(\cK)=s_p\,|x_p|^{\,o_p}\prod_{q\in\cK}|x_p-x_q|\ \max_{0\le x\le b}\left[\frac{E^{x}}{\prod_{q\in\cK}|x-x_q|}\,\frac{|W(x)|}{|x-x_p|\,|W(x_p)|}\right],
\ee
with $o_p$ the order of the pole. The algorithm repeatedly adds the pole with the largest $\cC_p(\cK)$, treating conjugate pairs as one object. The reference nodes in $W$ are refreshed in batches using the currently retained prefactor and held fixed within each batch, and the maximum is estimated on a finite grid over their range. Fixed batch boundaries preserve a nested ranking, so that increasing $n_{\rm poles}$ extends the previous choice. The final interpolant is then built with the optimal nodes for its own retained poles and degree.

Figure~\ref{fig:selectors} compares the rules at $c=2$. The \texttt{nearest} and \texttt{residue} rules improve only slowly with $n_{\rm poles}$, \texttt{contribution} does better at large $n_{\rm poles}$, and beyond a couple of dozen poles \texttt{contribution\_optimal} is the most accurate, by two to twelve orders of magnitude, reaching the truncation floor of the recursion with about $100$ poles out of several hundred.

\paragraph{Poles in the positivity region.} For $c\le1$ all Kac poles are real, and some of them lie at $x>0$, inside the region where the approximation is used. To interpolate on $x\geq0$ while preserving such a pole, it is split off exactly. Write each entry as $T(x)=E^xR(x)$ and extract the principal parts $\sum_{p,a}g_{p,a}/(x-x_p)^a$ of the rational function $R$ at all $x_p>0$. The damped regular remainder $E^x[R(x)-\sum_{p,a}g_{p,a}/(x-x_p)^a]$, whose remaining poles lie at $x\le0$, is interpolated as above. Adding back $E^x\sum_{p,a}g_{p,a}/(x-x_p)^a$ gives the form~\eqref{eq:interpform}, with the poles at $x>0$ included in $\chi_\ell$. The count $n_{\rm poles}$ then refers to the retained poles at $x\le0$.

At the minimal-model central charges, where Kac weights collide, the order of each merged pole is read from the merged Laurent coefficients of the recursion (section~\ref{sec:implementation}) rather than from the number of colliding weights, since cancellations between the colliding terms can lower it. The accuracy of the resulting tables is limited by $N$ and $n_{\rm poles}$ as at $c>1$, whereas inserting positive real poles directly into the equilibrium-density construction~\eqref{eq:optdensity} does not give the admissible smooth weight assumed in that construction. Because retained poles of odd order at $x>0$ make $\chi_\ell$ change sign, these tables are not suitable input for semidefinite programming. They provide accurate approximations of the $c\le1$ blocks for other numerical uses.

\paragraph{Accuracy of the interpolated blocks.} Besides finite-precision roundoff, a table has two approximation errors: the truncation of the $q$-series at order $N$ and the interpolation with $n_{\rm poles}<\text{all}$. The former is measured by comparing tables at different orders, and the command-line tool also estimates it from the tail of the series (\texttt{--q-gate}). It grows with $c$, so that $N=60$--$80$ is ample at moderate $c$ while large $c$ calls for a somewhat higher order.

Figure~\ref{fig:interperror} shows an example of the interpolation error. As a function of $\De$ it equioscillates across the region where the nodes lie and decays beyond it, the signature of near-optimal nodes~\cite{Chang:2025mwt}, and its maximum decreases exponentially in $n_{\rm poles}$ until it reaches the truncation floor, beyond which one must increase $N$ instead. For the entry shown, the values $n_{\rm poles}\geq48$ used in this paper give errors below $10^{-40}$. The tables behind the $\L=19$ bounds reproduce every derivative entry to about $15$ digits, and the large-spin towers with $12$ poles reproduce each entry to $4$--$5$ digits, or $23$ digits relative to their largest entry.

Higher derivatives are less accurate, and the relevant test is the stability of the bounds themselves (section~\ref{sec:moments-rational}). An inaccurate block does not simply weaken a bound but can produce a spurious exclusion (section~\ref{sec:gapbounds}), so we recommend monitoring every bound in $n_{\rm poles}$ and checking the final functional against directly evaluated blocks. The checks applied by \vbs\ are described in section~\ref{sec:results}. They do not bound the $q$-series truncation error, which is estimated separately.

\subsection{Block families and bases}
\label{sec:families}

Besides the sphere Virasoro blocks of section~\ref{sec:preliminaries}, \vb\ computes several other families of blocks, selected with \texttt{--family} and summarized in table~\ref{tab:families}. The sphere families share the following two-dimensional bases. Given chiral functions $k_h(z;\{h_i\})$ and $k_{\hb}(\zb;\{\hb_i\})$ for the holomorphic and antiholomorphic halves, the library forms
\be
\label{eq:sigmarule}
G(z,\zb)&=k_h(z;\{h_i\})\,k_{\hb}(\zb;\{\hb_i\})+\s\,k_{\hb}(z;\{\hb_i\})\,k_h(\zb;\{h_i\}),\nn\\
\s&=\begin{cases}0&\texttt{ordered\_B},\\1-\de_{\ell,0}&\texttt{scalar\_G},\\\pm\eta&\texttt{parity\_projected\_B},\end{cases}
\ee
with $h=(\De+\ell)/2$, $\hb=(\De-\ell)/2$. The \texttt{ordered\_B} basis is the plain product $B(h,\hb)$ and accepts scalar or spinning externals. The \texttt{scalar\_G} basis is~\eqref{eq:scalarG} and requires scalar externals and integer $\ell$, and the parity-projected basis $B_\pm=B\pm\eta\,\cP[B]$ adds the block with $h\leftrightarrow\hb$ and $h_i\leftrightarrow\hb_i$ exchanged, weighted by the phase
\be
\eta=p_{\rm exch}\prod_{i=1}^4p_i,\qquad p_i,p_{\rm exch}\in\{1,-1,i,-i\},
\ee
where the $p_i$ are the parities of the external operators (\texttt{--parity-ext}) and $p_{\rm exch}$ that of the exchanged operator (\texttt{--parity-exchanged}, default $1$). The sign $\pm$ is set by \texttt{--parity-project}, and only one of $B_\pm$ is written per run. The imaginary phases are needed for fermionic operators. For the Virasoro and global families the exchanged spin may be half-integer in the \texttt{ordered\_B} and \texttt{parity\_projected\_B} bases, while the Neveu--Schwarz and Ramond families currently require integer exchanged spins. The chiral bases (\texttt{chiral\_F} and its relatives) output $k_h(z)$ alone as a function of $h$.

\paragraph{Global blocks (\texttt{global}).} The $c\to\oo$ limit of the Virasoro block is the global $\mathrm{SL}(2)$ block, for which the library uses the closed form
\be
k_h(z)=a_0^{-h}\,z^{h+\de_z}\,{}_2F_1(h+h_2-h_1,\,h+h_3-h_4;\,2h;\,z),
\ee
with the same normalization constant and shift as in~\eqref{eq:libnorm}. The exponential base of the tables is $E=4\r(\half)/a_0$ with $\r(z)=(1-\sqrt{1-z})/(1+\sqrt{1-z})$ the radial coordinate of~\cite{Hogervorst:2013sma}, so that $\r(\half)=3-2\sqrt2$ and $E=(3-2\sqrt2)/4$ at the default $a_0=16$ (the choice $a_0=4$ reproduces the base $3-2\sqrt2$ of \texttt{scalar\_blocks}). The poles of the global block in $\De$ sit at $\De=\pm\ell-p$, $p=0,1,2,\ldots$, from the two chiral halves. Removable poles, whose residue vanishes for the given external weights (all even $p$ for pairwise identical externals), are dropped exactly.

The option \texttt{--order} $=M$ restricts the candidate poles to $0\leq p<M$ (default $M=30$) but does not truncate the hypergeometric function, which is evaluated at the working precision. An explicit \texttt{--npoles} selects from these poles by the rule of section~\ref{sec:interp}. When it is omitted, the number of poles is set from the working precision and the candidate range is enlarged to match it. The available bases are \texttt{global\_scalar\_G}, \texttt{global\_ordered\_B}, \texttt{global\_parity\_projected\_B}, and \texttt{global\_chiral\_F}, formed as in~\eqref{eq:sigmarule}.

At a removable exchanged weight we take the analytic limit in $h$ with the external weights held fixed. When only one numerator terminates, this retains a nonzero hypergeometric tail. For example, at $h=0$ with $h_2=h_1$ the hypergeometric factor is $\frac12[1+(1-z)^{-(h_3-h_4)}]$, which becomes $1$ when $h_3=h_4$ as well.

\paragraph{$\cN=1$ Neveu--Schwarz blocks (\texttt{ns}).} For four Neveu--Schwarz superprimaries (lower components) of chiral weights $h_i$ exchanging a Neveu--Schwarz superprimary of chiral weight $h$, the library computes the elliptic recursion of Hadasz, Jask\'olski, and Suchanek~\cite{Hadasz:2006qb,Hadasz:2007nt}. The parametrization is
\be
\hat c=\frac{2c}{3}=1-2\p{\b-\frac1\b}^2,\qquad h=\frac{\hat c-1}{16}+\frac{\a^2}{8},\qquad \hat h_{r,s}=\frac{\hat c-1}{16}+\frac{(r\b-s/\b)^2}{8},
\ee
with $r+s$ even and the null vector of $\hat h_{r,s}$ at level $rs/2$. The block splits into two sectors, $H^{(0)}$ containing the integer levels and $H^{(1/2)}$ the half-integer levels (\texttt{--super-sector even}/\texttt{odd}), which are coupled by the recursion,
\be
H^{(\e)}=g^{(\e)}+\sum_{\substack{r+s\text{ even}\\ rs\text{ even}}}\frac{(16q)^{rs/2}K_{r,s}}{h-\hat h_{r,s}}\,H^{(\e)}\big(\hat h_{r,s}+\tfrac{rs}2\big)+\sum_{\substack{r+s\text{ even}\\ rs\text{ odd}}}\frac{(16q)^{rs/2}K_{r,s}}{h-\hat h_{r,s}}\,H^{(\e')}\big(\hat h_{r,s}+\tfrac{rs}2\big),
\ee
where $\e'$ is the other sector, the seeds are $g^{(0)}=\th_3(q^2)$ and $g^{(1/2)}=0$, and $K_{r,s}$ is the product of the supersymmetric analogs of $A_{r,s}$ and of two fusion polynomials (starred or unstarred according to the parity of $rs$ and the sector), for which we refer to~\cite{Hadasz:2006qb,Hadasz:2007nt} and to the source.

The series is in $q^{1/2}$, and \texttt{--order} counts powers of $q$. The prefactor is the Virasoro one~\eqref{eq:elliptic} with every occurrence of $c-1$ replaced by $c-\frac32$, so that the exponential base is again $E=16q(\half)/a_0=e^{-\pi}$ at $a_0=16$, and the even sector starts as $F^{(0)}\sim a_0^{-h}z^{h+\de_z}$, while the odd sector starts at level $\frac12$. The two-dimensional bases \texttt{ns\_scalar\_G}, \texttt{ns\_ordered\_B}, and \texttt{ns\_parity\_projected\_B} are formed by~\eqref{eq:sigmarule} with the same sector on both halves. The two sectors are output separately and can be combined with independent coefficients.

The candidate poles are the $\hat h_{r,s}$ with $rs\leq2N+4$ at nome order $N$, slightly beyond the recursion's own truncation at $rs\leq2N$, and omitting \texttt{--npoles} keeps the resulting selectable poles. Removable poles and extra positive candidate poles with no principal part at this order are omitted. For $c\le\frac32$ some Kac poles lie at $x>0$, and as for the Virasoro blocks at $c\le1$ they are split off as exact principal parts. The singular values of $c$ (the $\cN=1$ minimal models, $c=\frac32$, rational $\b^2$) are handled at $c$ itself by a Laurent recursion, with merged-pole tables sampled from the same recursion. At $c=\frac32$ with all $h_i=\frac18$ the two sectors are known in closed form, $H^{(0)}=\th_3(q^2)$ and $H^{(1/2)}=\th_2(q^2)/h$, which is one of the tests of the implementation.

\paragraph{$\cN=1$ Ramond blocks (\texttt{ramond}).} For four Ramond externals exchanging a Neveu--Schwarz superprimary, the library implements the elliptic recursion of~\cite{Hadasz:2008dt}, with Ramond momenta $\g_i=\sqrt{c/24-h_i}$ and $b=i\b$. The four blocks obtained by choosing the signs of the Ramond three-point structures in the $(3,4)$ and $(1,2)$ fusion polynomials are selected with \texttt{--ramond-branch} $\in\{\texttt{++},\texttt{+-},\texttt{-+},\texttt{--}\}$, and the default \texttt{++} is the branch appropriate to identical bosonic Ramond components. The seeds are $g^{(0)}=1$ and $g^{(1/2)}=0$, and otherwise the poles, sectors, prefactor, and bases are as for the Neveu--Schwarz blocks.

Mixed Neveu--Schwarz/Ramond external configurations and Ramond intermediate modules are not yet implemented.

\paragraph{Torus characters (\texttt{torus-character}).} The library tabulates the reduced partition function $\hat Z=\sqrt{\tau_2}\,|\eta(\tau)|^2Z(\tau,\bar\tau)$, to which a non-degenerate Virasoro primary of weights $(h,\hb)$ contributes
\be
\hat\chi_{h,\hb}=\sqrt{\tau_2}\;q_\tau^{\,h-\frac{c-1}{24}}\,\bar q_\tau^{\,\hb-\frac{c-1}{24}},\qquad q_\tau=e^{2\pi i\tau},
\ee
and the vacuum contributes $\sqrt{\tau_2}\,(q_\tau\bar q_\tau)^{-\frac{c-1}{24}}(1-q_\tau)(1-\bar q_\tau)$. These tables describe non-degenerate modules. At $c>1$ every non-vacuum chiral module is non-degenerate, but a chiral current with $(h,\hb)=(j,0)$ or $(0,j)$ has the vacuum module on one side, so its character carries a factor $1-\bar q_\tau$ or $1-q_\tau$ and is not the $\De\to|j|$ limit of $\hat\chi_{h,\hb}$. For the vacuum, the currents, and the degenerate modules at singular $c$, the library writes the irreducible character at the given weights as a point table (section~\ref{sec:output}). The modular bootstrap of \vbs\ includes the vacuum and current characters as separate positivity constraints. The tables are Taylor coefficients in the coordinate $v=\frac{\tau-i}{\tau+i}$ (\texttt{vvb}) around the self-dual point $\tau=i$, $v=0$, on which the modular $S$ transformation acts as $(v,\bar v)\to(-v,-\bar v)$, so that the $S$-odd sector consists of the entries with $m+n$ odd. Since $q_\tau^{a}=e^{-2\pi a}\exp[-4\pi a\,v/(1-v)]$, each entry is exactly $e^{-2\pi(\De-\De_0)}$ times a polynomial of degree $m+n$ in $\De-\De_0$, with $\De_0=|j|$ for spin $j=h-\hb$. There is no series truncation and no pole list. For $j\neq0$ the pair $(j,-j)$ is summed.

\paragraph{Torus one-point blocks (\texttt{torus-onepoint}).} The torus one-point function of a scalar primary $\Phi$ of weights $(d,d)$ decomposes as $\<\Phi\>_\tau=\sum_\cO C_{\cO\Phi\cO}\,q_\tau^{h-c/24}\bar q_\tau^{\hb-c/24}\,\cF(d;h;c;q_\tau)\bar\cF$, with the torus one-point block~\cite{Hadasz:2009db,Poghossian:2009mk,Cho:2017oxl}
\be
\cF(d;h;c;q)&=\prod_{n\geq1}(1-q^n)^{-1}\,f(h,q),\qquad f(h,q)=1+\sum_{r,s\geq1}\frac{q^{rs}R_{r,s}}{h-h_{r,s}}\,f(h_{r,s}+rs,q),\nn\\
R_{r,s}&=A_{r,s}\,P_{r,s}[d,h_{r,s}]\,P_{r,s}[d,h_{r,s}+rs],
\ee
with $A_{r,s}$ and $P_{r,s}$ as in~\eqref{eq:Ars}--\eqref{eq:Prs}. The tables are again in the $v$ coordinates around $\tau=i$. Multiplication by $\tau_2^{d+1/2}|\eta(\tau)|^2$ cancels the oscillator products and gives the tabulated contribution
\be
\tau_2^{d+1/2}q_\tau^{h-\frac{c-1}{24}}\bar q_\tau^{\hb-\frac{c-1}{24}}f(h,q_\tau)f(\hb,\bar q_\tau).
\ee
The base is $e^{-2\pi}$, the poles sit at $\De=2h_{r,s}\mp j$, and the singular values of $c$ are handled by the Laurent method. For $j\ne0$ the two spin orientations are summed. At generic $c$, for $j=0$ a non-removable chiral pole at $h=0$ produces a double pole at $\De=0$, at the default anchor $\De_0=0$, so the first interpolation node is lifted off it, and a positive shift can be chosen with \texttt{--gap} or \texttt{--torus-scalar-shift}.

As for the sphere tables, all real poles inside the positivity region are split off as exact principal parts and recombined after interpolation, including their merged orders at singular $c$, so the interpolated family is available also for $c\le1$. The remaining retained poles are chosen nearest to the positivity region first (\texttt{--npoles}, the positive poles come on top), and each remainder entry $(m,n)$ is interpolated on its own $|\cK_{\rm rem}|+m+n+1$ nodes, preserving its large-$\De$ power.

\paragraph{Large-spin towers.} For imposing positivity along a family of operators of fixed twist $\tau$ and arbitrary spin, the library tabulates blocks in which the spin is the continuous variable: $h=L+\tau/2$, $\hb=\tau/2$, $\De=L+\tau$, with
\be
G_\tau(L;z,\zb)=k_{L+\tau/2}(z)\,k_{\tau/2}(\zb)+\s\,k_{\tau/2}(z)\,k_{L+\tau/2}(\zb),
\ee
where $k$ is the Virasoro chiral block (or the global one) and $\s$ is as in~\eqref{eq:sigmarule}, with $\s=1$ in the \texttt{scalar\_G} basis. Such a table is requested with \texttt{--large-spin-lmin}~$=L_{\min}$ and the twists with \texttt{--large-spin-twists}. The interpolation variable is $x=L-L_{\min}$, and the base per unit of $L$ is $16q(\half)/a_0$ for Virasoro blocks and $4\r(\half)/a_0$ for global blocks. The poles are at $L=h_{r,s}-\tau/2$ in the Virasoro case and $L=-p/2-\tau/2$ in the global case, with $0\leq p<M$ at order $M$ and removable poles omitted. These tables are used in section~\ref{sec:momentbounds} to impose positivity along the double-twist towers $\tau_n=2\De_\f+2n$ for $L\geq L_{\min}$, with fewer retained poles than for the fixed-spin blocks.

\begin{table}[!htbp]
\centering
\footnotesize
\setlength{\tabcolsep}{4pt}
\renewcommand{\arraystretch}{1.1}
\begin{tabular}{@{}lllccc@{}}
\toprule
family & bases & coordinates & variable & base $E$ & \texttt{npoles} \\\midrule
\texttt{virasoro} & \texttt{scalar\_G}, \texttt{ordered\_B}, & sphere, all & $\De-\De_0$ & $16q(\half)/a_0$ & all \\
 & \texttt{parity\_projected\_B}, \texttt{chiral\_F} & & & & \\
\texttt{global} & \texttt{global\_*} (same four) & sphere, all & $\De-\De_0$ & $4\r(\half)/a_0$ & chosen \\
\texttt{ns}, \texttt{ramond} & \texttt{ns\_*}, \texttt{ramond\_*} (same four) & sphere, all & $\De-\De_0$ & $16q(\half)/a_0$ & all \\
\texttt{virasoro}, \texttt{global} & large-spin towers, 2d bases & sphere, all & $L-L_{\min}$ & $E_{\rm sphere}$ & chosen \\
\texttt{torus-character} & \texttt{torus\_G} & \texttt{vvb} & $\De-|j|$ & $e^{-2\pi}$ & -- \\
\texttt{torus-onepoint} & \texttt{torus\_onept\_G} & \texttt{vvb} & $\De-\De_0$ & $e^{-2\pi}$ & chosen\\
\bottomrule
\end{tabular}
\caption{Block families produced by \vb, with their bases, admissible coordinates, interpolation variable, exponential base, and default number of retained poles. ``Sphere, all'' means $z\zb$, $y\yb$, $q\qb$, the folded $xt$, $ws$, $q\qb$-$ws$, and their radial slices, subject to the basis restrictions in section~\ref{sec:output}, and chiral bases use $z$, $y$, $q$ with $h$ in place of $\De$. ``All'' means that every pole is kept and ``chosen'' that the number of poles is set automatically by default. Every family also provides point tables at a single exchanged weight.}
\label{tab:families}
\end{table}

\FloatBarrier

\subsection{Code use and output}
\label{sec:output}

\vb\ is invoked from the command line. The command
\begin{lyxcode}
virasoro\_blocks table --family virasoro --basis scalar\_G \symbol{92}\\
\ \ --central-charge 10 --spin-ranges 0:2:10 --lambda 12 \symbol{92}\\
\ \ --delta-ext 0.15 0.15 0.15 0.15 --order 30 --npoles 30 \symbol{92}\\
\ \ --precision 120 --format json -o out/
\end{lyxcode}
produces, for each even spin $0\leq\ell\leq10$, a table of the derivatives~\eqref{eq:taylortable} of the blocks~\eqref{eq:scalarG} up to order $\L=12$ at $c=10$ with four identical external scalars of dimension $\De_\f=0.15$, set by \texttt{--delta-ext}, computed at recursion order $N=30$ and interpolated with 30 retained poles, working at 120 decimal digits. The main options are:
\begin{itemize}
\item \texttt{--family}: the sphere families are \texttt{virasoro}, \texttt{global}, \texttt{ns}, and \texttt{ramond}, and the default \texttt{auto} infers the family from the basis. The torus families are \texttt{torus-character} and \texttt{torus-onepoint} (section~\ref{sec:families}). They require \texttt{--surface torus}, with the one-point external weight $d$ given by \texttt{--torus-onept-ext}, and large-spin towers are requested with the \texttt{--large-spin-*} options.
\item \texttt{--basis}: \texttt{scalar\_G}, \texttt{ordered\_B}, \texttt{parity\_projected\_B}, or the chiral block \texttt{chiral\_F} alone (a function of $h$ only), with corresponding prefixed versions for the other families.
\item The central charge and external quantum numbers: \texttt{--central-charge}, with either \texttt{--delta-ext} and \texttt{--spin-ext}, or \texttt{--h-ext} and \texttt{--hbar-ext}.
\item \texttt{--spin-ranges}: the internal spins, given as \texttt{start:step:stop} ranges.
\item \texttt{--lambda}, \texttt{--order}, \texttt{--npoles}, \texttt{--precision}: the derivative cutoff $\L$, the recursion order $N$, the number of retained poles (or \texttt{all}), and the working precision in decimal digits.
\item \texttt{--pole-selection}, \texttt{--sample-via}: the pole-selection rule and the way block values at the nodes are computed (section~\ref{sec:interp}).
\item \texttt{--coordinates}: the cross-ratio coordinates in which derivatives are taken (see below).
\item \texttt{--chiral-norm}, \texttt{--z-shift}, \texttt{--zbar-shift}, \texttt{--u-shift}: the normalization constant $a_0$ and the shifts $\de_z$, $\de_{\zb}$ of~\eqref{eq:libnorm}.
\item \texttt{--gap}: raises the shift $\De_0(\ell)$ in~\eqref{eq:interpform} above the unitarity bound.
\item \texttt{--regularize}, \texttt{--kac-limit}: the treatment of singular central charges and of blocks on Kac weights (section~\ref{sec:implementation}).
\item \texttt{--target-digits}: a digit target that raises the working precision when needed.
\item \texttt{--num-threads}: the number of worker processes over which the spins are distributed.
\item \texttt{--format}: \texttt{json}, \texttt{m} (a Mathematica association in the style of \texttt{scalar\_blocks}), or \texttt{both}.
\end{itemize}
The command \texttt{virasoro\_blocks info} lists the available families, bases, coordinates, and operations, and \texttt{virasoro\_blocks evaluate} evaluates a single block numerically at a point in cross-ratio space. The command \texttt{virasoro\_blocks point-table} writes the derivative table at a single exchanged weight, for example an isolated operator. At a degenerate weight it gives the block or character of the irreducible module, including the vacuum and chiral-current characters on the torus. The same functionality is available from \texttt{Python} through the module \texttt{virasoro\_blocks.api}.

\paragraph{Coordinates.} Several choices of cross-ratio coordinates are available:
\begin{itemize}
\item $z,\zb$ (\texttt{zzb}): the coordinates defined above, with crossing-symmetric point $z=\zb=\half$ and crossing acting as $z\to1-z$.
\item $y,\yb$ (\texttt{yyb}): the coordinates~\cite{Mazac:2016qev} $z=\frac{(1+y)^2}{2(1+y^2)}$, i.e.\ $y=\frac{2z-1}{1+2\sqrt{z(1-z)}}$, which uniformize the cut $z$-plane. The crossing-symmetric point is $y=\yb=0$ and crossing acts as $y\to-y$.
\item $q,\qb$ (\texttt{qqb}): the elliptic nome~\eqref{eq:nome}, with crossing-symmetric point $q=\qb=e^{-\pi}$.
\item $x,t$ (\texttt{xt}), $w,s$ (\texttt{ws}), and their nome analogs (\texttt{qqb\_ws}): the ``folded'' coordinates $x=\frac{z+\zb-1}{2}$, $t=\p{\frac{z-\zb}{2}}^2$, and similarly for $y,\yb$ and $q,\qb$, which make the symmetry under $z\leftrightarrow\zb$ manifest. Blocks that are odd under $z\leftrightarrow\zb$ are divided by $\frac{z-\zb}{2}$ before expanding, as in~\cite{Erramilli:2020rlr}.
\item The \texttt{\_radial} variants of the folded coordinates retain only the derivatives along the diagonal $z=\zb$.
\end{itemize}
At a diagonal expansion point, unfolded tables of definite $z\leftrightarrow\zb$ parity contain the entries with $m+n\leq\L$ and $n\leq m$, since the remaining entries follow by exchanging the two derivative indices (up to the parity sign). The \texttt{ordered\_B} basis retains the full $m+n\leq\L$ triangle. Independent expansion centers may be specified with \texttt{--z-center} and \texttt{--zbar-center} in unfolded coordinates, and off the diagonal every basis retains the full triangle.

In folded coordinates $\s$ must be real, and the cutoff is $m+2n\leq\L_\pm$, with $\L_+=\L$ for even exchange parity and $\L_-=\L-1$ for odd exchange parity, following the conventions of \texttt{blocks\_3d}. A \texttt{scalar\_G} block has even parity at every $\ell$, while for \texttt{parity\_projected\_B} the effective parity includes the chosen projection and the external and internal parity phases. Folded tables use the crossing-symmetric expansion point and do not support raw \texttt{ordered\_B}.

\paragraph{Output.} Each output file describes one spin and contains, for every retained derivative $(m,n)$, the coefficients of the polynomial $P^{m,n}_\ell(x)$ in ascending powers of $x=\De-\De_0(\ell)$, together with the data needed to reconstruct the approximation~\eqref{eq:interpform}: the base $E$ (\texttt{exp\_base}), the shift $\De_0(\ell)$ (\texttt{variable\_minus\_x}), the list of retained poles $x_i$ (\texttt{pole\_list\_x}), and the values of each entry at a common set of output nodes (\texttt{sample\_block\_values}). Constant factors such as $E^{\De_0}$ are absorbed into the polynomial coefficients, so that the value of an entry at dimension $\De$ is recovered as
\be
T^{m,n}_\ell(\De)\approx E^{x}\,\frac{\sum_k a^{m,n}_k x^k}{\prod_{i\in\cK}(x-x_i)},\qquad x=\De-\De_0(\ell).
\ee
High-precision numerical data are written as decimal strings at the working precision. Integer controls and indices remain JSON integers. The file also records the external weights, the central charge, the normalization $a_0$, the shifts $\de_z,\de_{\zb}$, the parity conventions used, the coordinates, the pole-selection rule and the requested $n_{\rm poles}$, the number of digits the table was verified to deliver, and the full command line, so that a table is self-describing.

The Mathematica format writes the polynomial data and metadata (without the node values) as an association with keys such as \texttt{zzbDeriv[m,n]}, \texttt{shiftedPoles}, \texttt{expBase}, and \texttt{shift}, in files named following the \texttt{scalar\_blocks} conventions. The torus tables are written only as JSON. For the torus characters the tables are exact (polynomial in $\De$ times $e^{-2\pi\De}$) with no pole list, while the torus one-point blocks have Kac poles and are interpolated like the sphere blocks.

\subsection{Implementation details}
\label{sec:implementation}

\vb\ requires \texttt{Python} 3.10 or later. Its dependencies are \texttt{mpmath}~\cite{mpmath}, \texttt{numpy}~\cite{Harris:2020numpy}, and the \texttt{python-flint} bindings~\cite{pythonflint} to \texttt{FLINT} and \texttt{Arb}~\cite{Hart:2010flint,Johansson:2017arb}. The numerically intensive steps (the elliptic integrals and theta functions, power-series arithmetic, the recursion, the interpolation nodes, and the block values at them) run in the ball arithmetic of \texttt{Arb}, whose error estimates are used in constructing interpolation nodes and, near singular central charges, in setting the internal precision of the recursion from the measured loss. The working precision of the output is the user's and is never raised silently.

The recursion exploits the reality of the summed $q$-series for real central charge and weights, even though individual Kac weights and residues are complex for $1<c<25$. In this regime the pairs $(r,s)$ and $(s,r)$ give complex-conjugate data, and the library stores only $r\leq s$ and conjugates the rest. Pairing the factors $(p,k)$ and $(-p,-k)$ in the fusion polynomials~\eqref{eq:Prs} removes the external-momentum square roots. The products can still be complex in this strip. Real arithmetic is used for $c\ge25$ and $c<1$ as well. For $c\le1$, where $b=i\b$ with $\b$ real, the recursion is written in terms of $\b$.

Tables for different spins can be computed independently, so their computation is distributed over worker processes (\texttt{--num-threads}). The recursion tables, nome evaluations, and chiral evaluators are cached, and long power series are multiplied by Kronecker substitution. Preparing the tables and the semidefinite program for one $\L=19$ moment computation in section~\ref{sec:momentbounds} (parameters in appendix~\ref{app:params}) takes one to two minutes on four cores, far less than the subsequent \sdpb\ solve.

\paragraph{Singular central charges.} Individual terms of~\eqref{eq:hrecursion} can be singular even when their sum is finite. To compute the limit, these terms must be combined before the central charge is specialized. The singularities occur when a factor $mb+n/b$ in~\eqref{eq:Ars} vanishes, so that $b^2=-n/m$ is rational. These are the generalized minimal-model central charges $c=1-6(p-p')^2/pp'$ at $c<1$, the values at $c>25$ such as $c=26$ ($b^2=2/3$) and $c=28$ ($b^2=1/2$), and the branch points $c=1,25$ ($b^2=\mp1$). \vb\ detects them exactly (\texttt{--regularize auto}) by reading the central charge as an exact rational. A value near a locus is evaluated at that value rather than at the locus.

At a singular charge $c_0$, the recursion is expanded as a Laurent series in a local variable $t$, with $c=c_0+\sigma t^p$, where $p=2$ at the branch points $c_0=1,25$ and $p=1$ elsewhere. At $c_0=1$ the expansion is taken on the side $c=1-t^2$, where all coefficients are real. The constant coefficients of factors $mb+n/b$ that vanish at $c_0$ are set to exact zero, while their higher powers in $t$ are retained. The constant term gives the finite limit, and a surviving negative power signals a genuine singularity and raises an error.

Contributions from colliding Kac poles are combined into a rational function of the exchanged weight with the appropriate higher-order poles. Both the exact and the order-reduced tables use this merged representation, for the symbolic numerators and denominators, the block values at the interpolation nodes, and the principal parts used to rank poles. All of these are evaluated at $c_0$ itself rather than extrapolated from nearby $c$, and extrapolation serves only as a test.

For sphere Virasoro blocks, when the exchanged weight sits exactly on a Kac weight and the external weights satisfy the required null-vector decoupling conditions, direct block evaluation returns the block of the irreducible module, given by the constant term in $t$ of the recursion evaluated on the Kac curve $h=h_{r,s}(c)$, with the external weights kept on the fusion rule that decouples the null vector. The limit $c\to c_0$ at fixed weight agrees with it at the minimal models and at $h=0$, but at $c=1$ and $25$, where Kac poles of the same level merge, it retains finite contributions of the null submodule. The option \texttt{--kac-limit} selects either convention. The $\De$ tables carry the continuation in the weight, so they can differ from direct evaluation and point tables at such isolated points. For torus one-point blocks at a degenerate weight and singular $c$, direct evaluation and point tables instead give the trace over the irreducible module, while the $\De$ tables keep the Verma-module continuation.

The table polynomials can have degree in the hundreds and lose precision in the monomial basis. Every table is therefore checked against direct block evaluation at the same order $N$ between its interpolation nodes and, for the Virasoro, $\cN=1$, and torus one-point tables, also at large $\De$ beyond them. Torus one-point tables are also checked on both sides of representative poles inside the positivity region. Point tables are checked against a reference computed at higher precision. The digits delivered are recorded, and an optional digit target (\texttt{--target-digits}, with \texttt{--max-precision}) raises the precision when needed or stops with an error.

The test suite checks Virasoro blocks against the closed-form Ising blocks at $c=\half$~\cite{Belavin:1984vu}, including mixed correlators~\cite{Maloney:2016kee}; the vacuum block against its global-block expansion through level six~\cite{Perlmutter:2015iya}; and the large-$c$ limit against the global block and its level-one coefficient $\frac{(h+h_2-h_1)(h+h_3-h_4)}{2h}$. It also checks the vanishing of the identity pole for pairwise identical external operators, compares the Laurent and extrapolation treatments of singular $c$, and compares the exact rational representation with direct numerical evaluation at central charges from $c<0$ to $c=100$, singular values included.

For Neveu--Schwarz blocks, the tests use the closed forms of~\cite{Hadasz:2007nt} at $c=3/2$. Ramond blocks were checked against the block decomposition of the tricritical Ising model at $c=\frac7{10}$. Torus one-point blocks are compared with a brute-force Gram-matrix computation and with their known limits ($f=1$ at $d=0$, $f\to{}_2F_1(d,1-d;2h;q/(q-1))$ as $c\to\oo$).

At the level of full correlators, the crossing equation for the Ising four-point function $\<\s\s\s\s\>$ is verified using the tables, and the bootstrap examples of section~\ref{sec:results} reproduce known exact results in Liouville theory and in the theories of appendix~\ref{app:models}. The BPZ Ising closed forms, the explicit low-level vacuum expansion of~\cite{Perlmutter:2015iya}, and the $\cN=1$ closed forms of~\cite{Hadasz:2007nt} provide independent references. We have not compared against another numerical implementation of the recursion, such as that of~\cite{Chen:2017yze}.

\section{Results}
\label{sec:results}

We apply \vb\ to two bootstrap problems for the four-point function of an identical scalar Virasoro primary $\f$ of dimension $\De_\f$ in a unitary CFT with central charge $c>1$. Crossing symmetry of $\<\f\f\f\f\>$ gives
\be
\label{eq:crossing}
\sum_{\cO}\l_{\cO}^2\,F^{-}_{\De_\cO,\ell_\cO}(z,\zb)=0,\qquad F^{-}_{\De,\ell}(z,\zb)\equiv G_{\De,\ell}(z,\zb)-G_{\De,\ell}(1-z,1-\zb),
\ee
where the sum runs over the scalar primaries and the paired nonzero-spin channels in the $\f\x\f$ OPE, including the identity, with $\l_\cO^2\geq0$ by unitarity. We fix $\l_{\vac}^2=1$, and the multiplicities implied by~\eqref{eq:scalarG} are specified in appendix~\ref{app:models}. Here the blocks are taken in the normalization~\eqref{eq:libnorm} with $\de_z=\de_{\zb}=-\De_\f$, which includes the kinematic prefactor $(z\zb)^{-\De_\f}$ of the four-point function, so that no factors of $u=z\zb$ or $v=(1-z)(1-\zb)$ appear. Only even spins appear.

The companion package \vbs\ turns block tables into semidefinite programs. It builds the vector of Taylor coefficients of $F^-$ at $z=\zb=\half$ with $m+n$ odd and $n\leq m$, which has $21$ components at $\L=11$ and $55$ at $\L=19$, assembles a polynomial matrix program in $x=\De-\De_0(\ell)$ with one positivity constraint per spin, and writes it for \sdpb~\cite{Simmons-Duffin:2015qma,Landry:2019qug}. It provides builders for the standard single-correlator problems, namely gap feasibility, OPE-coefficient maximization and two-sided bounds, fixed-OPE feasibility, and the moment ratios of section~\ref{sec:momentbounds}. It also sets up the genus-one modular bootstrap of the reduced partition function using the torus characters of section~\ref{sec:families}, a combined torus-plus-sphere problem with an isolated relevant scalar, and the crossing-invariant slice $\zb=\half$ of the four-point equation, on which the parity-projected block factorizes. Without conserved currents other than $T$, its modular bootstrap reproduces the scalar-gap curve of~\cite{Collier:2016cls}.

\vbs\ checks the returned functionals by combining the polynomial coefficients in exact rational arithmetic and determining their sign between isolated real roots, and by evaluating physical blocks at additional spins and tower twists. A small cluster layer (\texttt{vb-submit}, \texttt{vb-collect}, \texttt{vb-certify}) expands parameter sweeps into SLURM jobs from a site-independent profile.

One step of the assembly is specific to Virasoro blocks. For $1<c<25$ the retained poles include complex-conjugate pairs, whose product $(x-x_i)(x-\bar x_i)$ is positive on the real axis, so positivity is unaffected, but \sdpb's preprocessing expects a damped-rational prefactor with real poles. Moving the pairs into the polynomial would preserve positivity but destroy the conditioning that the optimal interpolation achieves. \vbs\ therefore bypasses this preprocessing. It computes the sample points, sample scalings, and orthonormal bilinear bases itself from the full prefactor $\chi_\ell(x)$, complex pairs included, and carries the block values recorded at the table nodes to the sample points by Lagrange interpolation in product form, so that the high-degree polynomial is never evaluated in the monomial basis. The constraint matrices are then written in \sdpb's internal \texttt{sdp} format, which also avoids evaluating the polynomial coefficients at the solver precision~\cite{Chang:2025mwt}.

Since a table records its nodes, node values, and parameters, the same files can be reused for problems with a different gap, a different objective, or a subset of the spins, and \vbs\ checks that a table's normalization, pole selection, and $n_{\rm poles}$ agree with the problem before using it. The identity contribution is the vacuum block evaluated directly at $\De=0$, using the Laurent regularization at singular central charges. In the moment computations we also impose positivity along double-twist towers $\tau_n=2\De_\f+2n$ with continuous spin $\ell\geq\ell_{\min}$, using the large-spin tables of section~\ref{sec:families}. The central charge enters only through the blocks, including the vacuum block. The numerical parameters are collected in appendix~\ref{app:params}.

\subsection{Gap bounds assuming only scalars}
\label{sec:gapbounds}

The simplest single-correlator bootstrap problem, an upper bound on the dimension of the lightest scalar in the $\f\x\f$ OPE, is not useful in two dimensions without further assumptions. In our experiments the scalar gap admits no bound at the derivative orders we can reach, plausibly because the blocks of operators of small twist, and in the limit holomorphic currents, are nearly degenerate with the identity block at the crossing-symmetric point. Following~\cite{Collier:2017shs}, we therefore consider theories whose Virasoro primary spectrum consists only of scalars. In~\cite{Collier:2017shs} strong numerical evidence was found that an OPE $\f\x\f$ containing only scalar Virasoro primaries in a $c>1$ CFT must be governed by the structure constants of Liouville theory, whose spectrum is a continuum starting at
\be
\De=\frac{c-1}{12}\qquad\p{h=\hb=\frac{Q^2}{4}},
\ee
with structure constants given by the DOZZ formula~\cite{Dorn:1994xn,Zamolodchikov:1995aa}.

\begin{figure}[t]
\centering
\includegraphics[width=0.95\textwidth]{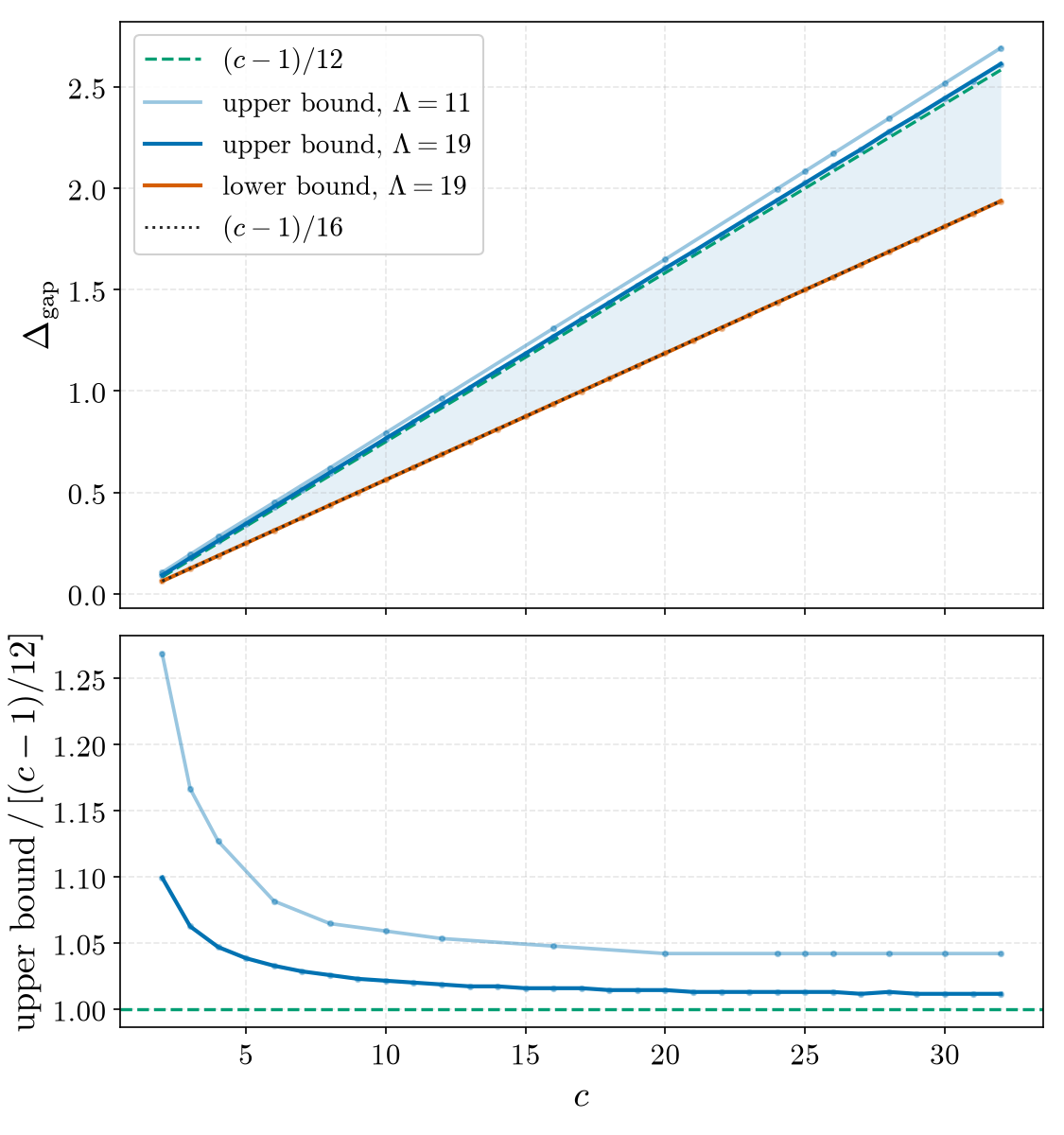}
\caption{Top: the window not excluded by the truncated scalar-gap problem, as a function of the central charge, for four-point functions of an identical scalar $\f$ with $\De_\f=\De_\gap$ in theories whose Virasoro primary spectrum contains only scalars. The upper bound is shown at $\L=11$ (light) and $\L=19$ (dark). The lower bound and the shading are at $\L=19$. The green dashed line is the Liouville threshold $\De=(c-1)/12$, and the dotted line $\De=(c-1)/16$ is drawn over the lower bound. Bottom: the upper bound divided by $(c-1)/12$. The bisection tolerance is $(c-1)/1200$ at $\L=11$ and $(c-1)/6000$ at $\L=19$. The central charges $c=25$, $26$, and $28$ are singular values of the recursion, where the blocks are regularized.}
\label{fig:scalaronly}
\end{figure}

We take the external scalar to be the lightest scalar, $\De_\f=\De_\gap$, and ask for which values of $\De_\gap$ the crossing equation~\eqref{eq:crossing} admits a solution with only scalar primaries of dimension $\De\geq\De_\gap$ (together with the identity). The allowed region is a window in $\De_\gap$, and we determine both bounds by bisection, at $\L=11$ and $19$, for $2\leq c\leq 32$. The result is shown in figure~\ref{fig:scalaronly}.

The upper bound lies above the Liouville threshold $(c-1)/12$ by about $27\%$ at $c=2$, a relative excess that decreases with $c$ to $4\%$ at $c=20$ and stays there, at the bisection resolution, up to $c=32$. At $\L=19$ the excess falls to $10\%$ at $c=2$ and to $1.2\%$ at large $c$, as the bottom panel of figure~\ref{fig:scalaronly} shows. The conjectured uniqueness of the scalar-only Liouville solution~\cite{Collier:2017shs} makes $(c-1)/12$, where its continuum begins, the expected limiting upper bound of this numerical window. The lower bound coincides, within the bisection tolerance, with
\be
\De_\gap=\frac{c-1}{16}=\frac{3}{4}\cdot\frac{c-1}{12}
\ee
at every $c$. This is the threshold found in~\cite{Collier:2017shs} (and in unpublished work of B.~van Rees cited there), where it was observed that the scalar-only crossing equations admit solutions for external dimensions $\De_\f\geq(c-1)/16$ but none for $\De_\f<(c-1)/16$, independently of any gap assumption.

The explanation given in~\cite{Collier:2017shs} is the analytic continuation of the Liouville four-point function in the external momenta. Writing $\De_\f=2\a(Q-\a)$ and $\De=2\a_s(Q-\a_s)$, the Liouville four-point function is an integral over $\a_s\in\frac Q2+i\R$ of DOZZ structure constants times blocks, and it can be continued from $\a\in\frac Q2+i\R$ to real $\a<\frac Q2$ as long as no pole of $C(\a,\a,\a_s)$ crosses the contour, which holds for $2\a>\frac Q2$. At $\a=\frac Q4$, or $\De_\f=\frac{3Q^2}8=\frac{c-1}{16}$, the pole at $\a_s=2\a$ hits the contour, and below it the continued correlator picks up a residue that violates unitarity. For $(c-1)/16<\De_\f\leq(c-1)/12$ this gives a scalar-only crossing solution with non-negative internal spectral density and continuum $\De\geq(c-1)/12\geq\De_\f$. The lower endpoint is understood as a limit.

The external insertions below the continuum threshold are non-normalizable, and the Liouville solution has no vacuum contribution. Its spectral density can, however, be multiplied by any constant $A>0$, since its crossing equation is homogeneous. Adding an identity contribution with coefficient $1$ then gives a relative crossing residual of order $A^{-1}$, which falls below the numerical feasibility tolerance for sufficiently large $A$. The rescaled Liouville solutions therefore pass the feasibility test of the truncated problem with unit identity coefficient. This explains why the observed lower bound tracks $(c-1)/16$ throughout the scan.

With fewer retained poles ($n_{\rm poles}=32$) the lower bound sat visibly above $(c-1)/16$, e.g.\ at $0.0703$ instead of $0.0623$ for $c=2$, and moved onto it at $n_{\rm poles}=48$, an example of the spurious exclusions discussed in section~\ref{sec:interp}. Table~\ref{tab:scalaronly} lists a subset of the numerical values.

\begin{table}[tb]
\centering
\renewcommand{\arraystretch}{1.0}
\begin{tabular}{@{}ccccccc@{}}
\toprule
& \multicolumn{2}{c}{upper bound} & & \multicolumn{2}{c}{lower bound} & \\
$c$ & $\L=11$ & $\L=19$ & $(c-1)/12$ & $\L=11$ & $\L=19$ & $(c-1)/16$\\\midrule
2 & 0.10572 & 0.09162 & 0.08333 & 0.06228 & 0.06248 & 0.0625\\
4 & 0.28176 & 0.26176 & 0.25 & 0.18685 & 0.18745 & 0.1875\\
8 & 0.62110 & 0.59839 & 0.58333 & 0.43598 & 0.43737 & 0.4375\\
16 & 1.30969 & 1.26987 & 1.25 & 0.93425 & 0.93723 & 0.9375\\
24 & 1.99734 & 1.94170 & 1.91667 & 1.43252 & 1.43708 & 1.4375\\
32 & 2.69207 & 2.61342 & 2.58333 & 1.93079 & 1.93693 & 1.9375\\
\bottomrule
\end{tabular}
\caption{Upper and lower bounds of figure~\ref{fig:scalaronly}, given by the midpoints of the final bisection brackets and compared with $(c-1)/12$ and $(c-1)/16$.}
\label{tab:scalaronly}
\end{table}

\subsection{Moment bounds at different central charges}
\label{sec:momentbounds}

At the crossing-symmetric point, each exchanged Virasoro representation makes a positive contribution to the correlator. Normalizing these contributions gives a probability distribution whose mean measures the typical exchanged dimension. Following the ``moment bootstrap'' of~\cite{Chiang:2026nmd,Poland:2024hvb}, we bound this mean using crossing symmetry. Define the damped moments
\be
\label{eq:moments}
\nu_{m,n}\equiv\sum_{\cO}\l_\cO^2\,G_{\De_\cO,\ell_\cO}(\thalf,\thalf)\,\De_\cO^m\,\ell_\cO^{\,n},
\ee
where $p_\cO=\l_\cO^2G_{\De_\cO,\ell_\cO}(\half,\half)$ is the contribution of a scalar primary or a paired nonzero-spin channel. The normalized weights are $\widehat p_\cO=p_\cO/\nu_{0,0}$. OPE convergence at $|q|=e^{-\pi}$ makes every moment finite. The weights are independent of the block normalization when $\l_\cO^2$ is expressed consistently. These are Virasoro analogs of the global-block moments of~\cite{Chiang:2026nmd}, and organizing the OPE into Virasoro primaries allows comparisons at fixed $c$. The library accepts polynomial weights in $\De$, $\ell$, $h$, and $\hb$, but here we use only the first $\De$-moment.

The zeroth moment is proportional to the correlator itself, whose bounds~\cite{Paulos:2021jxx} are closely related to the spectral-function bootstrap~\cite{Collier:2017shs}. It can be unbounded above at fixed $\De_\f$, motivating the normalized observables
\be
\label{eq:avgdelta}
\<\De\>\equiv\frac{\nu_{1,0}}{\nu_{0,0}},\qquad\<\De\>_*\equiv\frac{\nu_{1,0}}{\nu_{0,0}-\nu_{0,0}^{\vac}},\qquad \nu_{0,0}^{\vac}\equiv G_{0,0}(\thalf,\thalf).
\ee
The first is the mean of $\widehat p_\cO$, and the second is its mean conditioned on a non-vacuum representation. The identity Virasoro primary contributes the whole vacuum module, including the stress tensor and its descendants, so $\<\De\>_*$ subtracts the full vacuum block, which depends on $c$ and $\De_\f$. This differs from the standard global-block convention of~\cite{Chiang:2026nmd}, where the identity contributes $1$.

Bounds on $\<\De\>$ follow from a single semidefinite program. Adding to the crossing functional dual variables for the linear constraints $\nu_{0,0}=\sum_\cO p_\cO$ and $\nu_{1,0}=\sum_\cO\De_\cO p_\cO$, one rescales all spectral weights, including the identity coefficient, by $1/\nu_{0,0}$ and extremizes the rescaled $\nu_{1,0}$ with the rescaled $\nu_{0,0}=1$. Crossing remains homogeneous under this rescaling. This linear-fractional problem has a dual that is directly an \sdpb\ problem~\cite{Chiang:2026nmd}, and $\<\De\>_*$ is treated in the same way with the identity term moved to the other side. We assume only unitarity, $\De\geq\ell$ for all even $\ell$, imposed for $\ell\leq24$ and along double-twist towers at large spin. We check stability under the block and spin truncations and evaluate the functionals at higher spins and additional tower twists (appendix~\ref{app:params}). Positivity is therefore checked only on a finite set of spins and twists. The problems require only the tables for a single $\De_\f$, whose $(0,0)$ entries are the values $G_{\De,\ell}(\half,\half)$.

\newlength{\momentfigwidth}
\setlength{\momentfigwidth}{1.1\textwidth}
\newcommand{\momentfig}[5][0.86]{%
\begin{figure}[p]
\centering
\makebox[\textwidth][c]{\begin{adjustbox}{max width=\momentfigwidth,max totalheight=#1\textheight}%
\begin{tabular}{@{}c@{}}\includegraphics[width=\momentfigwidth]{figures/#2}\\[2pt]\includegraphics[width=\momentfigwidth]{figures/#3}\end{tabular}%
\end{adjustbox}}
\caption{#4}
\label{#5}
\end{figure}}

We have computed these bounds for $0<\De_\f\leq4$, at $\L=19$ for $c=8,4,2$ and the rational values $c=3/2,6/5,81/70,8/7$, and at $\L=11$ near $c=1$. At $c=8,4,2$ no compact unitary CFT with only Virasoro symmetry is known, and we compare with Liouville theory and with theories with extended chiral algebras, namely free bosons, products of Ising models, and the tricritical Ising model (TIM) times a spectator $\cT$. At the rational values many exactly solvable theories are available, and near $c=1$ the whole $c=1$ moduli space is available for comparison. The exact values of $\<\De\>$ for the known theories were computed as described in appendix~\ref{app:models} and are distributed as data.

In all plots the dark solid curves are the bounds, the shaded region is allowed, and the dashed curve is Liouville theory at that $c$. Each point is a scalar Virasoro primary of a known theory. When several primaries share a dimension, any linear combination is an allowed external operator and the point becomes an interval, drawn as a vertical bar. The free-boson curves show single vertex operators. Every known theory that we have checked lies inside the allowed region at the rational and generic central charges, as it must, and a remarkable number lie on or near its boundary.

\subsubsection{Larger central charges: \texorpdfstring{$c=8,4,2$}{c=8,4,2}}
\label{sec:moments-larger}

Figures~\ref{fig:momentc8},~\ref{fig:momentc4}, and~\ref{fig:momentc2} show the bounds at $c=8$, $4$, and $2$. With the identity excluded (top panels) the allowed region is wide, and its upper bound has a kink at $\De_\f=\frac18$ at all three central charges, below which it drops rapidly to zero, since a very light external operator forces the non-identity part of the correlator to be light as well. Above $\De_\f\approx1$ the upper bound rises almost linearly up to $\De_\f=4$, without further features.

The kink is saturated by an Ising spin field tensored with a spectator identity. Any CFT that factorizes as $\mathrm{Ising}\x\cT$ with $\cT$ unitary of central charge $c-\half$ contains the scalar $\s\otimes\vac$ of dimension $\frac18$, whose four-point function is the Ising one but whose decomposition into Virasoro blocks at central charge $c$, obtained by the triangular decomposition of product theories in appendix~\ref{app:models}, depends on $c$. Its exact $\<\De\>_*$ coincides with the bound at the kink to four decimals at all three central charges, as it does at $c=6/5$ and $3/2$ (section~\ref{sec:moments-rational}). With the identity included, $\s\otimes\vac$ also saturates the upper bound at $c=4$ and $8$ and lies just inside it at $c=2$, and at $c=8$ so does $\e\otimes\vac$ in $\mathrm{TIM}\x\cT$ at $\De_\f=\frac15$.

To the resolution of the plots the lower bound follows Liouville theory up to $\De_\f\approx2$ at $c=8$, $1.2$ at $c=4$, and $0.9$ at $c=2$, beyond which the bound falls below the Liouville curve. At $\De_\f=4$ it lies $7\%$, $20\%$, and $40\%$ below Liouville at $c=8$, $4$, and $2$. At large enough $c$ or small enough $\De_\f$ the Liouville four-point function thus lies on the lower bound. The departure at smaller $c$ and larger $\De_\f$ may mean that the bounds have not converged in $\L$, or that a solution with a lighter spectral density exists there.

\momentfig[0.81]{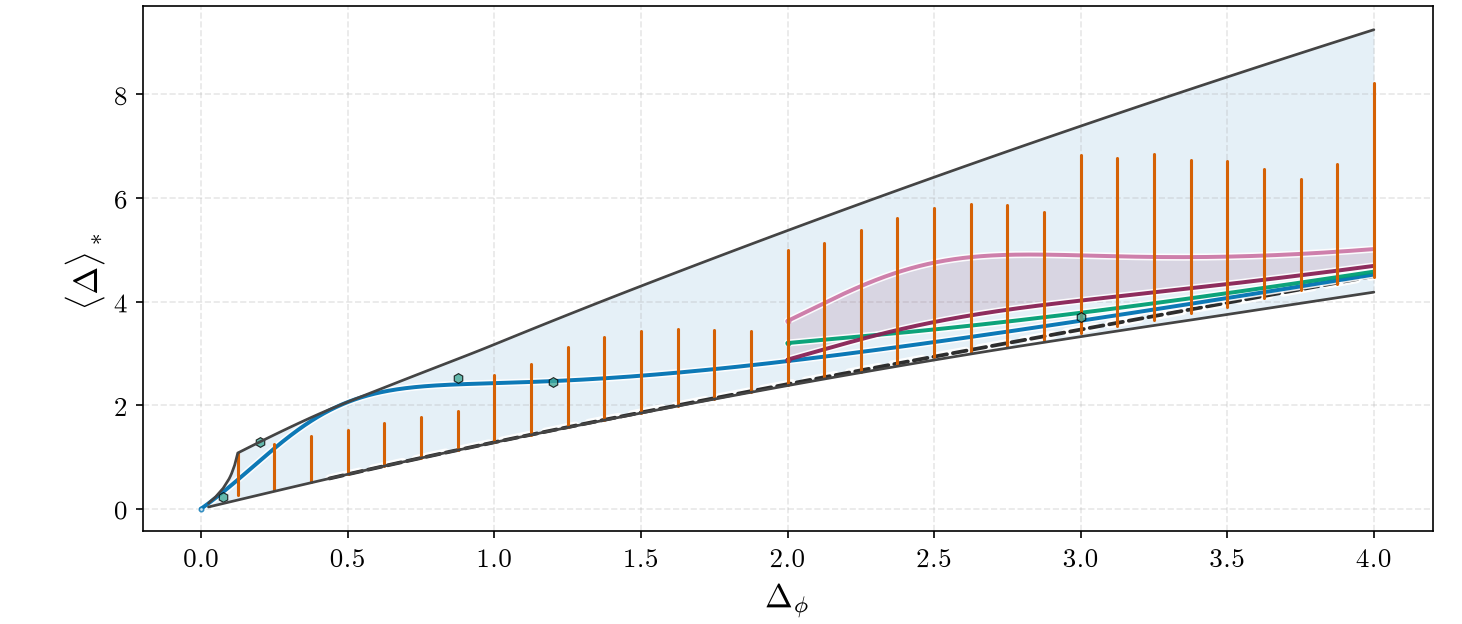}{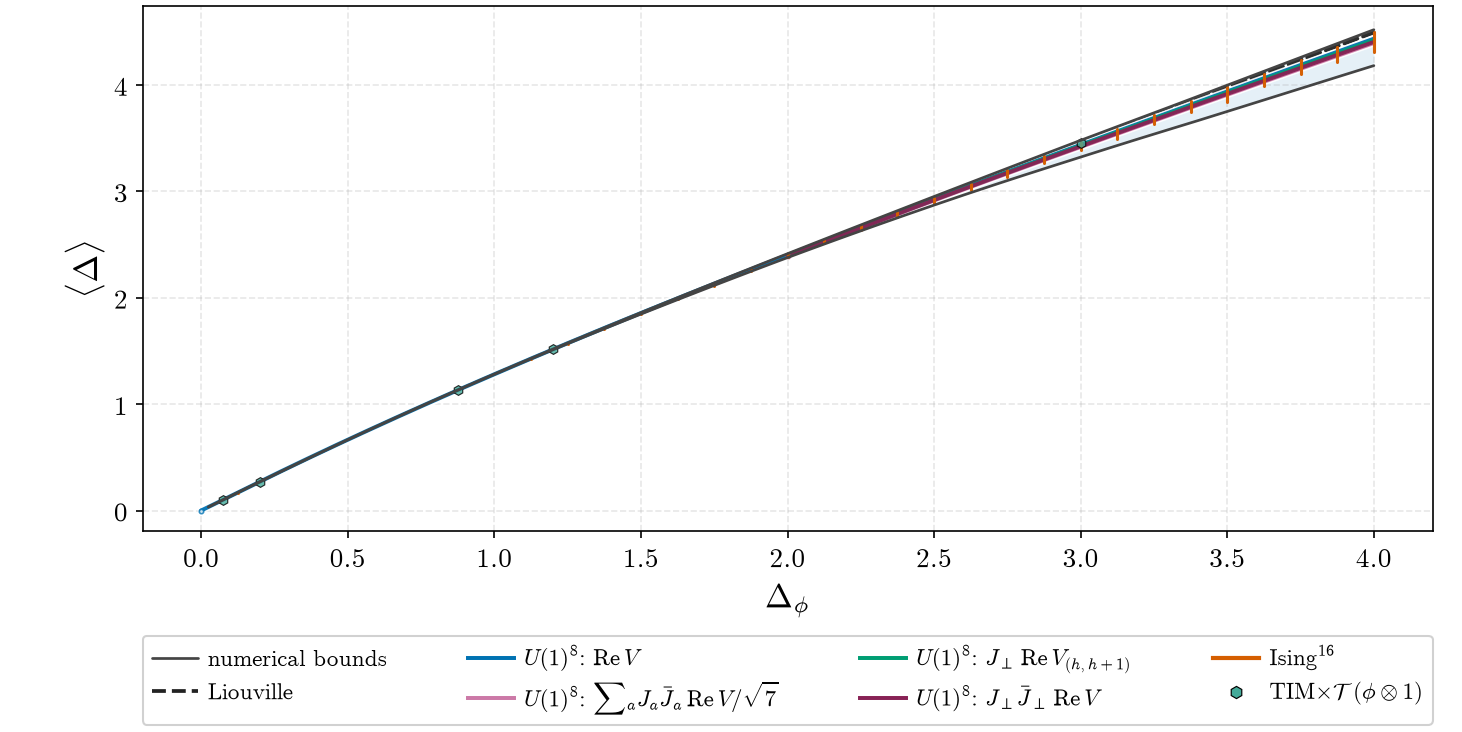}{Upper and lower bounds (solid) on the normalized first moment at $c=8$, $\L=19$, as a function of $\De_\f$. Top: identity excluded, $\<\De\>_*$. Bottom: identity included, $\<\De\>$. The dashed curve is Liouville theory, computed from the DOZZ structure constants. The solid curves are the free bosons $U(1)^8$ (the vertex operator $\Re V$ and its current dressings), and shading between the two bilinear-current curves gives the range over their mixtures. Orange bars give the numerically estimated ranges over real mixtures in $\mathrm{Ising}^{16}$ (appendix~\ref{app:models}). Hexagons mark $\f\otimes\vac$ in $\mathrm{TIM}\x\cT$. Open circles mark limiting endpoints of the free-boson curves.}{fig:momentc8}

\momentfig{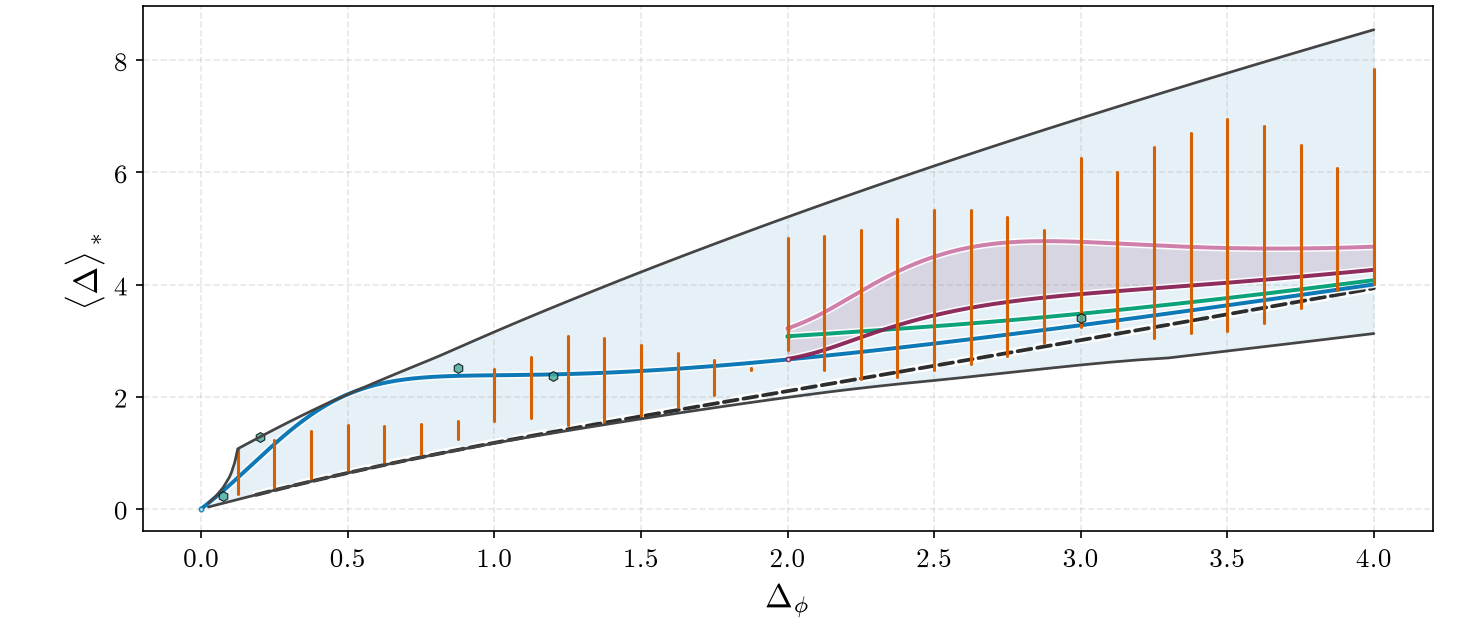}{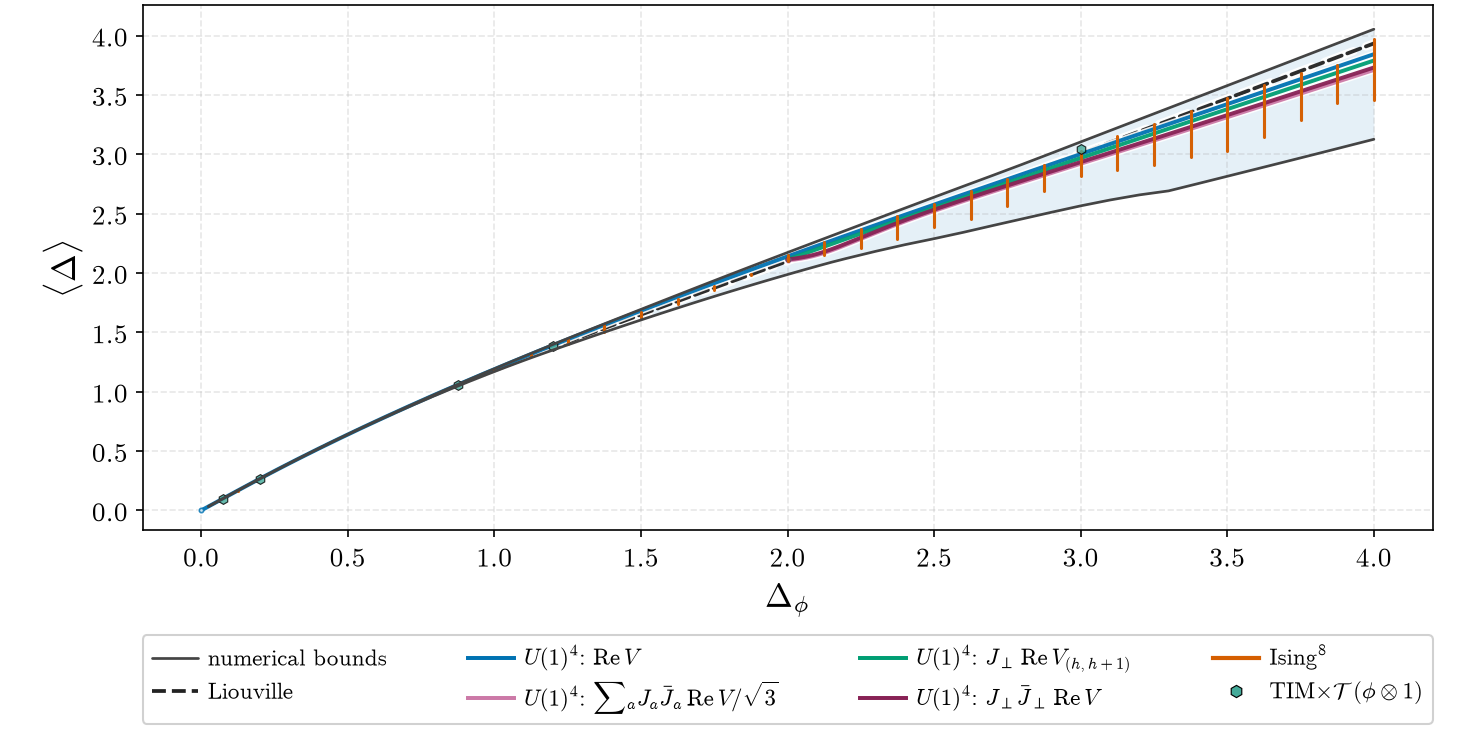}{Bounds on the normalized first moment at $c=4$, $\L=19$, with the identity excluded (top) and included (bottom), compared with Liouville theory (dashed), $U(1)^4$ (solid curves), $\mathrm{Ising}^8$ (orange mixture bars), and $\mathrm{TIM}\x\cT$ (hexagons), as in figure~\ref{fig:momentc8}.}{fig:momentc4}

\momentfig{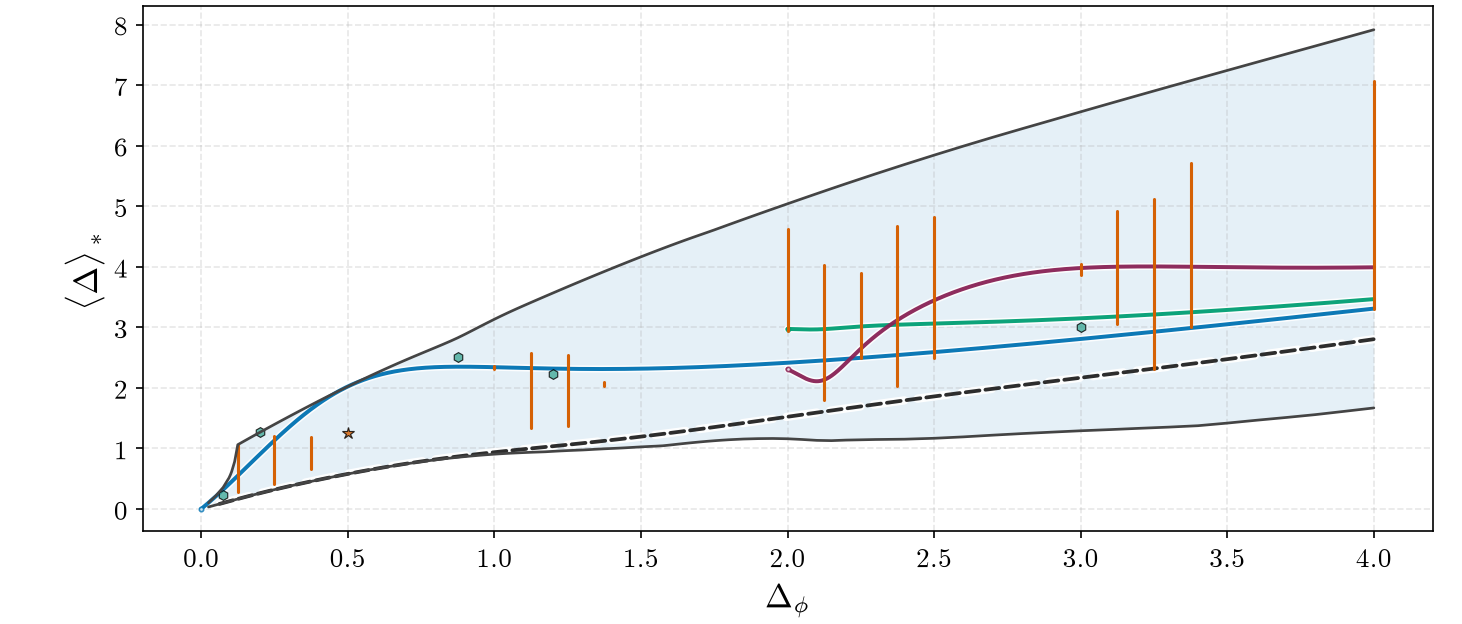}{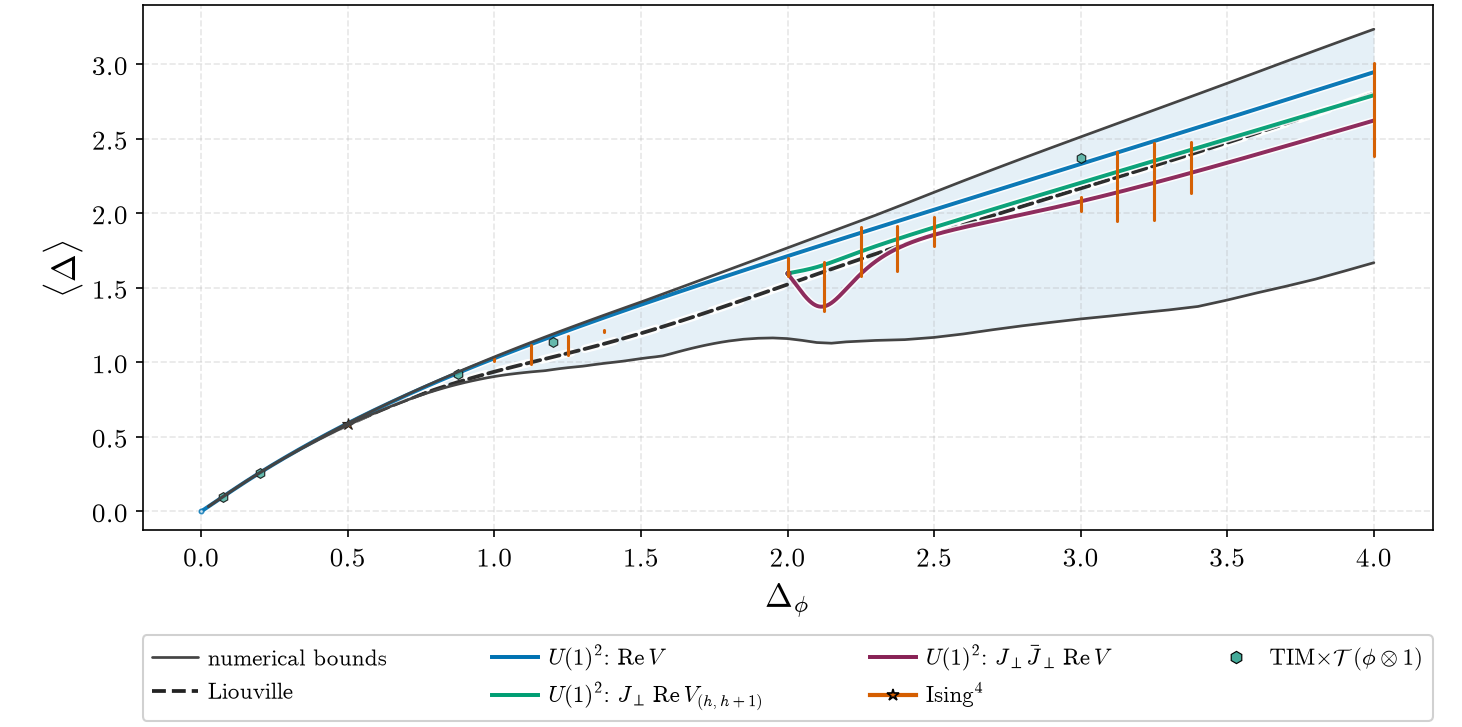}{Bounds on the normalized first moment at $c=2$, $\L=19$, with the identity excluded (top) and included (bottom), compared with Liouville theory (dashed), $U(1)^2$ (solid curves), $\mathrm{Ising}^4$ (orange mixture bars), and $\mathrm{TIM}\x\cT$ (hexagons), as in figure~\ref{fig:momentc8}. At $c=2$ the two bilinear-current curves coincide, so there is no shaded range. The star at $\De_\f=\half$ marks the isolated $\s^{\otimes4}$ operator of $\mathrm{Ising}^4$.}{fig:momentc2}

With the identity included (bottom panels) the allowed region is a narrow band whose width grows with $\De_\f$ and shrinks with $c$, so thin for $\De_\f\lesssim0.7$ that the two bounds are indistinguishable on the plot. A light external operator leaves little freedom in the spectral density once the identity is accounted for. Liouville theory lies inside the band at every computed point, and at $c=8$ it nearly touches both bounds. Numerically, the lower bounds with and without the identity coincide. At $c=8$ Liouville also runs along the upper bound for $0.62\lesssim\De_\f\lesssim0.8$, and stays within $1\%$ of it from $\De_\f\approx2$ to $4$.

With the identity excluded, the vertex operator $\Re V$ touches the upper bound near $\De_\f=\half$ at all three central charges, and at $c=4$ and $8$ the lowest Ising-product mixtures reach the lower bound, where it coincides with Liouville. Elsewhere these scalars lie inside the allowed region. With the identity included the band is so narrow at small $\De_\f$ that they lie close to both bounds, and at $c=8$ the vertex operator reaches the upper bound near $\De_\f=\half$ and the lower bound near $0.62$. A single real vertex operator has the same four-point function in every toroidal compactification, so the $U(1)^c$ curves apply to the whole Narain moduli space, including the $E_8$ lattice theory at $c=8$. Above $\De_\f=2$ at $c=8$ these scalars crowd just above the Liouville curve of the identity-excluded panel, close to the lower bound.

\subsubsection{Rational central charges: \texorpdfstring{$c=3/2,6/5,81/70,8/7$}{c=3/2, 6/5, 81/70, 8/7}}
\label{sec:moments-rational}

Figures~\ref{fig:momentc32},~\ref{fig:momentc65},~\ref{fig:momentc8170}, and~\ref{fig:momentc87} show the bounds at four rational central charges with exactly solvable theories. At $c=3/2$ there is a large collection of theories, including the products of the Ising model with the whole $c=1$ moduli space, whose scalars trace out curves as the radius varies, the $\cN=2$ minimal model at $k=2$~\cite{Boucher:1986bh}, the $\Z_{10}$ parafermions and their $E_6$-type modular invariant~\cite{Gepner:1986hr,Cappelli:1986hf}, the superconformal lines of~\cite{Dixon:1988qd}, and several orbifolds and products of minimal models~\cite{Belavin:1984vu,Friedan:1983xq}.

At $c=6/5$ we compare with $\mathrm{Ising}\x\mathrm{TIM}$, the $W_3$ minimal model $W_3(5,6)$~\cite{Zamolodchikov:1985wn} and its charge-conjugation orbifold, and the coset $\mathrm{SU}(3)_2/\mathrm{U}(1)^2$~\cite{Gepner:1987sm}. These four theories are closely related, since the coset is a $\Z_2$ orbifold of $\mathrm{Ising}\x\mathrm{TIM}$~\cite{Davenport:2012fcs,Bajnok:2016ocb} and the emergent $W_3$ symmetry of the Ising/tricritical-Ising interface found in~\cite{Antunes:2026kqb} connects the product theory to $W_3(5,6)$.\footnote{We thank Junchen Rong for discussions of the classification of unitary CFTs at $c=6/5$.} At $c=81/70$ we compare with the $\cN=1$ superconformal minimal model $\mathrm{SM}(5)$~\cite{Friedan:1984rv,Bershadsky:1985dq,Eichenherr:1985cx,Goddard:1986ee}, and at $c=8/7$, the smallest central charge above $1$ at which a compact unitary CFT is known~\cite{Antunes:2025huk}, with the $\Z_5$ parafermion theory~\cite{Fateev:1985mm} and its charge-conjugation orbifold.

The recurring kinks have a simple origin in product theories. At $c=3/2$ and $6/5$, $\s\otimes\vac$ in $\mathrm{Ising}\x\cT$ saturates the identity-excluded upper bound at $\De_\f=\frac18$, and $\e\otimes\vac$ in $\mathrm{TIM}\x\cT$ similarly saturates a kink at $\De_\f=\frac15$. The latter point also lies on or very near the bound at $c=8,4,2$. At $c=81/70$ and $8/7$ neither product exists, and a kink near $\De_\f\approx0.18$ remains unpopulated.

\momentfig[0.78]{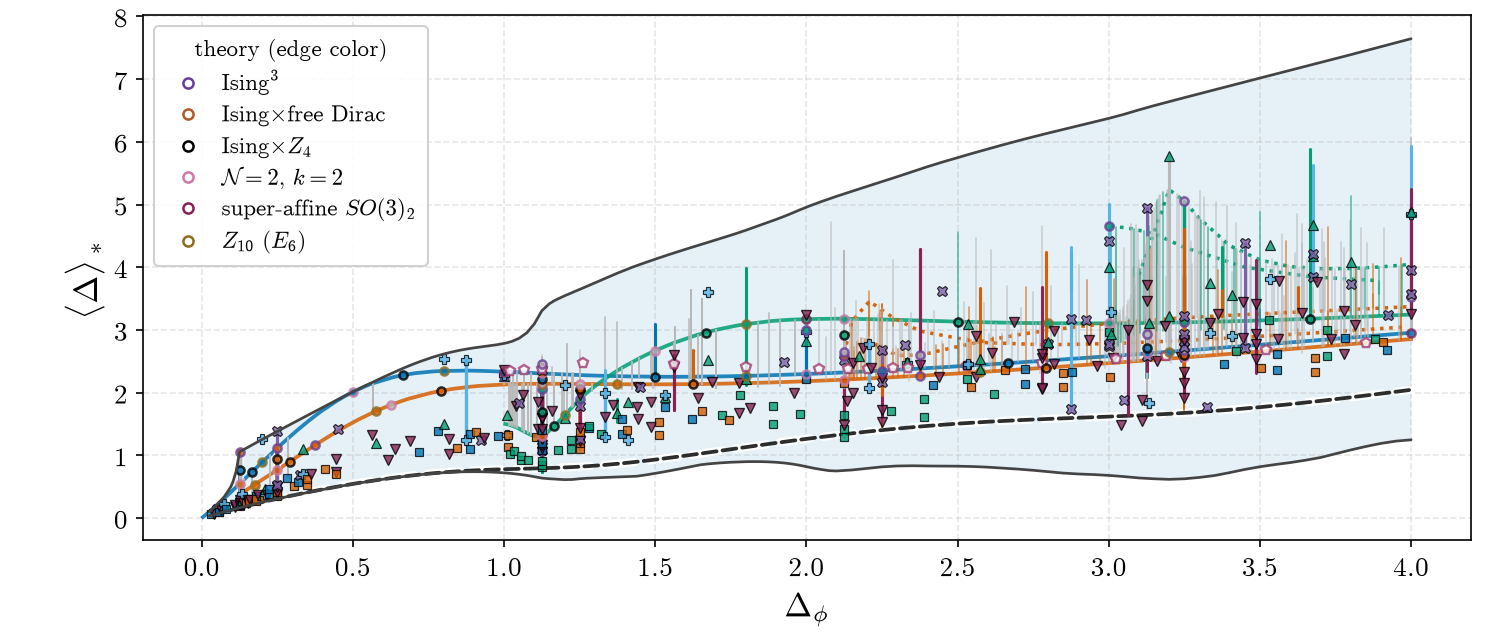}{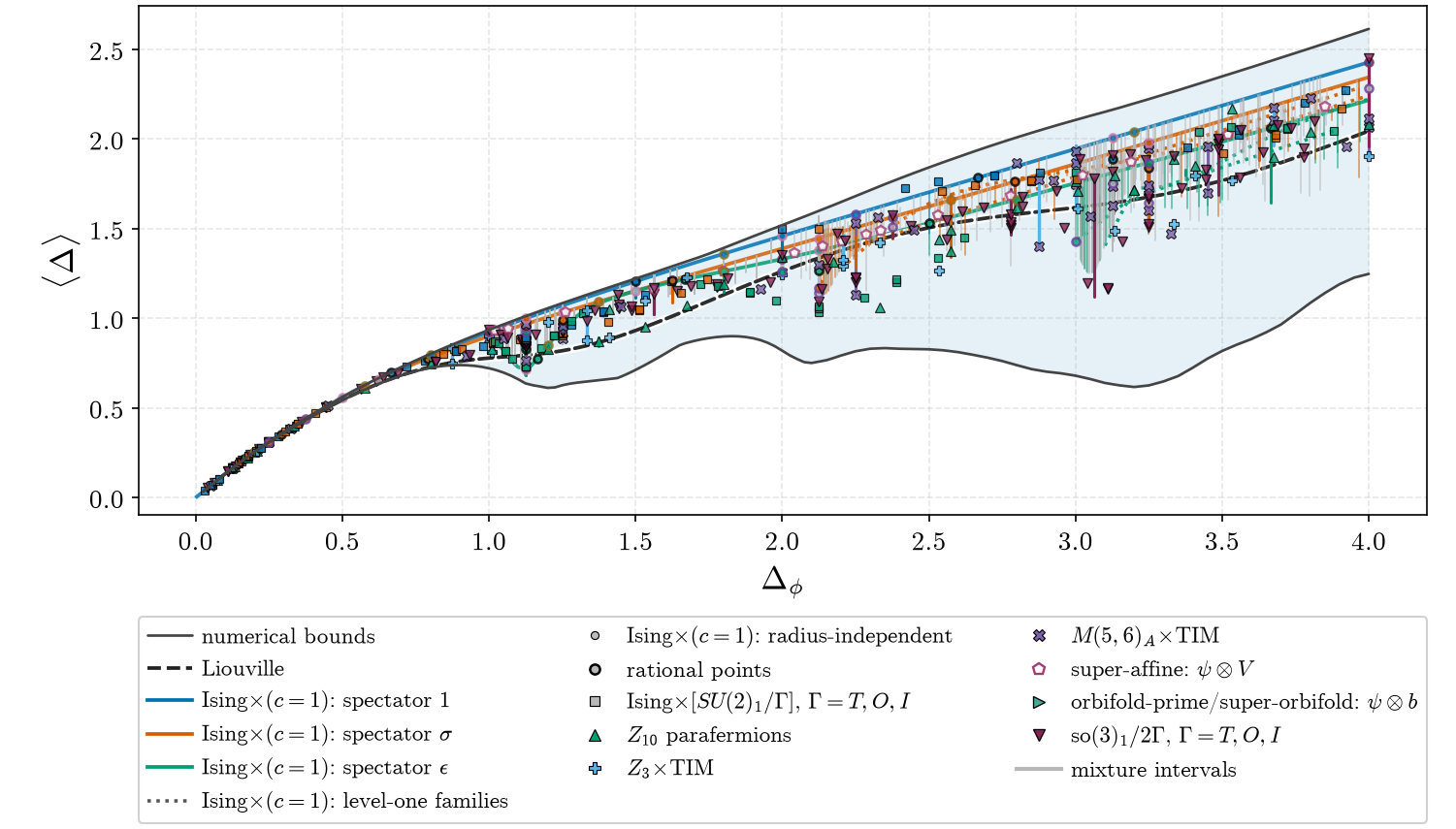}{Bounds on the normalized first moment at $c=3/2$, $\L=19$, with the identity excluded (top) and included (bottom), compared with Liouville theory (dashed) and the theories listed in the legends. Colored curves are the $\mathrm{Ising}\x S^1$ and $\mathrm{Ising}\x S^1/\Z_2$ families. For these families, marker faces and curve colors identify the Ising spectator, and marker edges identify the rational theories in the inset. Squares denote $\mathrm{Ising}\x[\mathrm{SU}(2)_1/\G]$, $\G=T,O,I$. Dotted curves are level-one descendant families. The $\psi\otimes V$ and $\psi\otimes b$ markers are Ramond-type twists on the super-affine and orbifold-prime/super-orbifold lines. Vertical bars give mixture intervals, colored by theory or spectator and gray when spectators mix.}{fig:momentc32}

\momentfig{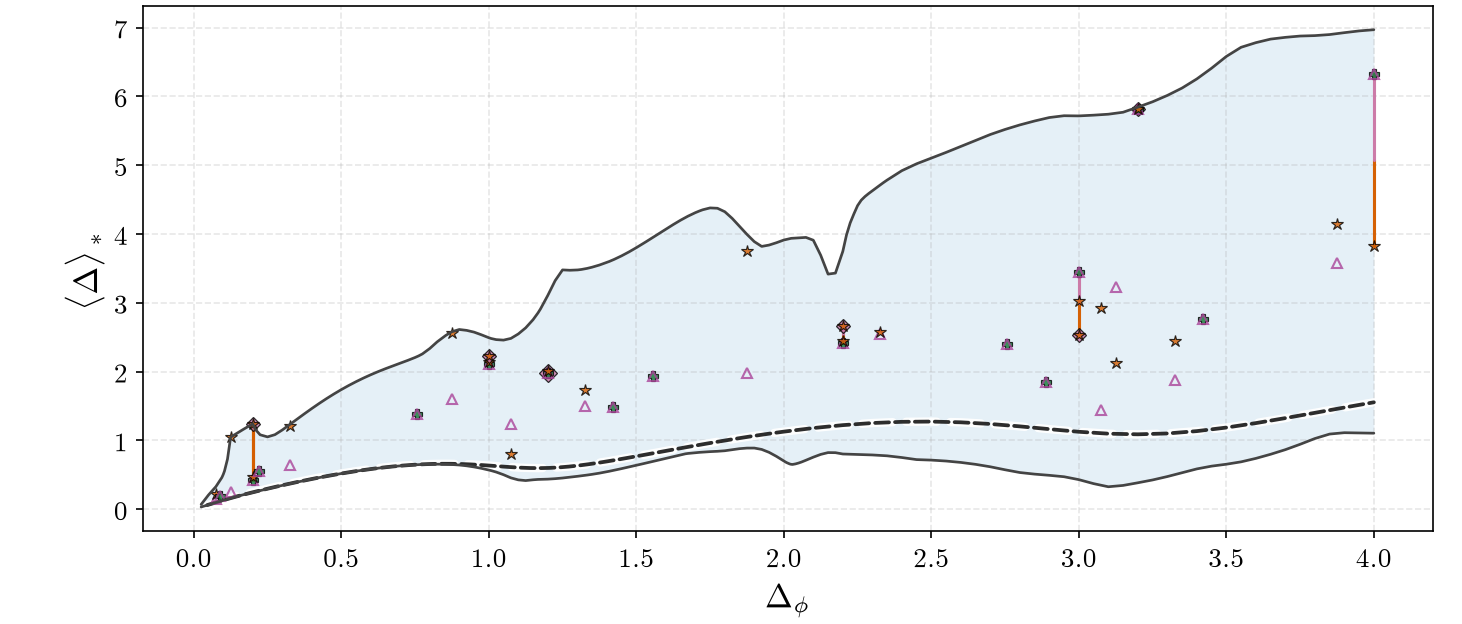}{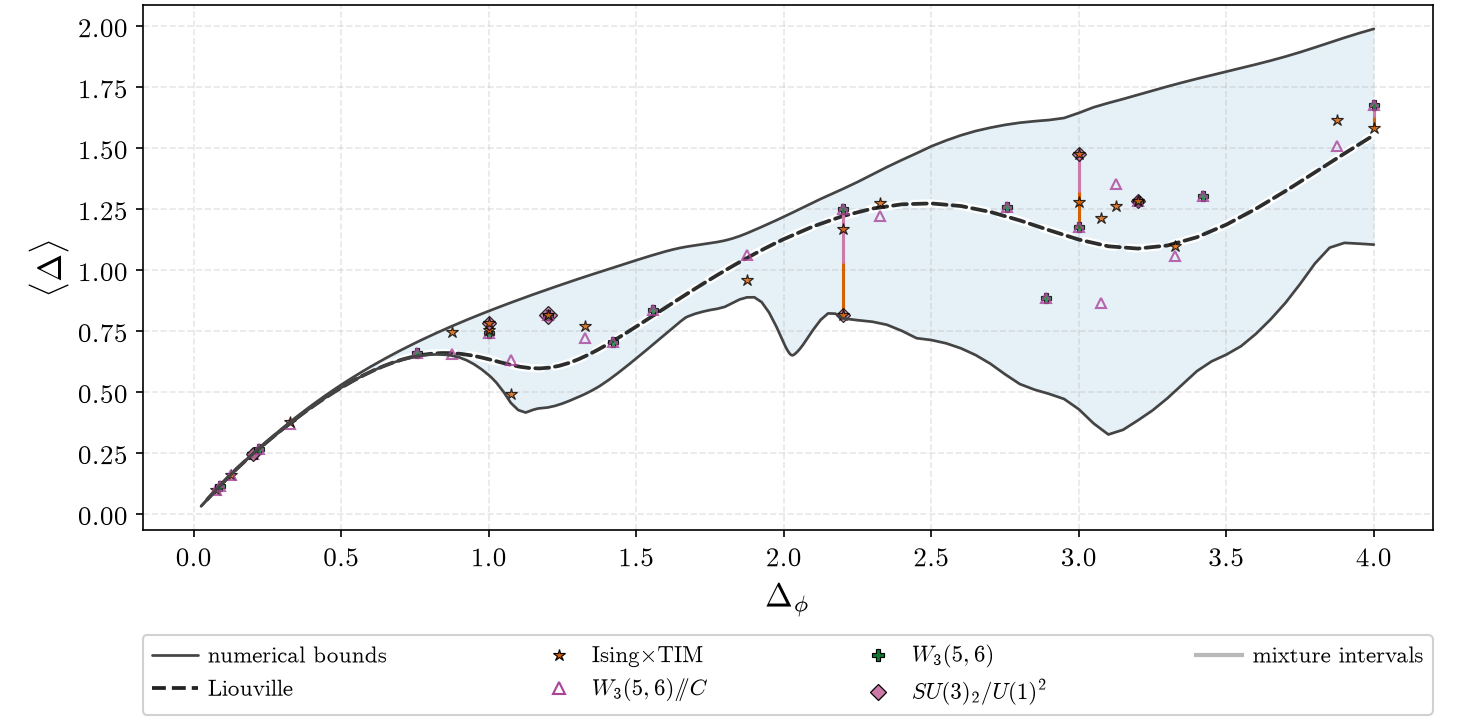}{Bounds on the normalized first moment at $c=6/5$, $\L=19$, with the identity excluded (top) and included (bottom), compared with Liouville theory (dashed), the product of the Ising and tricritical Ising models (stars), the $W_3$ minimal model $W_3(5,6)$ (crosses) and its charge-conjugation orbifold (triangles), and the coset $\mathrm{SU}(3)_2/\mathrm{U}(1)^2$ (diamonds).}{fig:momentc65}

\momentfig{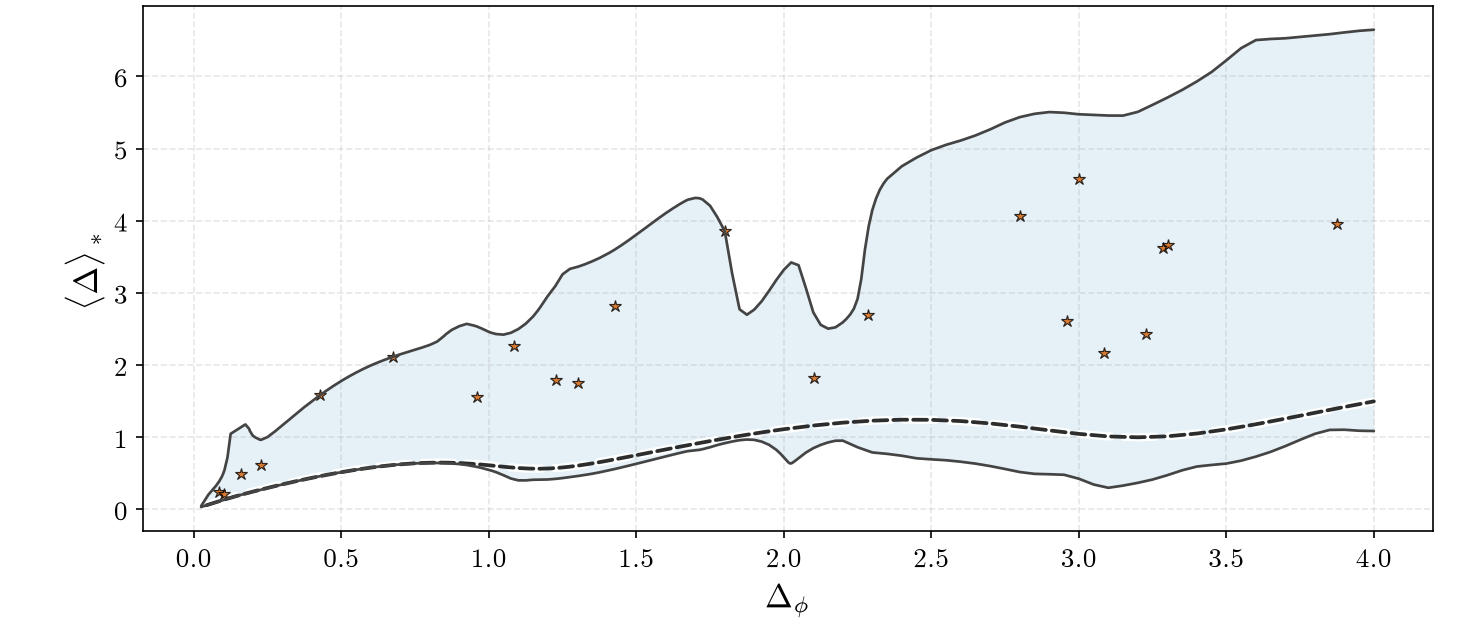}{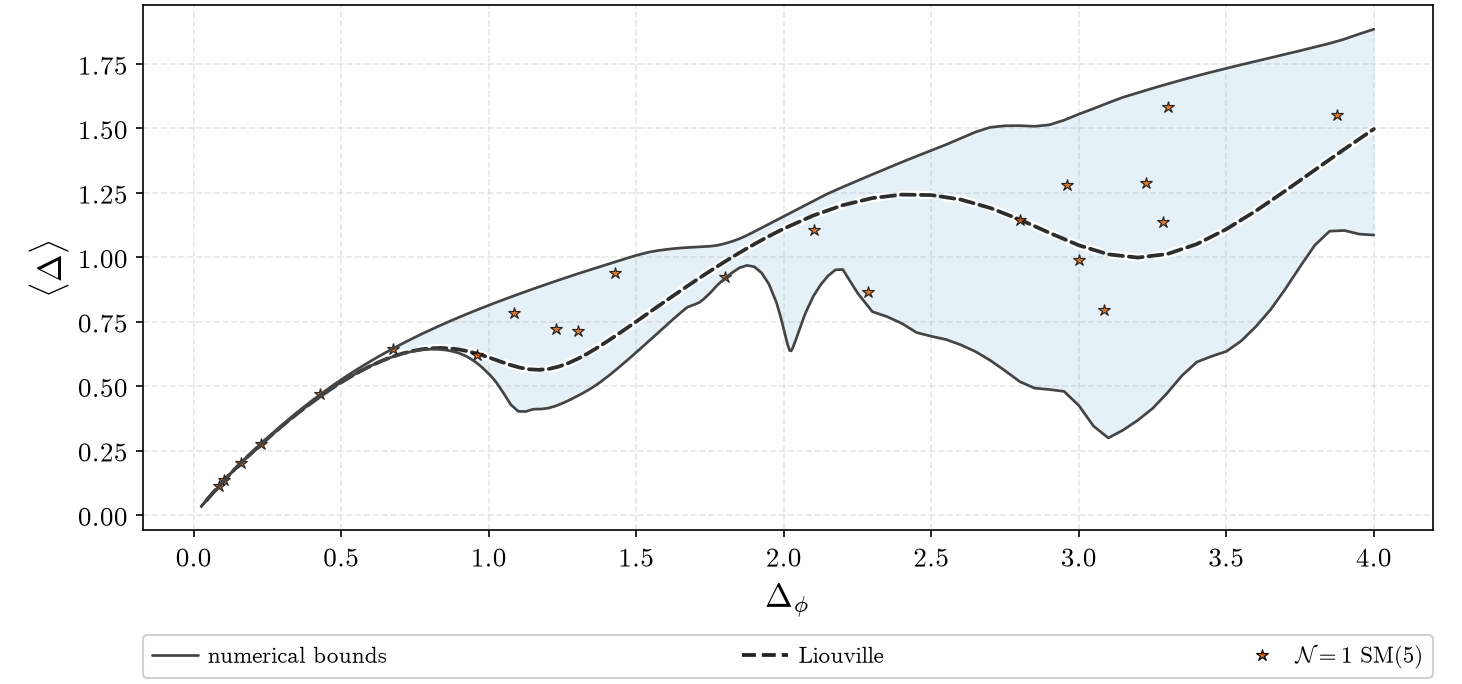}{Bounds on the normalized first moment at $c=81/70$, $\L=19$, with the identity excluded (top) and included (bottom), compared with Liouville theory (dashed) and the Neveu--Schwarz and Ramond sector scalars of the $\cN=1$ superconformal minimal model $\mathrm{SM}(5)$ (stars), computed from its coset description $\mathrm{SU}(2)_3\x\mathrm{SU}(2)_2/\mathrm{SU}(2)_5$.}{fig:momentc8170}

\momentfig{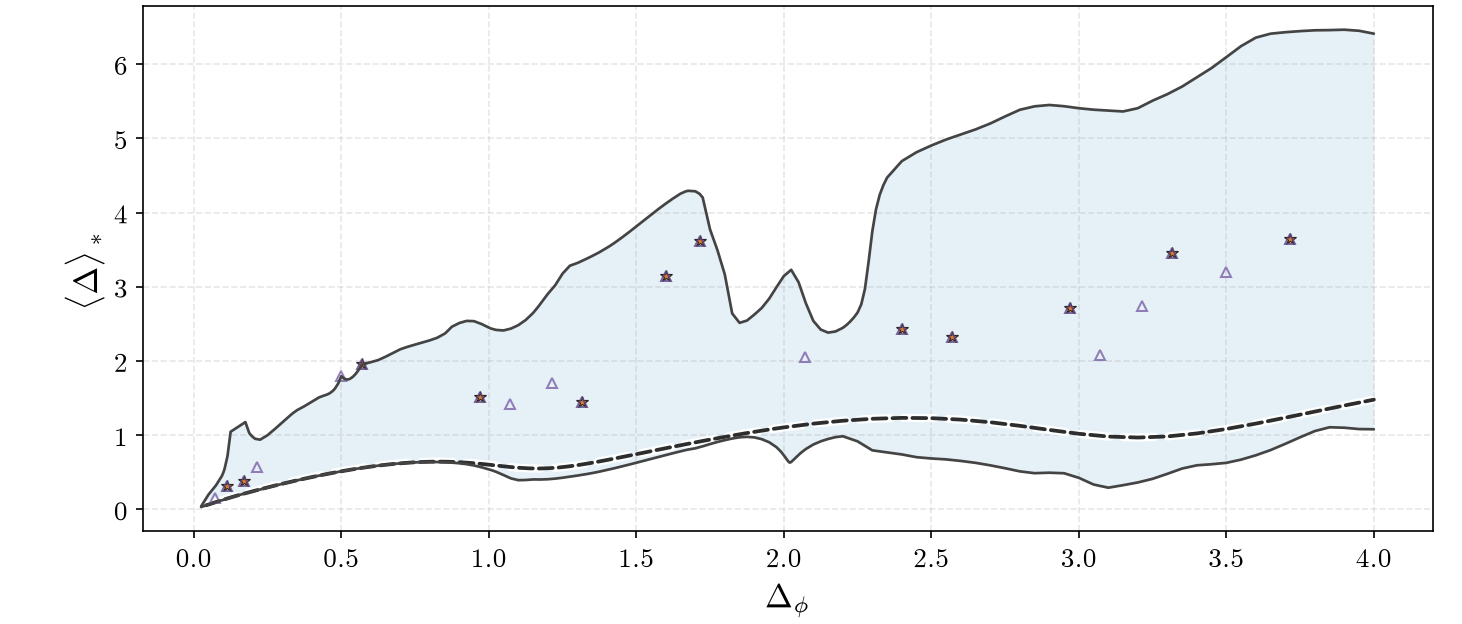}{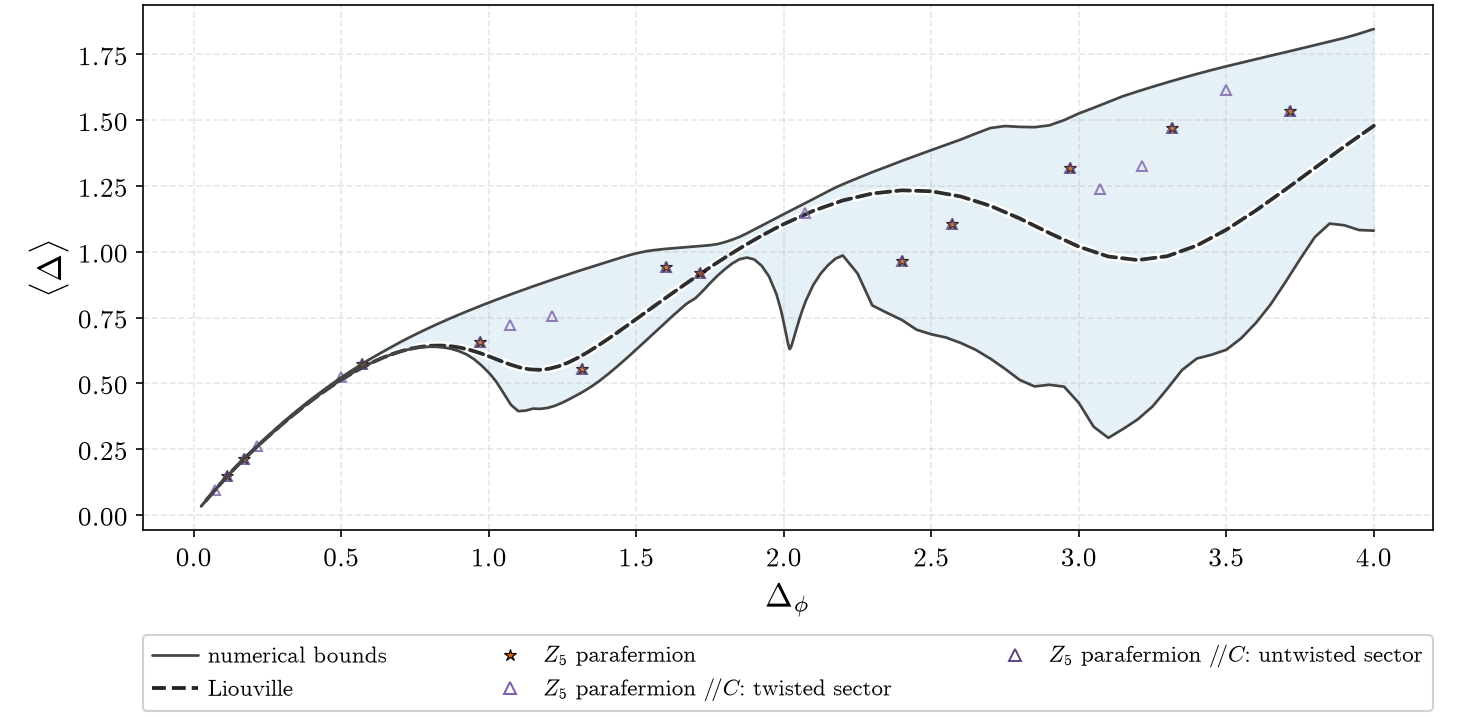}{Bounds on the normalized first moment at $c=8/7$, $\L=19$, with the identity excluded (top) and included (bottom), compared with Liouville theory (dashed), the scalar primaries of the $\Z_5$ parafermion theory (stars), and the untwisted and twisted sectors of its charge-conjugation orbifold (triangles).}{fig:momentc87}

Several other scalars saturate the identity-excluded upper boundary to within the plot resolution: the $\mathrm{SM}(5)$ scalars at $\De_\f\approx0.43$, $0.68$, $1.80$; the $\mathrm{Ising}\x\mathrm{TIM}$ scalars at $\De_\f\approx0.33$, $0.88$; and the $\Z_5$ parafermion scalar at $\De_\f\approx0.57$ and its orbifold scalar at $\De_\f=\half$. At $c=3/2$, the $\mathrm{Ising}\x S^1$ line with the Ising identity as spectator touches the upper bound near $\De_\f=\half$, and $\f_{1,2}\otimes\vac$ in the product $M(5,6)_A\x\mathrm{TIM}$ of the tetracritical and tricritical Ising models saturates it at $\De_\f=\frac14$. Above this point the bound leaves the known theories behind. With the identity included, the lightest scalars lie on the nearly degenerate band and several heavier scalars approach its upper bound.

The lower boundary follows Liouville at small $\De_\f$, departing below it from $\De_\f\approx0.9$ at all four central charges. Its pronounced dip at $1\lesssim\De_\f\lesssim1.5$, broader at $c=3/2$, is not reached by any known theory, although at $c=6/5$ the scalar $\e_{\rm I}\otimes\s_{\rm T}$ at $\De_\f=\frac{43}{40}$ lies just above its falling edge. The identity-excluded upper boundary also has an unpopulated peak near $\De_\f\approx1.7$--$1.8$ at the three smallest central charges. The $\Z_5$ scalar at $12/7$ lies well below the peak, while the $\mathrm{SM}(5)$ scalar at $9/5$ saturates the falling edge. Whether the unpopulated features reveal unknown theories or disappear as $\L$ grows is an open question.

Above $\De_\f=2$, the bounds develop nearby peaks and dips, strongest at the two smallest central charges and weaker at $c=3/2$. Section~\ref{sec:moments-c1} relates these features to the degenerate external dimensions at $c=1$. Farther out, at the three smallest central charges the lower bound has a sharp minimum at $\De_\f\approx3.1$, below a shallower minimum of the Liouville curve near $3.2$. At $c=3/2$ the minimum of the bound is shallow and the Liouville curve rises monotonically. The only known scalar above $\De_\f=2$ close to a bound is the level-one primary of the product $\e_{\rm I}\e_{\rm T}$ of the Ising and tricritical Ising energy operators at $c=6/5$, with $\De_\f=16/5$ (appendix~\ref{app:models}). It occurs in all four theories at this central charge and lies $0.4\%$ below the identity-excluded upper bound.

These features are stable in the block parameters. For $1.8\leq\De_\f\leq2$ at $c=8/7$ and $81/70$, the bounds with $n_{\rm poles}=60$ and $84$ agree to $3\x10^{-6}$, and at $c=8/7$ and $\De_\f=\half$ the identity-excluded upper bound agrees to eleven digits between $(n_{\rm poles},N)=(32,60)$ and $(60,100)$, while with substantially fewer poles \sdpb\ can fail to converge or report spurious bounds (section~\ref{sec:gapbounds}). The recurring saturation by known theories suggests that moment bounds can identify and characterize CFTs, in the spirit of~\cite{Chiang:2026nmd}.

\FloatBarrier

\subsubsection{The \texorpdfstring{$c\to1$}{c -> 1} limit}
\label{sec:moments-c1}

Figure~\ref{fig:momentcbetween} shows the bounds at $\L=11$ for several central charges between $c=8/7$ and $1.02$, together with the bounds at $c=1+10^{-6}$, which we use below for the $c\to1$ limit. As $c$ decreases the lower bound barely changes, and the identity-excluded upper bound deforms smoothly into the curve at $c=1+10^{-6}$. Its peak near $\De_\f\approx1.7$ at $c=8/7$ moves down to a spike at $\De_\f=\frac32$, the steep rise near $\De_\f\approx2.3$ moves to a spike at $\frac52$, a spike at $\De_\f=2$ is present at every $c$, and the notch at $\De_\f=\frac12$ narrows. With the identity included the upper bound decreases onto the curve at $c=1+10^{-6}$. The features of the $\L=19$ bounds near $\De_\f=2$ at the rational central charges sit at the same places, so they are the $c>1$ continuation of this structure.

\momentfig{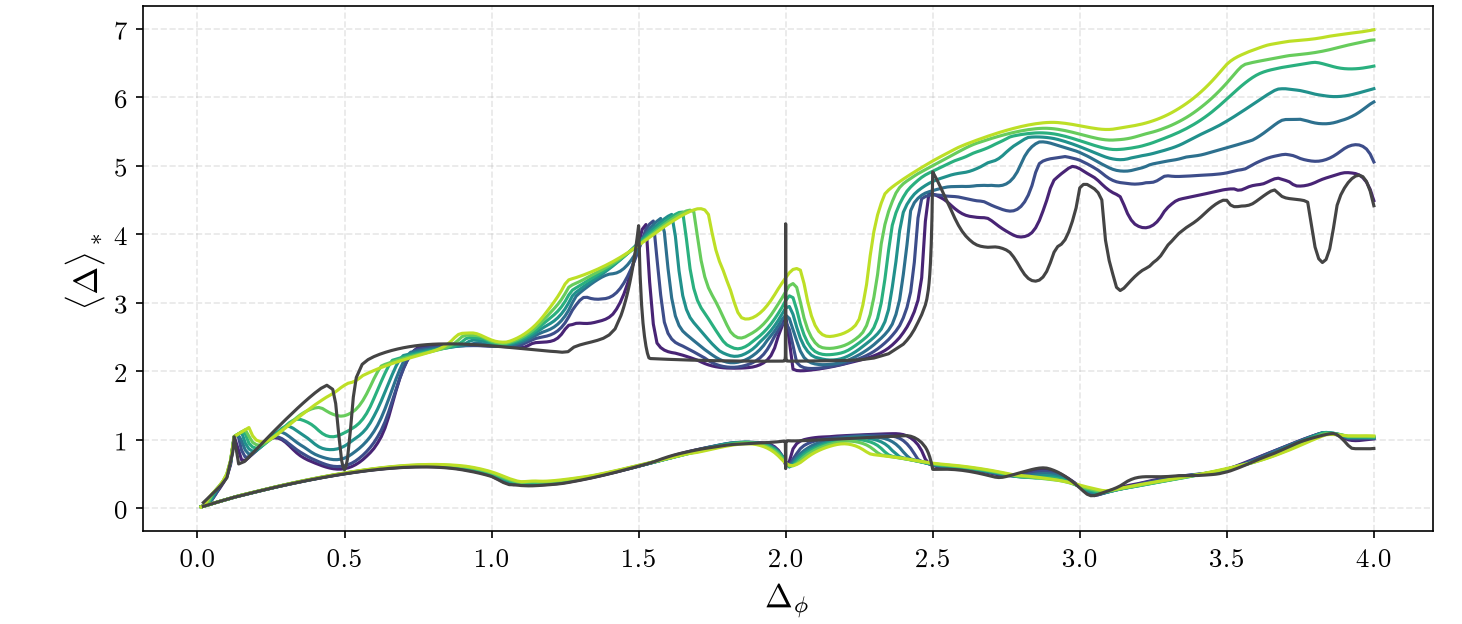}{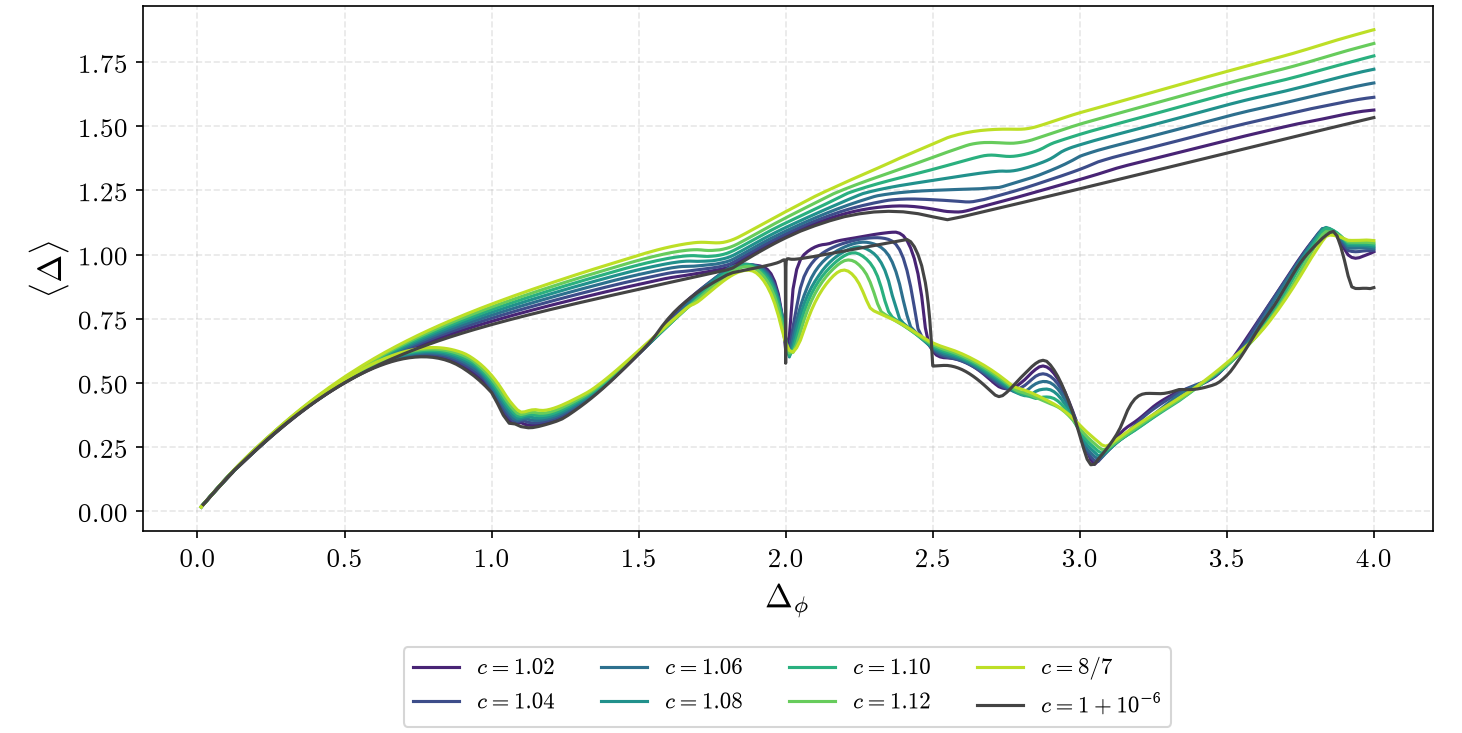}{Bounds on the normalized first moment at $\L=11$ for several central charges between $c=1.02$ and $8/7$, with the identity excluded (top) and included (bottom), together with the bounds computed at $c=1+10^{-6}$ (black), including the notch at $\De_\f=\frac12$. The parameters are given in table~\ref{tab:params}.}{fig:momentcbetween}

\momentfig[0.78]{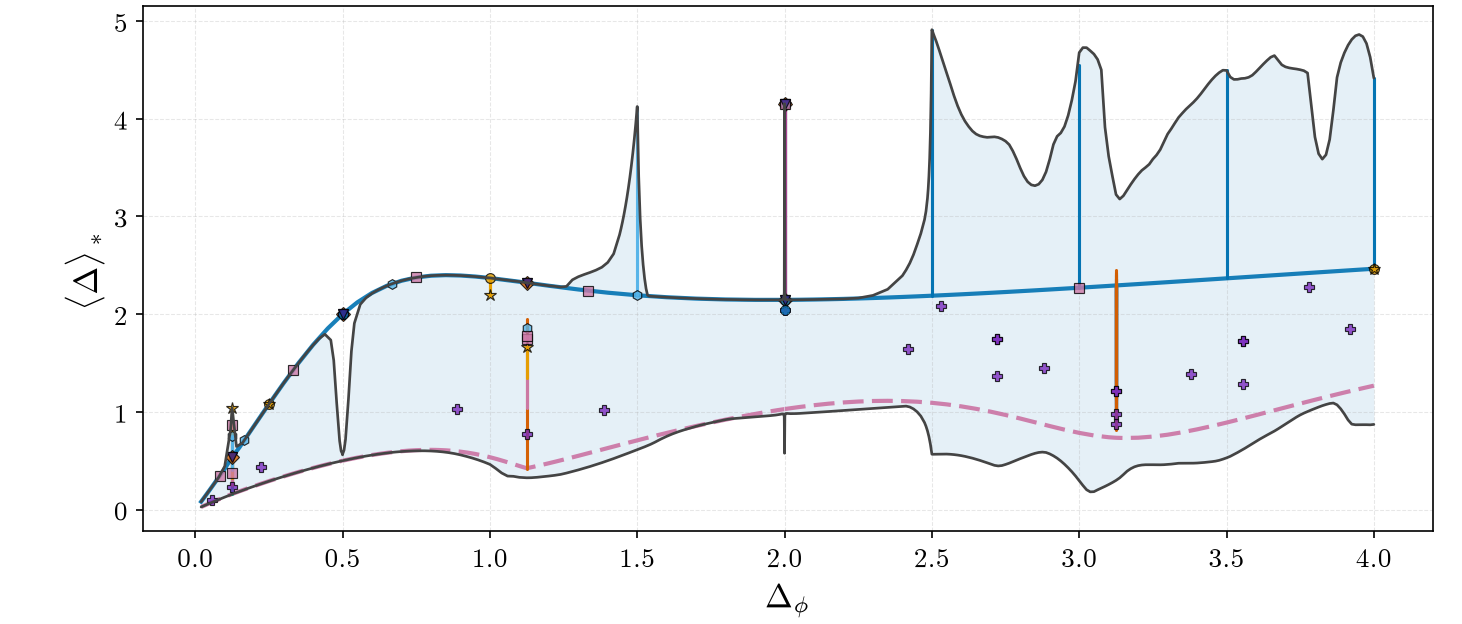}{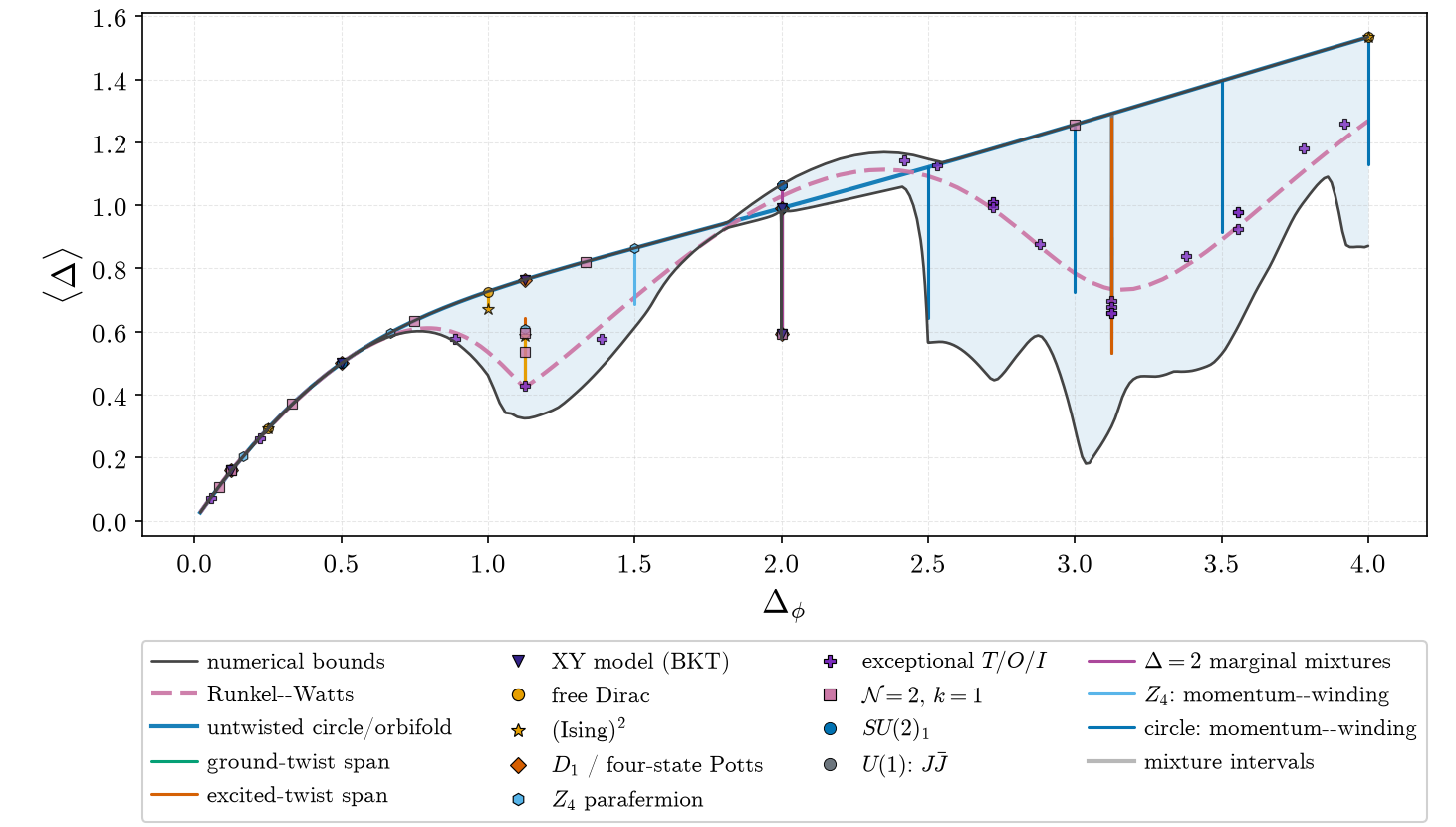}{Bounds on the normalized first moment as $c\to1$ at $\L=11$, computed at $c=1+10^{-6}$, with the identity excluded (top) and included (bottom), compared with the $c=1$ moduli space. The solid curve traces the compact-boson and $\Z_2$-orbifold branches, with rational points identified in the legend. Also shown are the exceptional orbifolds $\mathrm{SU}(2)_1/\G$, $\G=T,O,I$, and the Runkel--Watts theory (dashed). Twist spans run over the orbifold radius, and bar colors identify the theory or family. Inside the notch at $\De_\f=\frac12$ the circle branch lies above the bound.}{fig:momentc1}

The moduli space of compact unitary $c=1$ CFTs is believed to be completely known~\cite{Ginsparg:1987eb,Dijkgraaf:1987vp,Ginsparg:1988ui}; see~\cite{Thorngren:2021yso} for a modern account. It consists of the compact boson $S^1_R$ and the orbifold $S^1_R/\Z_2$, which meet at the Kosterlitz--Thouless point, and three isolated theories $\mathrm{SU}(2)_1/\G$ with $\G=T,O,I$ the tetrahedral, octahedral, and icosahedral subgroups of $\mathrm{SO}(3)$. Figure~\ref{fig:momentc1} compares the bounds at $c=1+10^{-6}$ with these theories and with the non-compact Runkel--Watts theory~\cite{Runkel:2001ng}, which is the $c\to1$ limit of both the unitary minimal models and Liouville theory~\cite{Schomerus:2003vv,Ribault:2015sxa}.

At $c=1$ every Kac pole becomes real, so we approach $c=1$ from $c=1+\epsilon$, where the poles near $\De=n^2/2$ lie a distance of order $\sqrt\epsilon$ from the real axis. The limits $c\to1$ and $\L\to\oo$ need not commute. At fixed $\L$ the bounds converge as $\epsilon\to0$, but slowly near the degenerate external dimensions $\De_\f=\frac12$ and $2$. At fixed $\epsilon$, raising $\L$ instead widens the regions around these dimensions where the bounds fall below the $c=1$ theories. At $c=1+10^{-5}$, a preliminary identity-excluded upper bound at $\L=19$ lies far below the $\L=11$ bound near $\De_\f=\frac12$ and $2$, while near $\De_\f=1$ the two agree to $5\x10^{-3}$. A $c=1$ correlator need not have a positive decomposition into $c=1+\epsilon$ blocks, so theories just above $c=1$ can be much more strongly constrained near these dimensions. Global blocks show a related effect, which makes the moment bounds discontinuous as $d\to2^+$~\cite{Chiang:2026nmd}. At $\L=11$, $\epsilon=10^{-6}$ already approximates the fixed-$\L$ limit away from the degenerate dimensions, and we use these bounds to compare with the $c=1$ theories.

At $\De_\f=\frac12$ and $2$ the external operator has the degenerate weight $h=\frac14$ or $1$ at $c=1$, so its OPE is truncated at $c=1$ but not at $c=1+\epsilon$. Near $\De_\f=\frac12$ the identity-excluded upper bound has a narrow notch. At its center the bound stays at $\<\De\>_*=0.560$ for $10^{-6}\leq\epsilon\leq10^{-3}$, well below the value $2$ of the $c=1$ theories, while its width shrinks roughly as $\epsilon^{1/4}$. If this trend continues, the notch shrinks to the single point $\De_\f=\frac12$, where the limiting bound is discontinuous. The features at $\De_\f=2$ are much narrower. The marginal operator $J\bar J$, which exists only at this external dimension, has $\<\De\>_*=\frac{54}{13}$ at the tip of the upper-bound spike and $\<\De\>=\frac{16}{27}$ just above the lower-bound dip. Neither external dimension appears in the Runkel--Watts spectrum.

With the identity excluded, the untwisted branch of the moduli space, made of the momentum and winding vertex operators of the circle and orbifold theories, saturates the upper bound for $\De_\f\lesssim2.2$ except near $\De_\f=\frac18$, $\frac12$, $\frac32$, and $2$. The spike at $\De_\f=\frac18$ is the Ising kink, reached by $\s\otimes\vac$ of $(\mathrm{Ising})^2$, a ground twist field of the orbifold branch. The spike at $\De_\f=\frac32$ rises from a broad base that disappears at higher $\L$. Its tip and the tip at $\De_\f=2$ are reached by mixtures of scalars of equal dimension, namely the momentum-winding mixtures of the $\Z_4$ parafermion theory and the marginal scalars. Beyond $\De_\f\approx2.2$ the upper bound rises well above the untwisted branch, with further spikes at $\De_\f=\frac52$, $3$, $\frac72$, and $4$. The Runkel--Watts theory follows the lower bound for $\De_\f\lesssim0.8$, and the other twist fields lie inside. With the identity included, the upper bound follows the untwisted branch for $\De_\f\lesssim1.8$ and again from $\De_\f\approx2.6$ to $4$, where many of the exceptional $T$, $O$, and $I$ operators also lie. The lower bound follows Runkel--Watts for $\De_\f\lesssim0.8$ and has pronounced dips near $\De_\f\approx1.2$ and $3.05$ below every known theory.

In the $c\to1$ limit, known theories thus populate much of the boundary allowed by a single crossing equation. The circle and orbifold branches trace the upper bound up to $\De_\f\approx2.2$. Beyond it their momentum-winding mixtures reach the spike tips at $\De_\f=\frac52$, $3$, $\frac72$, and $4$, so the upper bound cannot get much stronger there. The Runkel--Watts theory traces the lower bound up to $\De_\f\approx0.8$. It will be interesting to further study whether increasing $\L$ after taking $c\to1$ closes the remaining gaps.

\FloatBarrier

No compact unitary CFT is known for $1<c<8/7$. A recent modular-bootstrap search~\cite{Benjamin:2026lbj} found candidate partition functions that appear to form a continuous family, together with evidence for a stronger gap bound near $c=1$ than $\De_\gap\leq c/6+1/3$~\cite{Collier:2016cls}. A partition function alone does not establish a CFT, and the sphere bounds give a complementary test, since every scalar primary of a candidate must have $\<\De\>$ inside the allowed region at its own central charge. The unpopulated parts of the allowed regions, such as the dips at $1\lesssim\De_\f\lesssim1.5$, are natural places to look, although a new theory need not lie on the boundary. A candidate spectrum fixes the light dimensions and degeneracies, and sphere crossing can then constrain its OPE coefficients or rule it out.

\section{Outlook}
\label{sec:outlook}
\enlargethispage{\baselineskip}

\vb\ supplies high-precision Virasoro block tables for numerical bootstrap computations at any central charge. The moment bounds suggest tests of candidate theories and an analytic study of the boundary allowed by crossing.

Mixed-correlator crossing and global symmetries can sharpen these tests. General external weights and the ordered and parity-projected bases allow mixed-correlator systems, including spinning operators such as currents. A concrete target is the infrared fixed point of $N$ coupled large-$m$ minimal models~\cite{Antunes:2022vtb,Antunes:2024mfb}, with perturbative predictions for the central charge, low-lying dimensions, and leading Regge trajectory. Its $S_N$ symmetry, or the finite-group symmetry of~\cite{Antunes:2025erb}, can be imposed in a mixed-correlator bootstrap of the lightest singlet and non-singlet scalars, as in the coupled Potts replicas of~\cite{Kousvos:2024dlz}. A twist gap removes the degeneracy discussed in section~\ref{sec:gapbounds}. The coupled Fibonacci anyon chains of~\cite{Antunes:2025huk}, with a candidate irrational phase at $c\approx2.1$ and lattice estimates of the spectrum, offer a complementary target.

Modular invariance provides complementary constraints on the spectrum, while sphere crossing also constrains OPE coefficients. Both can be imposed in a common semidefinite program using the torus characters and sphere blocks, with torus one-point functions supplying further constraints on $C_{\cO\cO\f}$. The candidate modular spectra in $1<c<8/7$~\cite{Benjamin:2026lbj} are a natural test. Their lightest scalars must satisfy sphere crossing and the moment bounds at the same central charge. This could rule out candidates or substantially constrain their OPE data.

The numerical bounds also invite an analytic explanation. Known theories populate much of the allowed boundary, and Liouville theory approaches the lower bound over substantial ranges. Understanding this pattern would clarify what the bounds measure about the spectrum. Extending the computations beyond $\De_\f=4$ would also connect them to the heavy limit. For moments of global conformal primaries, one has $2^{k/2}\leq\lim_{\De_\f\to\oo}\<\De^k\>/\De_\f^k\leq2^{3k/2-1}$ for integer $k\geq1$, with saturation by generalized free-field constructions~\cite{Poland:2024hvb}. Do Virasoro moments at fixed $c$ obey the same asymptotics? The available analytic ingredients include maximally heavy observables and their dynamical phase transitions~\cite{Poland:2026uiy}, universal heavy asymptotics~\cite{Hartman:2014oaa,Collier:2019weq,Benjamin:2023qsc}, light-state dominance~\cite{Kraus:2018pax}, and holography~\cite{Fitzpatrick:2014vua,Grabovsky:2024jwf,Abajian:2023jye}.

The superconformal blocks allow analogous bootstrap studies of $\cN=1$ theories and Ramond-sector correlators. Computations at $c<1$ permit non-unitary bootstrap studies~\cite{Esterlis:2016psv,Huang:2025qkk} and studies near minimal models, while the rational representation and high-order series can aid analytic investigations of large-$c$, large-$h$, and Regge limits~\cite{Fitzpatrick:2014vua,Chen:2017yze,Collier:2018exn}.

\par\vspace{3.5ex plus 1.3ex minus .7ex}
\noindent\begin{minipage}{\textwidth}
\section*{Acknowledgments}

We thank Li-Yuan Chiang, Wei Li, Gordon Rogelberg, and Junchen Rong for discussions. This work was developed in collaboration with large language models from Anthropic and OpenAI. DP is supported by DOE grant DE-SC0017660. Computations in this work were performed on the Yale Bouchet computing cluster, supported by the facilities and staff of the Yale University Faculty of Sciences High Performance Computing Center.
\end{minipage}

\appendix

\section{Code availability}

The software \vb\ is freely available from GitHub at

\url{https://github.com/davidpoland/virasoro_blocks}\\
It is distributed under the MIT license, installs with \texttt{pip}, requires \texttt{Python}~3.10 or later, and uses \sdpb~3.x for bounds. The exact moments of the known theories used in section~\ref{sec:momentbounds} are included in the directory \texttt{data/known\_cft\_moments}.

\section{Details on numerics}
\label{app:params}
Table~\ref{tab:params} gives the \vb, \vbs, and \sdpb~\cite{Simmons-Duffin:2015qma,Landry:2019qug} parameters used for section~\ref{sec:results}. Unlisted \sdpb\ parameters take their default values.
\begin{table}[tb]
\centering
\footnotesize
\setlength{\tabcolsep}{4pt}
\renewcommand{\arraystretch}{1.1}
\begin{tabular}{@{}lcccc@{}}
\toprule
& & \multicolumn{3}{c}{moments} \\\cmidrule(l){3-5}
parameter & gap bound & $c\geq\frac87$ & $c=1+10^{-6}$ & $1<c\leq\frac87$ \\\midrule
$\L$ & 11; 19 & 19 & 11 & 11\\
spins & 0 & $0,2,\ldots,24$ & $0,2,\ldots,24$ & $0,2,\ldots,24$\\
double-twist towers & none & $\tau_{n<16}$, $\ell\geq20$ & $\tau_{n<16}$, $\ell\geq16$ & $\tau_{n<16}$, $\ell\geq16$\\
\addlinespace[2pt]
\texttt{order} ($N$) & 60; 120 & 80 & 60 & 60\\
\texttt{npoles} & 48; 84 & 60; 84 & 64; 128 & 64\\
tower \texttt{npoles} & -- & 12 & 12; 24 & 24\\
\texttt{precision} (digits) & 160; 200 & 200 & 160; 240 & 240\\
\addlinespace[2pt]
\texttt{pole-selection} & \multicolumn{4}{c}{\texttt{contribution\_optimal}}\\
\texttt{chiral-norm} ($a_0$) & \multicolumn{4}{c}{16}\\
\texttt{coordinates} & \multicolumn{4}{c}{\texttt{zzb}}\\
\texttt{regularize} & \multicolumn{4}{c}{\texttt{auto}}\\\midrule
\texttt{precision} (\sdpb, bits) & 768 & 1024 & 768; 1024 & 1024\\
\texttt{dualityGapThreshold} & default & $10^{-10}$ & $10^{-10}$ & $10^{-10}$\\
\texttt{maxIterations} & default & 1000 & 1000 & 1000\\
\texttt{find\char`\{Primal,Dual\char`\}Feasible} & \texttt{true} & \texttt{false} & \texttt{false} & \texttt{false}\\
\texttt{detect\char`\{Primal,Dual\char`\}FeasibleJump} & \texttt{true} & \texttt{false} & \texttt{false} & \texttt{false}\\
\bottomrule
\end{tabular}
\caption{Parameters used for the numerical computations. The double-twist towers $\tau_n=2\De_\f+2n$ are imposed for $0\leq n<16$ and continuous spin $\ell\geq\ell_{\min}$. Where two values are given, the second applies to the $\L=19$ gap bound, to the moments at $c\geq\frac87$ with $\De_\f\geq1.8$, and at $c=1+10^{-6}$ to $\De_\f>2$ and the scans near $\De_\f=\frac12$ and $2$. The scan near $\frac12$ uses $n_{\rm poles}=96$ instead of $128$. Functional checks also include physical blocks up to $\ell=64$ and additional tower twists.}
\label{tab:params}
\end{table}

\section{Exact moments of known theories}
\label{app:models}

To compute the moments of known theories shown in section~\ref{sec:momentbounds}, we decompose exact four-point functions $\<\f\f\f\f\>$ into Virasoro blocks at the total central charge $c$, in the normalization of section~\ref{sec:recursion}, and evaluate~\eqref{eq:moments}. These calculations use separate research code, and the resulting data are distributed with \vb. The moments are computed from the truncated list of Virasoro primaries with $\De\leq\De_{\rm cut}$. The block factor $G_{\De,\ell}(\half,\half)\sim e^{-\pi\De}$ suppresses high dimensions but does not by itself bound the omitted tail, so we assess convergence by varying the spectrum cutoff, nome order, and fit settings. The data record the crossing residual of each point and, where available, the change between adjacent cutoffs.

\paragraph{Conventions.} We use~\eqref{eq:libnorm} with $a_0=16$ and $\de_z=\de_{\zb}=-\De_\f$. Setting $\l_{\vac}^2=1$ makes $\sum_\cO\l_\cO^2G_{\De_\cO,\ell_\cO}$ the physical four-point function with unit identity OPE coefficient. A scalar primary therefore has $\l_\cO^2=C_{\f\f\cO}^2\,16^{\De_\cO}$. For $\ell>0$ we combine the two spin orientations into one channel with $\ell=|h-\hb|$ and coefficient $C_{\f\f\cO}^2\,16^{\De_\cO}$ when their coefficients agree. If they differ, the channel describes the parity-symmetrized correlator with the average coefficient $\frac12(C_{\f\f\cO}^2+C_{\f\f\bar\cO}^2)\,16^{\De_\cO}$.

The identity contributes the directly evaluated vacuum block $G_{0,0}(\half,\half)$, which depends on $c$ and $\De_\f$ and is not $1$. This dependence is retained when forming the ratios. The identity-excluded ratio subtracts only the Virasoro identity, so Virasoro primaries that are descendants in an extended algebra (e.g.\ the superconformal vacuum module at $c=81/70$) stay in both denominators. As a check, the Ising four-point function $\<\s\s\s\s\>$ at $c=\half$ gives $\l_\e^2=4$ from $C_{\s\s\e}^2=\frac14$, satisfies crossing to $10^{-60}$, and has $\nu_{0,0}=\sqrt2$, $\nu_{1,0}=\frac{\sqrt2-1}{2}$.

When several scalar primaries share a dimension, any normalized combination $\f=\sum_ac_a\f_a$ is an allowed external operator. The moments are then quartic forms in $c_a$ whose ratio we extremize numerically over the unit sphere, and the resulting interval is drawn as a vertical bar. For most mixture spaces the quartic forms are assembled exactly from the block decompositions of the component correlators, and for the rest they are fitted to sampled correlators and tested on held-out points. The largest $\mathrm{Ising}^{16}$ spaces have up to $7\times10^6$ primaries, and there we extremize over structured families instead. Coherent sums over all placements of product fields and of their permutation-invariant level-one descendants, together with operators on eight copies, give the lower ends of the intervals. The upper end with the identity excluded starts from operators $\f=\sum_Bc_B\f_B$ supported on disjoint blocks of copies. Their four-point function is $G_{\mathrm{GFF}}+\sum_Bc_B^4(G_B-G_{\mathrm{GFF}})$, with $G_{\mathrm{GFF}}$ the generalized free correlator, so only correlators of at most eight copies are needed. Letting the energy factors of these operators move to all sixteen copies raises the upper end further. The extremization is done numerically from many starting points, which does not prove global optimality.

\paragraph{Liouville theory.} Liouville primaries are labeled by a momentum $P\geq0$ with $\De(P)=\frac{c-1}{12}+2P^2$, and the external operator by $P_\f$ with $\De_\f=\De(P_\f)$. The four-point function is $\int_0^\oo\frac{dP}{\pi}\,C(P_\f,P_\f,P)^2\,\cF\bar\cF$ with the DOZZ structure constant~\cite{Dorn:1994xn,Zamolodchikov:1995aa} in the form (see~\cite{Collier:2017shs}, appendix B)
\be
C(P_1,P_2,P_3)=\frac{\Upsilon_b'(0)\prod_{i=1}^3\big[\Upsilon_b(2iP_i)\Upsilon_b(-2iP_i)\big]^{1/2}}{\Upsilon_b\p{\frac Q2+i\sum_iP_i}\prod_{i=1}^3\Upsilon_b\p{\frac Q2+i(\sum_jP_j-2P_i)}},
\ee
up to a $P$-independent normalization which cancels in the ratios, and with $\Upsilon_b$ evaluated from its integral representation in the strip $0<\Re x<Q$ together with the shift relations $\Upsilon_b(x+b)=\g(bx)b^{1-2bx}\Upsilon_b(x)$ and its $b\leftrightarrow b^{-1}$ counterpart, where $\g(x)=\Gamma(x)/\Gamma(1-x)$. Since the integrand of the integral representation decays like $e^{-\min(\Re x,\,Q-\Re x)\,t}$, the argument is first moved by the shift relations into the central band $\frac Q4\leq\Re x\leq\frac{3Q}4$, where a truncated quadrature is accurate. This matters for the normalization $\Upsilon_b'(0)$ and for the analytic continuation below, whose arguments approach the edge of the strip. The moments are then
\be
\nu_{n,0}=\int_0^{P_{\rm cut}}\frac{dP}{\pi}\,C(P_\f,P_\f,P)^2\,16^{\De(P)}\,\De(P)^n\,G_{\De(P),0}(\thalf,\thalf),
\ee
with the blocks evaluated directly from the recursion at each $P$ and the integral performed by Gauss--Legendre quadrature, with convergence checked by comparing two cutoffs $P_{\rm cut}$. Double precision is ample for the plots.

For $(c-1)/16\le\De_\f<(c-1)/12$ the external momentum is imaginary, $P_\f=i\ka$, and the same expression defines the analytically continued four-point function discussed in section~\ref{sec:gapbounds}, with the internal spectrum still the continuum $\De\geq(c-1)/12$. The arguments $\frac Q2-2\ka\pm iP$ stay inside the strip for $\ka<\frac Q4$ and reach its edge at $\De_\f=(c-1)/16$. We do not continue below $\De_\f=(c-1)/16$, so the Liouville curves in section~\ref{sec:momentbounds} start there. Since Liouville theory has no normalizable identity operator, its ``identity excluded'' moment is the same as its ``identity included'' one, and the dashed curve is identical in the two panels of each figure.

\paragraph{The Runkel--Watts theory.} The Runkel--Watts theory~\cite{Runkel:2001ng} is the $c\to1$ limit of the unitary minimal models and of Liouville theory~\cite{Schomerus:2003vv,Ribault:2015sxa}. Its structure constants follow from the $c=1$ limit of the DOZZ formula: they are the $c=1$ ($b=i$) limit of the $c\le1$ Liouville structure constants of~\cite{Zamolodchikov:2005fy} multiplied by a discontinuous step function. The step function is needed for unitarity, since without it the analytic structure constants solve crossing only along a contour off the real axis, with a non-positive spectral measure. With $\De=2P^2$, $\Upsilon_1(x)=G(x)G(2-x)$ in terms of the Barnes $G$-function, and
\be
C_{\rm an}(P_1,P_2,P_3)=\frac{\prod_{\pm,\pm}\Upsilon_1(1+P_1\pm P_2\pm P_3)}{\prod_i\Upsilon_1(1+2P_i)},
\ee
the Runkel--Watts spectral density is $\r(P)=\frac14\,\s(P_\f,P_\f,P)\,C_{\rm an}(P_\f,P_\f,P)\,C_{\rm an}(-P,P_\f,P_\f)$, where $\s\in\{0,1\}$ is the step function of~\cite{Ribault:2015sxa} (equal to 1 when $\prod_{\pm,\pm}\sin\pi(P_1\pm P_2\pm P_3)<0$), which we cross-check against the fractional-part rule of~\cite{Runkel:2001ng}. The moments in figure~\ref{fig:momentc1} are integrals of $\r(P)$ against $16^{2P^2}(2P^2)^nG^{c=1}_{2P^2,0}(\half,\half)$, with the $c=1$ blocks computed by the Laurent regularization of section~\ref{sec:implementation}, and the integration split at the discontinuities of $\s$ and at the degenerate momenta. As a check, at $\De_\f=\frac18$ we recover $\nu_{0,0}=\frac{\pi}{16K(\half)}$ and $\<\De\>=\frac{1}{2\pi}$.

The Runkel--Watts spectrum does not contain the external operators with $2P_\f\in\Z_{>0}$, which in our range are $\De_\f=\half$ and $2$. As $\De_\f$ approaches these values, the support of the spectral density shrinks onto the degenerate momenta $2P\in\Z_{\geq0}$. The curve is not defined at the two points themselves, but its limits from either side are smooth. The DOZZ moments evaluated with the code above at $c=1.01$ lie within $3\%$ of the Runkel--Watts curve.

\paragraph{The $c=1$ moduli space.} For the compact boson and its $\Z_2$ orbifold the four-point functions of the untwisted scalar primaries are known in closed form. The external operator is the real vertex operator $\f=\sqrt2\,\Re V_q$ with $\De_\f=q^2$. Its four-point function is a sum over the six neutral charge assignments of products of powers $|z|^{a}|1-z|^{b}$ fixed by the charges, and is independent of the radius, so the momentum and winding scalars of the whole circle and orbifold branches trace out a single universal curve in the $(\De_\f,\<\De\>)$ plane. The rational points on it ($\mathrm{SU}(2)_1$ at $\De_\f=\half$, the Dirac fermion, $(\mathrm{Ising})^2$, the $\Z_4$ parafermions, the $\cN=2$ minimal model, the Kosterlitz--Thouless point, and the four-state Potts model) are marked.

The four-point functions of the orbifold twist fields (dimension $\frac18$) are given by the partition function of the compact boson on the double cover torus with modulus $\tau=iK(1-z)/K(z)$~\cite{Dijkgraaf:1987vp,Dijkgraaf:1989hb}, $\<\s\s\s\s\>\propto|2^8z(1-z)|^{-1/12}Z_{S^1_R}(\tau,\bar\tau)$, which we decompose into $c=1$ Virasoro blocks character by character. The excited twist fields (dimensions $\frac98$ and $\frac{25}8$) are treated similarly. These depend on the radius, and we plot the range swept out as $R$ varies, extrapolating to the decompactification limit.

For the three exceptional theories $\mathrm{SU}(2)_1/\G$~\cite{Ginsparg:1987eb}, the twisted-sector scalars are obtained as class sums over the orbifold group. For order-two elements the relevant correlators are reflection-lattice sums~\cite{Dijkgraaf:1989hb}, and for the cyclic classes of order $3$, $4$, and $5$ we use the Painlev\'e~VI $\tau$-function representation of $c=1$ four-point functions with prescribed monodromy~\cite{Gamayun:2012ma,Gavrylenko:2018cai}, expanded as $\tau(z)=\sum_nC_n\,z^{(\s+n)^2-\th_0^2-\th_z^2}\cB_n(z)$ with Barnes-$G$ coefficients $C_n$ and $c=1$ blocks $\cB_n$. Each class sum is required to have unit identity coefficient, no fractional spins, and positive coefficients.

\paragraph{Theories with extended chiral algebras.} The theories at $c=8/7,81/70,6/5,3/2$, and the free bosons and products at $c=2,4,8$, have extended chiral algebras, or are products or orbifolds of simpler theories, and their four-point functions are known either in closed form or as solutions of ordinary differential equations. We decompose them into Virasoro blocks at the total central charge in one of three ways.
\begin{enumerate}
\item For product theories $\cA\x\cB$ (e.g.\ $\mathrm{Ising}\x\mathrm{TIM}$, $\mathrm{Ising}\x(c=1)$, $\Z_3\x\mathrm{TIM}$, $\mathrm{Ising}^{2c}$), the product of two chiral blocks $\cF_{c_1,h_1}(z)\cF_{c_2,h_2}(z)$ is expanded in blocks $\cF_{c_1+c_2,h_1+h_2+n}(z)$ by an exact triangular solve in the nome. In the elliptic representation~\eqref{eq:elliptic} the ratio of prefactors is a known power series in $q$, and since $H=1+O(q)$ the coefficient $a_n$ is fixed by the order-$q^n$ term once the lower ones are known.

The OPE coefficients of the minimal-model factors are obtained from an overdetermined solve of their own crossing equations, which reproduces the known structure constants to working precision. Compact-boson factors and the $c$ free bosons $U(1)^c$ are known in closed form, with Wick contraction for operators dressed by currents. For $\mathrm{Ising}\x\mathrm{SU}(2)_1/\G$ the Ising factor is taken in closed form, and the $c=1$ factor is the class sum described above. The two-dimensional spectrum is $\De=2(h_a+h_b)+n+m$, $\ell=|n-m|$, with squared OPE coefficients $C_a^2C_b^2a_na_m$, so that $\l^2=C_a^2C_b^2a_na_m16^\De$.
\item For theories with an extended chiral algebra whose blocks are known from a differential equation, such as the $\Z_k$ parafermions~\cite{Fateev:1985mm} and the coset $\mathrm{SU}(3)_2/\mathrm{U}(1)^2$ (Knizhnik--Zamolodchikov equations~\cite{Knizhnik:1984nr} for the parent WZW model divided by the $\mathrm{U}(1)$ factors), or the $W_3$ minimal model $W_3(5,6)$~\cite{Zamolodchikov:1985wn} (realized as the coset $\mathrm{SU}(3)_2\x\mathrm{SU}(3)_1/\mathrm{SU}(3)_3$, whose blocks are projections of the numerator Knizhnik--Zamolodchikov blocks onto the denominator blocks), each extended block is expanded in Virasoro blocks by a triangular solve. For numerically evaluated KZ blocks, the Taylor coefficients in $z$ are extracted by a discrete Fourier transform on a sampling circle. The relative coefficients are fixed either by monodromy invariance and exchange symmetry of the extended correlator, as in the generic $W_3(5,6)$ coset construction, or by least squares on the odd-derivative components of the crossing equation at $z=\zb=\half$, with the identity normalized to 1.
\item For theories whose correlators do not factorize into chiral blocks in a convenient basis (the orbifolds~\cite{Dixon:1985jw,Dixon:1986jc}, the permutation orbifolds via covering-space correlators~\cite{Lunin:2000yv,Borisov:1997nc}, and the coset description $\mathrm{SU}(2)_3\x\mathrm{SU}(2)_2/\mathrm{SU}(2)_5$ of $\mathrm{SM}(5)$~\cite{Goddard:1984vk}), we evaluate the exact four-point function on a grid of points in cross-ratio space and fit non-negative coefficients $\l^2_{\De,\ell}$ for a candidate spectrum of even-spin Virasoro primaries, read off from the characters of the theory, by non-negative least squares. A subset of the grid is held out to test the fit, the identity coefficient is either fixed to 1 or checked to come out equal to 1, and the resulting moments must be stable to $10^{-5}$ as the cutoff $\De_{\rm cut}$ on the candidate spectrum is raised.
\end{enumerate}
Some grid fits are done in double precision, so their stability checks also vary the sampling grid. For the few mixtures whose cross terms have no exact decomposition (among them the $c=1$ orbifold twist-field and octahedral class mixtures), the cross terms are fitted by signed least squares with the identity coefficient fixed, and unitarity is checked on the assembled spectrum.

\paragraph{Primaries built from descendants.} Above $\De_\f\approx2$ most scalar Virasoro primaries of these theories are not primaries of the chiral algebra but combinations of its descendants. In a product theory the simplest are the level-one primaries
\be
\big(h_bL^{\cA}_{-1}-h_aL^{\cB}_{-1}\big)\big(\bar h_b\bar L^{\cA}_{-1}-\bar h_a\bar L^{\cB}_{-1}\big)\,\f_a\f_b,
\ee
annihilated by $L_1=L_1^{\cA}+L_1^{\cB}$ and its antiholomorphic partner. For free bosons they are vertex operators dressed by currents, such as $J_a\bar J_b\,\Re V$ with $a,b$ transverse to the momentum, and in theories with an extended chiral algebra they are combinations of current, $W$, or supercurrent descendants. Their four-point functions follow from those of the primaries by the Ward identities of the chiral algebra, which act on the chiral blocks as differential operators or, for current descendants, through the Knizhnik--Zamolodchikov recursion. We then decompose these correlators by routes 1--3.

\paragraph{The $\mathrm{so}(3)_1/2\G$ orbifolds.} We use the conformal embedding $\mathrm{SU}(2)_2\x\mathrm{Ising}\subset\mathrm{SU}(2)_1\x\mathrm{SU}(2)_1=\mathrm{so}(4)_1$, with $\mathrm{SU}(2)_2=\mathrm{so}(3)_1$ the three-Majorana theory. Under the diagonal $\mathrm{SU}(2)$ action the four Majoranas split into a rotating triplet and an Ising singlet. Products of the $c=1$ monodromy correlators described above give correlators of the triplet twist field tensored with an Ising spectator. For the NS-type twists this spectator is the identity, so no Ising division is needed. For Ramond-type twists it is the Ising spin field, whose correlator is removed on the block decomposition by inversion order by order in the nome. We sum over conjugacy classes of the orbifold group as at $c=1$. Excited twist fields are obtained from the ground ones by acting with current and Virasoro modes. As a check, the $T$ orbifolds reproduce the exactly known $\mathrm{Ising}^3/\Z_3$~\cite{Borisov:1997nc} and its discrete-torsion partner~\cite{Vafa:1986wx}.

In all cases the final list of primaries and coefficients is checked against the crossing equation at $z=\zb=\half$. This is an independent test for route 1 and for the monodromy-based constructions of route 2, but only a fit diagnostic where the coefficients were themselves fitted to crossing. For route 3 the independent checks are the held-out grid points and the cutoff ladder. The data are distributed with \vb\ in the directory \texttt{data/known\_cft\_moments}, as a list of $(c,\De_\f,\<\De\>,\<\De\>_*)$ together with the residuals. It contains individual operators, mixture intervals, the $c=1$ catalog, and the continuous families shown in the plots through $\De_\f=4$, with a manifest listing the files.

\bibliographystyle{JHEP}
\bibliography{biblio}

\providecommand{\href}[2]{#2}\begingroup\raggedright\begin{thebibliography}{100}

\bibitem{Rattazzi:2008pe}
R.~Rattazzi, V.~S. Rychkov, E.~Tonni and A.~Vichi, \emph{{Bounding scalar
  operator dimensions in 4D CFT}},
  \href{http://dx.doi.org/10.1088/1126-6708/2008/12/031}{\emph{JHEP} {\bf 12}
  (2008) 031}, [\href{https://arxiv.org/abs/0807.0004}{{\tt 0807.0004}}].

\bibitem{Poland:2018epd}
D.~Poland, S.~Rychkov and A.~Vichi, \emph{{The Conformal Bootstrap: Theory,
  Numerical Techniques, and Applications}},
  \href{http://dx.doi.org/10.1103/RevModPhys.91.015002}{\emph{Rev. Mod. Phys.}
  {\bf 91} (2019) 015002}, [\href{https://arxiv.org/abs/1805.04405}{{\tt
  1805.04405}}].

\bibitem{Poland:2011ey}
D.~Poland, D.~Simmons-Duffin and A.~Vichi, \emph{{Carving Out the Space of 4D
  CFTs}}, \href{http://dx.doi.org/10.1007/JHEP05(2012)110}{\emph{JHEP} {\bf 05}
  (2012) 110}, [\href{https://arxiv.org/abs/1109.5176}{{\tt 1109.5176}}].

\bibitem{Simmons-Duffin:2015qma}
D.~Simmons-Duffin, \emph{{A Semidefinite Program Solver for the Conformal
  Bootstrap}}, \href{http://dx.doi.org/10.1007/JHEP06(2015)174}{\emph{JHEP}
  {\bf 06} (2015) 174}, [\href{https://arxiv.org/abs/1502.02033}{{\tt
  1502.02033}}].

\bibitem{Landry:2019qug}
W.~Landry and D.~Simmons-Duffin, \emph{{Scaling the semidefinite program solver
  SDPB}},  \href{https://arxiv.org/abs/1909.09745}{{\tt 1909.09745}}.

\bibitem{scalarblocks}
C.~Behan, P.~Kravchuk, W.~Landry and D.~Simmons-Duffin, ``{scalar\_blocks}.''
  \url{https://gitlab.com/bootstrapcollaboration/scalar_blocks}, 2019.

\bibitem{Erramilli:2020rlr}
R.~S. Erramilli, L.~V. Iliesiu, P.~Kravchuk, W.~Landry, D.~Poland and
  D.~Simmons-Duffin, \emph{{blocks{\_}3d: software for general 3d conformal
  blocks}}, \href{http://dx.doi.org/10.1007/JHEP11(2021)006}{\emph{JHEP} {\bf
  11} (2021) 006}, [\href{https://arxiv.org/abs/2011.01959}{{\tt 2011.01959}}].

\bibitem{Belavin:1984vu}
A.~A. Belavin, A.~M. Polyakov and A.~B. Zamolodchikov, \emph{{Infinite
  Conformal Symmetry in Two-Dimensional Quantum Field Theory}},
  \href{http://dx.doi.org/10.1016/0550-3213(84)90052-X}{\emph{Nucl. Phys. B}
  {\bf 241} (1984) 333--380}.

\bibitem{Hellerman:2009bu}
S.~Hellerman, \emph{{A Universal Inequality for CFT and Quantum Gravity}},
  \href{http://dx.doi.org/10.1007/JHEP08(2011)130}{\emph{JHEP} {\bf 08} (2011)
  130}, [\href{https://arxiv.org/abs/0902.2790}{{\tt 0902.2790}}].

\bibitem{Friedan:2013cba}
D.~Friedan and C.~A. Keller, \emph{{Constraints on 2d CFT partition
  functions}}, \href{http://dx.doi.org/10.1007/JHEP10(2013)180}{\emph{JHEP}
  {\bf 10} (2013) 180}, [\href{https://arxiv.org/abs/1307.6562}{{\tt
  1307.6562}}].

\bibitem{Collier:2016cls}
S.~Collier, Y.-H. Lin and X.~Yin, \emph{{Modular Bootstrap Revisited}},
  \href{http://dx.doi.org/10.1007/JHEP09(2018)061}{\emph{JHEP} {\bf 09} (2018)
  061}, [\href{https://arxiv.org/abs/1608.06241}{{\tt 1608.06241}}].

\bibitem{Afkhami-Jeddi:2019zci}
N.~Afkhami-Jeddi, T.~Hartman and A.~Tajdini, \emph{{Fast Conformal Bootstrap
  and Constraints on 3d Gravity}},
  \href{http://dx.doi.org/10.1007/JHEP05(2019)087}{\emph{JHEP} {\bf 05} (2019)
  087}, [\href{https://arxiv.org/abs/1903.06272}{{\tt 1903.06272}}].

\bibitem{Hartman:2019pcd}
T.~Hartman, D.~Maz{\'a}{\v{c}} and L.~Rastelli, \emph{{Sphere Packing and
  Quantum Gravity}},
  \href{http://dx.doi.org/10.1007/JHEP12(2019)048}{\emph{JHEP} {\bf 12} (2019)
  048}, [\href{https://arxiv.org/abs/1905.01319}{{\tt 1905.01319}}].

\bibitem{Benjamin:2026lbj}
N.~Benjamin, A.~L. Fitzpatrick, W.~Li and J.~Thaler, \emph{{Descending into the
  modular bootstrap}},
  \href{http://dx.doi.org/10.1007/JHEP09(2026)029}{\emph{JHEP} {\bf 09} (2026)
  029}, [\href{https://arxiv.org/abs/2604.01275}{{\tt 2604.01275}}].

\bibitem{Zamolodchikov:1984eqp}
A.~B. Zamolodchikov, \emph{{Conformal Symmetry in Two-Dimensions: An Explicit
  Recurrence Formula for the Conformal Partial Wave Amplitude}},
  \href{http://dx.doi.org/10.1007/BF01214585}{\emph{Commun. Math. Phys.} {\bf
  96} (1984) 419--422}.

\bibitem{Zamolodchikov:1987avt}
A.~B. Zamolodchikov, \emph{{Conformal symmetry in two-dimensional space:
  Recursion representation of conformal block}},
  \href{http://dx.doi.org/10.1007/BF01022967}{\emph{Theor. Math. Phys.} {\bf
  73} (1987) 1088--1093}.

\bibitem{Esterlis:2016psv}
I.~Esterlis, A.~L. Fitzpatrick and D.~Ramirez, \emph{{Closure of the Operator
  Product Expansion in the Non-Unitary Bootstrap}},
  \href{http://dx.doi.org/10.1007/JHEP11(2016)030}{\emph{JHEP} {\bf 11} (2016)
  030}, [\href{https://arxiv.org/abs/1606.07458}{{\tt 1606.07458}}].

\bibitem{Huang:2025qkk}
Y.-t. Huang, S.-C. Lee, H.~Liao and J.~Rumbutis, \emph{{Bootstrapping
  non-unitary CFTs}},  \href{https://arxiv.org/abs/2512.07706}{{\tt
  2512.07706}}.

\bibitem{Picco:2016ilr}
M.~Picco, S.~Ribault and R.~Santachiara, \emph{{A conformal bootstrap approach
  to critical percolation in two dimensions}},
  \href{http://dx.doi.org/10.21468/SciPostPhys.1.1.009}{\emph{SciPost Phys.}
  {\bf 1} (2016) 009}, [\href{https://arxiv.org/abs/1607.07224}{{\tt
  1607.07224}}].

\bibitem{He:2020rfk}
Y.~He, J.~L. Jacobsen and H.~Saleur, \emph{{Geometrical four-point functions in
  the two-dimensional critical $Q$-state Potts model: The interchiral conformal
  bootstrap}}, \href{http://dx.doi.org/10.1007/JHEP12(2020)019}{\emph{JHEP}
  {\bf 12} (2020) 019}, [\href{https://arxiv.org/abs/2005.07258}{{\tt
  2005.07258}}].

\bibitem{Nivesvivat:2023kfp}
R.~Nivesvivat, S.~Ribault and J.~L. Jacobsen, \emph{{Critical loop models are
  exactly solvable}},
  \href{http://dx.doi.org/10.21468/SciPostPhys.17.2.029}{\emph{SciPost Phys.}
  {\bf 17} (2024) 029}, [\href{https://arxiv.org/abs/2311.17558}{{\tt
  2311.17558}}].

\bibitem{Lin:2015wcg}
Y.-H. Lin, S.-H. Shao, D.~Simmons-Duffin, Y.~Wang and X.~Yin, \emph{{$
  \mathcal{N} $ = 4 superconformal bootstrap of the K3 CFT}},
  \href{http://dx.doi.org/10.1007/JHEP05(2017)126}{\emph{JHEP} {\bf 05} (2017)
  126}, [\href{https://arxiv.org/abs/1511.04065}{{\tt 1511.04065}}].

\bibitem{Lin:2016gcl}
Y.-H. Lin, S.-H. Shao, Y.~Wang and X.~Yin, \emph{{(2, 2) superconformal
  bootstrap in two dimensions}},
  \href{http://dx.doi.org/10.1007/JHEP05(2017)112}{\emph{JHEP} {\bf 05} (2017)
  112}, [\href{https://arxiv.org/abs/1610.05371}{{\tt 1610.05371}}].

\bibitem{Collier:2017shs}
S.~Collier, P.~Kravchuk, Y.-H. Lin and X.~Yin, \emph{{Bootstrapping the
  Spectral Function: On the Uniqueness of Liouville and the Universality of
  BTZ}}, \href{http://dx.doi.org/10.1007/JHEP09(2018)150}{\emph{JHEP} {\bf 09}
  (2018) 150}, [\href{https://arxiv.org/abs/1702.00423}{{\tt 1702.00423}}].

\bibitem{Besken:2021eps}
M.~Besken, \emph{{Virasoro conformal bootstrap with $c > 1$}},
  \href{https://arxiv.org/abs/2106.08999}{{\tt 2106.08999}}.

\bibitem{Ribault:2014hia}
S.~Ribault, \emph{{Conformal field theory on the plane}},
  \href{https://arxiv.org/abs/1406.4290}{{\tt 1406.4290}}.

\bibitem{Ribault:2024rvk}
S.~Ribault, \emph{{Exactly solvable conformal field theories}},
  \href{http://dx.doi.org/10.21468/SciPostPhysLectNotes.130}{\emph{SciPost
  Phys. Lect. Notes} {\bf 130} (2026) 1},
  [\href{https://arxiv.org/abs/2411.17262}{{\tt 2411.17262}}].

\bibitem{Perlmutter:2015iya}
E.~Perlmutter, \emph{{Virasoro conformal blocks in closed form}},
  \href{http://dx.doi.org/10.1007/JHEP08(2015)088}{\emph{JHEP} {\bf 08} (2015)
  088}, [\href{https://arxiv.org/abs/1502.07742}{{\tt 1502.07742}}].

\bibitem{Cho:2017oxl}
M.~Cho, S.~Collier and X.~Yin, \emph{{Recursive Representations of Arbitrary
  Virasoro Conformal Blocks}},
  \href{http://dx.doi.org/10.1007/JHEP04(2019)018}{\emph{JHEP} {\bf 04} (2019)
  018}, [\href{https://arxiv.org/abs/1703.09805}{{\tt 1703.09805}}].

\bibitem{Kos:2013tga}
F.~Kos, D.~Poland and D.~Simmons-Duffin, \emph{{Bootstrapping the $O(N)$ vector
  models}}, \href{http://dx.doi.org/10.1007/JHEP06(2014)091}{\emph{JHEP} {\bf
  06} (2014) 091}, [\href{https://arxiv.org/abs/1307.6856}{{\tt 1307.6856}}].

\bibitem{Chang:2025mwt}
C.-H. Chang, V.~Dommes, P.~Kravchuk, D.~Poland and D.~Simmons-Duffin,
  \emph{{Accurate bootstrap bounds from optimal interpolation}},
  \href{http://dx.doi.org/10.1007/JHEP05(2026)275}{\emph{JHEP} {\bf 05} (2026)
  275}, [\href{https://arxiv.org/abs/2509.14307}{{\tt 2509.14307}}].

\bibitem{Antunes:2022vtb}
A.~Antunes and C.~Behan, \emph{{Coupled Minimal Conformal Field Theory Models
  Revisited}},
  \href{http://dx.doi.org/10.1103/PhysRevLett.130.071602}{\emph{Phys. Rev.
  Lett.} {\bf 130} (2023) 071602},
  [\href{https://arxiv.org/abs/2211.16503}{{\tt 2211.16503}}].

\bibitem{Antunes:2024mfb}
A.~Antunes and C.~Behan, \emph{{Coupled minimal models revisited II:
  Constraints from permutation symmetry}},
  \href{http://dx.doi.org/10.21468/SciPostPhys.18.4.132}{\emph{SciPost Phys.}
  {\bf 18} (2025) 132}, [\href{https://arxiv.org/abs/2412.21107}{{\tt
  2412.21107}}].

\bibitem{Ludwig:1987gs}
A.~W.~W. Ludwig and J.~L. Cardy, \emph{{Perturbative Evaluation of the
  Conformal Anomaly at New Critical Points with Applications to Random
  Systems}}, \href{http://dx.doi.org/10.1016/0550-3213(87)90362-2}{\emph{Nucl.
  Phys. B} {\bf 285} (1987) 687--718}.

\bibitem{Dotsenko:1998gyp}
V.~Dotsenko, J.~L. Jacobsen, M.-A. Lewis and M.~Picco, \emph{{Coupled Potts
  models: Self-duality and fixed point structure}},
  \href{http://dx.doi.org/10.1016/S0550-3213(99)00097-8}{\emph{Nucl. Phys. B}
  {\bf 546} (1999) 505--557},
  [\href{https://arxiv.org/abs/cond-mat/9812227}{{\tt cond-mat/9812227}}].

\bibitem{Fendley:2008fp}
P.~Fendley and J.~L. Jacobsen, \emph{{Critical points in coupled Potts models
  and critical phases in coupled loop models}},
  \href{http://dx.doi.org/10.1088/1751-8113/41/21/215001}{\emph{J. Phys. A}
  {\bf 41} (2008) 215001}, [\href{https://arxiv.org/abs/0803.2618}{{\tt
  0803.2618}}].

\bibitem{Antunes:2025erb}
A.~Antunes and N.~Suchel, \emph{{Taxonomy of coupled minimal models from finite
  groups}},  \href{https://arxiv.org/abs/2512.23664}{{\tt 2512.23664}}.

\bibitem{Antunes:2025huk}
A.~Antunes and J.~Rong, \emph{{Irrational CFTs from coupled anyon chains with
  non-invertible symmetries?}},
  \href{http://dx.doi.org/10.21468/SciPostPhys.21.1.005}{\emph{SciPost Phys.}
  {\bf 21} (2026) 005}, [\href{https://arxiv.org/abs/2507.14280}{{\tt
  2507.14280}}].

\bibitem{mpmath}
F.~Johansson et~al., ``{mpmath: a Python library for arbitrary-precision
  floating-point arithmetic (version 1.3.0)}.'' \url{https://mpmath.org/},
  2023.

\bibitem{Hart:2010flint}
W.~B. Hart, \emph{{Fast Library for Number Theory: An Introduction}},  in
  \emph{Mathematical Software -- ICMS 2010}, vol.~6327 of \emph{Lecture Notes
  in Computer Science}, pp.~88--91, Springer, 2010.
\newblock \href{http://dx.doi.org/10.1007/978-3-642-15582-6_18}{DOI}.

\bibitem{Johansson:2017arb}
F.~Johansson, \emph{{Arb: efficient arbitrary-precision midpoint-radius
  interval arithmetic}},
  \href{http://dx.doi.org/10.1109/TC.2017.2690633}{\emph{IEEE Transactions on
  Computers} {\bf 66} (2017) 1281--1292}.

\bibitem{pythonflint}
F.~Johansson et~al., ``{python-flint: Python bindings for FLINT and Arb}.''
  \url{https://github.com/flintlib/python-flint}, 2024.

\bibitem{Hadasz:2006qb}
L.~Hadasz, Z.~Jaskolski and P.~Suchanek, \emph{{Recursion representation of the
  Neveu-Schwarz superconformal block}},
  \href{http://dx.doi.org/10.1088/1126-6708/2007/03/032}{\emph{JHEP} {\bf 03}
  (2007) 032}, [\href{https://arxiv.org/abs/hep-th/0611266}{{\tt
  hep-th/0611266}}].

\bibitem{Hadasz:2007nt}
L.~Hadasz, Z.~Jaskolski and P.~Suchanek, \emph{{Elliptic recurrence
  representation of the N = 1 Neveu-Schwarz blocks}},
  \href{http://dx.doi.org/10.1016/j.nuclphysb.2007.12.015}{\emph{Nucl. Phys. B}
  {\bf 798} (2008) 363--378}, [\href{https://arxiv.org/abs/0711.1619}{{\tt
  0711.1619}}].

\bibitem{Belavin:2007gz}
A.~Belavin, V.~Belavin, A.~Neveu and A.~Zamolodchikov, \emph{{Bootstrap in
  Supersymmetric Liouville Field Theory. I. NS Sector}},
  \href{http://dx.doi.org/10.1016/j.nuclphysb.2007.04.018}{\emph{Nucl. Phys. B}
  {\bf 784} (2007) 202--233}, [\href{https://arxiv.org/abs/hep-th/0703084}{{\tt
  hep-th/0703084}}].

\bibitem{Hadasz:2008dt}
L.~Hadasz, Z.~Jaskolski and P.~Suchanek, \emph{{Elliptic recurrence
  representation of the N=1 superconformal blocks in the Ramond sector}},
  \href{http://dx.doi.org/10.1088/1126-6708/2008/11/060}{\emph{JHEP} {\bf 11}
  (2008) 060}, [\href{https://arxiv.org/abs/0810.1203}{{\tt 0810.1203}}].

\bibitem{Hadasz:2009db}
L.~Hadasz, Z.~Jaskolski and P.~Suchanek, \emph{{Recursive representation of the
  torus 1-point conformal block}},
  \href{http://dx.doi.org/10.1007/JHEP01(2010)063}{\emph{JHEP} {\bf 01} (2010)
  063}, [\href{https://arxiv.org/abs/0911.2353}{{\tt 0911.2353}}].

\bibitem{Poghossian:2009mk}
R.~Poghossian, \emph{{Recursion relations in CFT and N=2 SYM theory}},
  \href{http://dx.doi.org/10.1088/1126-6708/2009/12/038}{\emph{JHEP} {\bf 12}
  (2009) 038}, [\href{https://arxiv.org/abs/0909.3412}{{\tt 0909.3412}}].

\bibitem{Chiang:2026nmd}
L.-Y. Chiang, D.~Poland and G.~Rogelberg, \emph{{Moments in the CFT
  Landscape}},  \href{https://arxiv.org/abs/2603.18140}{{\tt 2603.18140}}.

\bibitem{Poland:2024hvb}
D.~Poland and G.~Rogelberg, \emph{{Moments and saddles of heavy CFT
  correlators}}, \href{http://dx.doi.org/10.1007/JHEP10(2025)100}{\emph{JHEP}
  {\bf 10} (2025) 100}, [\href{https://arxiv.org/abs/2501.00092}{{\tt
  2501.00092}}].

\bibitem{Maldacena:2015iua}
J.~Maldacena, D.~Simmons-Duffin and A.~Zhiboedov, \emph{{Looking for a bulk
  point}}, \href{http://dx.doi.org/10.1007/JHEP01(2017)013}{\emph{JHEP} {\bf
  01} (2017) 013}, [\href{https://arxiv.org/abs/1509.03612}{{\tt 1509.03612}}].

\bibitem{Ghosal:2020sis}
P.~Ghosal, G.~Remy, X.~Sun and Y.~Sun, \emph{{Probabilistic conformal blocks
  for Liouville CFT on the torus}},
  \href{http://dx.doi.org/10.1215/00127094-2023-0031}{\emph{Duke Math. J.} {\bf
  173} (2024) 1085--1175}, [\href{https://arxiv.org/abs/2003.03802}{{\tt
  2003.03802}}].

\bibitem{Kac:1978ge}
V.~G. Kac, \emph{{Contravariant Form for Infinite Dimensional Lie Algebras and
  Superalgebras}},  in \emph{{7th International Group Theory Colloquium: The
  Integrative Conference on Group Theory and Mathematical Physics}},
  pp.~441--445, 1978.

\bibitem{Feigin:1988se}
B.~L. Feigin and D.~B. Fuchs, \emph{{Verma modules over the Virasoro algebra}},
  {\emph{Lect. Notes Math.} {\bf 1060} (1984) 230--245}.

\bibitem{Collier:2018exn}
S.~Collier, Y.~Gobeil, H.~Maxfield and E.~Perlmutter, \emph{{Quantum Regge
  Trajectories and the Virasoro Analytic Bootstrap}},
  \href{http://dx.doi.org/10.1007/JHEP05(2019)212}{\emph{JHEP} {\bf 05} (2019)
  212}, [\href{https://arxiv.org/abs/1811.05710}{{\tt 1811.05710}}].

\bibitem{Dorn:1994xn}
H.~Dorn and H.~J. Otto, \emph{{Two and three point functions in Liouville
  theory}}, \href{http://dx.doi.org/10.1016/0550-3213(94)00352-1}{\emph{Nucl.
  Phys. B} {\bf 429} (1994) 375--388},
  [\href{https://arxiv.org/abs/hep-th/9403141}{{\tt hep-th/9403141}}].

\bibitem{Zamolodchikov:1995aa}
A.~B. Zamolodchikov and A.~B. Zamolodchikov, \emph{{Structure constants and
  conformal bootstrap in Liouville field theory}},
  \href{http://dx.doi.org/10.1016/0550-3213(96)00351-3}{\emph{Nucl. Phys. B}
  {\bf 477} (1996) 577--605}, [\href{https://arxiv.org/abs/hep-th/9506136}{{\tt
  hep-th/9506136}}].

\bibitem{Friedan:1983xq}
D.~Friedan, Z.-a. Qiu and S.~H. Shenker, \emph{{Conformal Invariance, Unitarity
  and Two-Dimensional Critical Exponents}},
  \href{http://dx.doi.org/10.1103/PhysRevLett.52.1575}{\emph{Phys. Rev. Lett.}
  {\bf 52} (1984) 1575--1578}.

\bibitem{Goddard:1986ee}
P.~Goddard, A.~Kent and D.~I. Olive, \emph{{Unitary Representations of the
  Virasoro and Supervirasoro Algebras}},
  \href{http://dx.doi.org/10.1007/BF01464283}{\emph{Commun. Math. Phys.} {\bf
  103} (1986) 105--119}.

\bibitem{SaffTotik:1997}
E.~B. Saff and V.~Totik, \emph{{Logarithmic Potentials with External Fields}},
  vol.~316 of \emph{Grundlehren der mathematischen Wissenschaften}.
\newblock Springer, Berlin, 1997,
  \href{http://dx.doi.org/10.1007/978-3-662-03329-6}{10.1007/978-3-662-03329-6}.

\bibitem{Hogervorst:2013sma}
M.~Hogervorst and S.~Rychkov, \emph{{Radial Coordinates for Conformal Blocks}},
  \href{http://dx.doi.org/10.1103/PhysRevD.87.106004}{\emph{Phys. Rev. D} {\bf
  87} (2013) 106004}, [\href{https://arxiv.org/abs/1303.1111}{{\tt
  1303.1111}}].

\bibitem{Mazac:2016qev}
D.~Mazac, \emph{{Analytic bounds and emergence of AdS$_{2}$ physics from the
  conformal bootstrap}},
  \href{http://dx.doi.org/10.1007/JHEP04(2017)146}{\emph{JHEP} {\bf 04} (2017)
  146}, [\href{https://arxiv.org/abs/1611.10060}{{\tt 1611.10060}}].

\bibitem{Harris:2020numpy}
C.~R. Harris et~al., \emph{{Array programming with NumPy}},
  \href{http://dx.doi.org/10.1038/s41586-020-2649-2}{\emph{Nature} {\bf 585}
  (2020) 357--362}, [\href{https://arxiv.org/abs/2006.10256}{{\tt
  2006.10256}}].

\bibitem{Maloney:2016kee}
A.~Maloney, H.~Maxfield and G.~S. Ng, \emph{{A conformal block Farey tail}},
  \href{http://dx.doi.org/10.1007/JHEP06(2017)117}{\emph{JHEP} {\bf 06} (2017)
  117}, [\href{https://arxiv.org/abs/1609.02165}{{\tt 1609.02165}}].

\bibitem{Chen:2017yze}
H.~Chen, C.~Hussong, J.~Kaplan and D.~Li, \emph{{A Numerical Approach to
  Virasoro Blocks and the Information Paradox}},
  \href{http://dx.doi.org/10.1007/JHEP09(2017)102}{\emph{JHEP} {\bf 09} (2017)
  102}, [\href{https://arxiv.org/abs/1703.09727}{{\tt 1703.09727}}].

\bibitem{Paulos:2021jxx}
M.~F. Paulos and Z.~Zheng, \emph{{Bounding 3d CFT correlators}},
  \href{http://dx.doi.org/10.1007/JHEP04(2022)102}{\emph{JHEP} {\bf 04} (2022)
  102}, [\href{https://arxiv.org/abs/2107.01215}{{\tt 2107.01215}}].

\bibitem{Boucher:1986bh}
W.~Boucher, D.~Friedan and A.~Kent, \emph{{Determinant Formulae and Unitarity
  for the N=2 Superconformal Algebras in Two-Dimensions or Exact Results on
  String Compactification}},
  \href{http://dx.doi.org/10.1016/0370-2693(86)90260-1}{\emph{Phys. Lett. B}
  {\bf 172} (1986) 316}.

\bibitem{Gepner:1986hr}
D.~Gepner and Z.-a. Qiu, \emph{{Modular Invariant Partition Functions for
  Parafermionic Field Theories}},
  \href{http://dx.doi.org/10.1016/0550-3213(87)90348-8}{\emph{Nucl. Phys. B}
  {\bf 285} (1987) 423}.

\bibitem{Cappelli:1986hf}
A.~Cappelli, C.~Itzykson and J.~B. Zuber, \emph{{Modular invariant partition
  functions in two dimensions}},
  \href{http://dx.doi.org/10.1016/0550-3213(87)90155-6}{\emph{Nucl. Phys. B}
  {\bf 280} (1987) 445--465}.

\bibitem{Dixon:1988qd}
L.~J. Dixon, P.~H. Ginsparg and J.~A. Harvey, \emph{{Beauty and the Beast:
  Superconformal Symmetry in a Monster Module}},
  \href{http://dx.doi.org/10.1007/BF01217740}{\emph{Commun. Math. Phys.} {\bf
  119} (1988) 221--241}.

\bibitem{Zamolodchikov:1985wn}
A.~B. Zamolodchikov, \emph{{Infinite Additional Symmetries in Two-Dimensional
  Conformal Quantum Field Theory}},
  \href{http://dx.doi.org/10.1007/BF01036128}{\emph{Theor. Math. Phys.} {\bf
  65} (1985) 1205--1213}.

\bibitem{Gepner:1987sm}
D.~Gepner, \emph{{New Conformal Field Theories Associated with Lie Algebras and
  their Partition Functions}},
  \href{http://dx.doi.org/10.1016/0550-3213(87)90176-3}{\emph{Nucl. Phys. B}
  {\bf 290} (1987) 10--24}.

\bibitem{Davenport:2012fcs}
S.~C. Davenport, E.~Ardonne, N.~Regnault and S.~H. Simon, \emph{{Spin-singlet
  Gaffnian wave function for fractional quantum Hall systems}},
  \href{http://dx.doi.org/10.1103/PhysRevB.87.045310}{\emph{Phys. Rev. B} {\bf
  87} (2013) 045310}, [\href{https://arxiv.org/abs/1210.8143}{{\tt
  1210.8143}}].

\bibitem{Bajnok:2016ocb}
Z.~Bajnok, J.~Balog, K.~Ito, Y.~Satoh and G.~Z. Toth, \emph{{On the
  mass-coupling relation of multi-scale quantum integrable models}},
  \href{http://dx.doi.org/10.1007/JHEP06(2016)071}{\emph{JHEP} {\bf 06} (2016)
  071}, [\href{https://arxiv.org/abs/1604.02811}{{\tt 1604.02811}}].

\bibitem{Antunes:2026kqb}
A.~Antunes and J.~Rong, \emph{{A new Ising/tricritical-Ising interface: From
  ${W}_3$ symmetry to Rydberg atoms}},
  \href{https://arxiv.org/abs/2605.23848}{{\tt 2605.23848}}.

\bibitem{Friedan:1984rv}
D.~Friedan, Z.-a. Qiu and S.~H. Shenker, \emph{{Superconformal Invariance in
  Two-Dimensions and the Tricritical Ising Model}},
  \href{http://dx.doi.org/10.1016/0370-2693(85)90819-6}{\emph{Phys. Lett. B}
  {\bf 151} (1985) 37--43}.

\bibitem{Bershadsky:1985dq}
M.~A. Bershadsky, V.~G. Knizhnik and M.~G. Teitelman, \emph{{Superconformal
  Symmetry in Two-Dimensions}},
  \href{http://dx.doi.org/10.1016/0370-2693(85)90818-4}{\emph{Phys. Lett. B}
  {\bf 151} (1985) 31--36}.

\bibitem{Eichenherr:1985cx}
H.~Eichenherr, \emph{{Minimal Operator Algebras in Superconformal Quantum Field
  Theory}}, \href{http://dx.doi.org/10.1016/0370-2693(85)90817-2}{\emph{Phys.
  Lett. B} {\bf 151} (1985) 26--30}.

\bibitem{Fateev:1985mm}
V.~A. Fateev and A.~B. Zamolodchikov, \emph{{Parafermionic Currents in the
  Two-Dimensional Conformal Quantum Field Theory and Selfdual Critical Points
  in Z(n) Invariant Statistical Systems}}, {\emph{Sov. Phys. JETP} {\bf 62}
  (1985) 215--225}.

\bibitem{Ginsparg:1987eb}
P.~H. Ginsparg, \emph{{Curiosities at c = 1}},
  \href{http://dx.doi.org/10.1016/0550-3213(88)90249-0}{\emph{Nucl. Phys. B}
  {\bf 295} (1988) 153--170}.

\bibitem{Dijkgraaf:1987vp}
R.~Dijkgraaf, E.~P. Verlinde and H.~L. Verlinde, \emph{{C = 1 Conformal Field
  Theories on Riemann Surfaces}},
  \href{http://dx.doi.org/10.1007/BF01224132}{\emph{Commun. Math. Phys.} {\bf
  115} (1988) 649--690}.

\bibitem{Ginsparg:1988ui}
P.~H. Ginsparg, \emph{{Applied Conformal Field Theory}},  in \emph{{Les Houches
  Summer School in Theoretical Physics: Fields, Strings, Critical Phenomena}},
  9, 1988.
\newblock \href{https://arxiv.org/abs/hep-th/9108028}{{\tt hep-th/9108028}}.

\bibitem{Thorngren:2021yso}
R.~Thorngren and Y.~Wang, \emph{{Fusion category symmetry. Part II.
  Categoriosities at c = 1 and beyond}},
  \href{http://dx.doi.org/10.1007/JHEP07(2024)051}{\emph{JHEP} {\bf 07} (2024)
  051}, [\href{https://arxiv.org/abs/2106.12577}{{\tt 2106.12577}}].

\bibitem{Runkel:2001ng}
I.~Runkel and G.~M.~T. Watts, \emph{{A Nonrational CFT with c = 1 as a limit of
  minimal models}},
  \href{http://dx.doi.org/10.1088/1126-6708/2001/09/006}{\emph{JHEP} {\bf 09}
  (2001) 006}, [\href{https://arxiv.org/abs/hep-th/0107118}{{\tt
  hep-th/0107118}}].

\bibitem{Schomerus:2003vv}
V.~Schomerus, \emph{{Rolling tachyons from Liouville theory}},
  \href{http://dx.doi.org/10.1088/1126-6708/2003/11/043}{\emph{JHEP} {\bf 11}
  (2003) 043}, [\href{https://arxiv.org/abs/hep-th/0306026}{{\tt
  hep-th/0306026}}].

\bibitem{Ribault:2015sxa}
S.~Ribault and R.~Santachiara, \emph{{Liouville theory with a central charge
  less than one}}, \href{http://dx.doi.org/10.1007/JHEP08(2015)109}{\emph{JHEP}
  {\bf 08} (2015) 109}, [\href{https://arxiv.org/abs/1503.02067}{{\tt
  1503.02067}}].

\bibitem{Kousvos:2024dlz}
S.~R. Kousvos, A.~Piazza and A.~Vichi, \emph{{Exploring replica-Potts CFTs in
  two dimensions}},
  \href{http://dx.doi.org/10.1007/JHEP11(2024)030}{\emph{JHEP} {\bf 11} (2024)
  030}, [\href{https://arxiv.org/abs/2405.19416}{{\tt 2405.19416}}].

\bibitem{Poland:2026uiy}
D.~Poland and G.~Rogelberg, \emph{{Maximally heavy dynamics in the causal
  diamond}},  \href{https://arxiv.org/abs/2603.28853}{{\tt 2603.28853}}.

\bibitem{Hartman:2014oaa}
T.~Hartman, C.~A. Keller and B.~Stoica, \emph{{Universal Spectrum of 2d
  Conformal Field Theory in the Large c Limit}},
  \href{http://dx.doi.org/10.1007/JHEP09(2014)118}{\emph{JHEP} {\bf 09} (2014)
  118}, [\href{https://arxiv.org/abs/1405.5137}{{\tt 1405.5137}}].

\bibitem{Collier:2019weq}
S.~Collier, A.~Maloney, H.~Maxfield and I.~Tsiares, \emph{{Universal dynamics
  of heavy operators in CFT$_{2}$}},
  \href{http://dx.doi.org/10.1007/JHEP07(2020)074}{\emph{JHEP} {\bf 07} (2020)
  074}, [\href{https://arxiv.org/abs/1912.00222}{{\tt 1912.00222}}].

\bibitem{Benjamin:2023qsc}
N.~Benjamin, J.~Lee, H.~Ooguri and D.~Simmons-Duffin, \emph{{Universal
  asymptotics for high energy CFT data}},
  \href{http://dx.doi.org/10.1007/JHEP03(2024)115}{\emph{JHEP} {\bf 03} (2024)
  115}, [\href{https://arxiv.org/abs/2306.08031}{{\tt 2306.08031}}].

\bibitem{Kraus:2018pax}
P.~Kraus and A.~Sivaramakrishnan, \emph{{Light-state Dominance from the
  Conformal Bootstrap}},
  \href{http://dx.doi.org/10.1007/JHEP08(2019)013}{\emph{JHEP} {\bf 08} (2019)
  013}, [\href{https://arxiv.org/abs/1812.02226}{{\tt 1812.02226}}].

\bibitem{Fitzpatrick:2014vua}
A.~L. Fitzpatrick, J.~Kaplan and M.~T. Walters, \emph{{Universality of
  Long-Distance AdS Physics from the CFT Bootstrap}},
  \href{http://dx.doi.org/10.1007/JHEP08(2014)145}{\emph{JHEP} {\bf 08} (2014)
  145}, [\href{https://arxiv.org/abs/1403.6829}{{\tt 1403.6829}}].

\bibitem{Grabovsky:2024jwf}
D.~Grabovsky, \emph{{Heavy states in 3d gravity and 2d CFT}},
  \href{http://dx.doi.org/10.1007/JHEP07(2024)287}{\emph{JHEP} {\bf 07} (2024)
  287}, [\href{https://arxiv.org/abs/2403.13757}{{\tt 2403.13757}}].

\bibitem{Abajian:2023jye}
J.~Abajian, F.~Aprile, R.~C. Myers and P.~Vieira, \emph{{Holography and
  correlation functions of huge operators: spacetime bananas}},
  \href{http://dx.doi.org/10.1007/JHEP12(2023)058}{\emph{JHEP} {\bf 12} (2023)
  058}, [\href{https://arxiv.org/abs/2306.15105}{{\tt 2306.15105}}].

\bibitem{Zamolodchikov:2005fy}
A.~B. Zamolodchikov, \emph{{Three-point function in the minimal Liouville
  gravity}}, \href{http://dx.doi.org/10.1007/s11232-005-0003-3}{\emph{Theor.
  Math. Phys.} {\bf 142} (2005) 183--196},
  [\href{https://arxiv.org/abs/hep-th/0505063}{{\tt hep-th/0505063}}].

\bibitem{Dijkgraaf:1989hb}
R.~Dijkgraaf, C.~Vafa, E.~P. Verlinde and H.~L. Verlinde, \emph{{The Operator
  Algebra of Orbifold Models}},
  \href{http://dx.doi.org/10.1007/BF01238812}{\emph{Commun. Math. Phys.} {\bf
  123} (1989) 485}.

\bibitem{Gamayun:2012ma}
O.~Gamayun, N.~Iorgov and O.~Lisovyy, \emph{{Conformal field theory of
  Painlev{\'e} VI}},
  \href{http://dx.doi.org/10.1007/JHEP10(2012)038}{\emph{JHEP} {\bf 10} (2012)
  038}, [\href{https://arxiv.org/abs/1207.0787}{{\tt 1207.0787}}].

\bibitem{Gavrylenko:2018cai}
P.~Gavrylenko and R.~Santachiara, \emph{{Crossing invariant correlation
  functions at $c=1$ from isomonodromic $\tau$ functions}},
  \href{http://dx.doi.org/10.1007/JHEP11(2019)119}{\emph{JHEP} {\bf 11} (2019)
  119}, [\href{https://arxiv.org/abs/1812.10362}{{\tt 1812.10362}}].

\bibitem{Knizhnik:1984nr}
V.~G. Knizhnik and A.~B. Zamolodchikov, \emph{{Current Algebra and Wess-Zumino
  Model in Two-Dimensions}},
  \href{http://dx.doi.org/10.1016/0550-3213(84)90374-2}{\emph{Nucl. Phys. B}
  {\bf 247} (1984) 83--103}.

\bibitem{Dixon:1985jw}
L.~J. Dixon, J.~A. Harvey, C.~Vafa and E.~Witten, \emph{{Strings on
  Orbifolds}},
  \href{http://dx.doi.org/10.1016/0550-3213(85)90593-0}{\emph{Nucl. Phys. B}
  {\bf 261} (1985) 678--686}.

\bibitem{Dixon:1986jc}
L.~J. Dixon, J.~A. Harvey, C.~Vafa and E.~Witten, \emph{{Strings on Orbifolds.
  2.}}, \href{http://dx.doi.org/10.1016/0550-3213(86)90287-7}{\emph{Nucl. Phys.
  B} {\bf 274} (1986) 285--314}.

\bibitem{Lunin:2000yv}
O.~Lunin and S.~D. Mathur, \emph{{Correlation functions for M**N / S(N)
  orbifolds}}, \href{http://dx.doi.org/10.1007/s002200100431}{\emph{Commun.
  Math. Phys.} {\bf 219} (2001) 399--442},
  [\href{https://arxiv.org/abs/hep-th/0006196}{{\tt hep-th/0006196}}].

\bibitem{Borisov:1997nc}
L.~Borisov, M.~B. Halpern and C.~Schweigert, \emph{{Systematic approach to
  cyclic orbifolds}},
  \href{http://dx.doi.org/10.1142/S0217751X98000044}{\emph{Int. J. Mod. Phys.
  A} {\bf 13} (1998) 125--168},
  [\href{https://arxiv.org/abs/hep-th/9701061}{{\tt hep-th/9701061}}].

\bibitem{Goddard:1984vk}
P.~Goddard, A.~Kent and D.~I. Olive, \emph{{Virasoro Algebras and Coset Space
  Models}}, \href{http://dx.doi.org/10.1016/0370-2693(85)91145-1}{\emph{Phys.
  Lett. B} {\bf 152} (1985) 88--92}.

\bibitem{Vafa:1986wx}
C.~Vafa, \emph{{Modular Invariance and Discrete Torsion on Orbifolds}},
  \href{http://dx.doi.org/10.1016/0550-3213(86)90379-2}{\emph{Nucl. Phys. B}
  {\bf 273} (1986) 592--606}.

\end{thebibliography}\endgroup

\end{document}